\documentclass[%
 reprint,
 amsmath,amssymb,
 aps,
]{revtex4-1}

\usepackage{graphicx} 
\usepackage{dcolumn}
\usepackage{bm}
\usepackage[font=small,format=plain,labelformat=simple,labelsep=period,justification=raggedright]{caption}
\usepackage{amsmath}
\usepackage{here}
\usepackage{natbib}
\usepackage{color}
\usepackage{url}
\usepackage{comment}
\usepackage[english]{babel}
\usepackage{caption}

\begin{document}
\preprint{APS/123-QED}
\title{Metastability due to a branching-merging structure in a simple network of an exclusion process}

\author{Hiroki Yamamoto$^{1}$}
\email{h18m1140@hirosaki-u.ac.jp}
\thanks{}%
\author{Daichi Yanagisawa$^{2}$}%
\author{Katsuhiro Nishinari$^{2,3}$}%
\affiliation{%
$^{1}$ School of Medicine, Hirosaki University, 5 Zaifu-cho Hirosaki city, Aomori, 036-8562, Japan\\
$^{2}$ Department of Aeronautics and Astronautics, School of Engineering, The University of Tokyo,\\
7-3-1 Hongo, Bunkyo-ku, Tokyo 113-8656, Japan\\
$^{3}$ Research Center for Advanced Science and Technology, The University of Tokyo,\\
4-6-1 Komaba, Meguro-ku, Tokyo 153-8904, Japan
}%

\date{\today}

\begin{abstract}
We investigate a simple network, which has a branching-merging structure, using the totally asymmetric simple exclusion process, considering conflicts at the merging point. For both periodic and open boundary conditions, the system exhibits metastability. Specifically, for open boundary conditions, we observe two types of metastability: hysteresis and a nonergodic phase. We analytically determine the tipping points, that is, the critical conditions under which a small disturbance can lead to the collapse of metastability. Our findings provide novel insights into metastability induced by branching-merging structures, which exist in all network systems in various fields. 
\end{abstract}

\maketitle


\section{INTRODOCTION}
In many systems, such as ecological networks, climates, politics, financial markets, and traffic networks, multiple states, which are either undesired or positive, have been identified~\cite{doi:10.1126/science.1225244, doi:10.1073/pnas.1907493117,macy2021polarization}. Understanding the transition between states is essential to anticipate undesired or positive changes. Various indicators have been proposed to detect tipping points, where a small perturbation can induce a drastic change between states. An example of a tipping point in physics is the phase transition point.


The totally asymmetric simple exclusion process (TASEP) is a paradigmatic model in the field of nonequilibrium statistical physics that has been investigated in various fields~\cite{chowdhury2000statistical,chowdhury2005physics,schadschneider2010stochastic} since being introduced by MacDonald and Gibbs~\cite{macdonald1968kinetics,macdonald1969concerning}. 
The TASEP exhibits a phase transition between low-density (LD) and high-density (HD) phases with a jump in density. This transition occurs when the input probability (or rate) $\alpha$ is equal to output probability (or rate) $\beta$; that is, $\alpha=\beta$ is considered as a tipping point. However, during the phase transition, the flow, which is a significant performance measure, does not change discontinuously. 

Thus far, in the extensions of the TASEP and related models~\cite{PhysRevE.55.5597,barlovic1998metastable,PhysRevLett.86.2498,PhysRevE.66.046113,PhysRevE.68.026135,Nishinari_2004,PhysRevE.72.035102,nishimura2006metastable,Sakai_2006,Kanai_2006,moussa2007metastable,HU2007397,zhu2007modified,Nishinari_2010, MIURA2020125152, PhysRevE.79.036104,NAKATA20105353} exhibiting metastability, a jump in flow has been observed. The metastable states observed in these studies were dependent on the initial conditions; therefore, metastability in the TASEP can also be interpreted as nonergodicity~\cite{PhysRevLett.91.238302,PhysRevLett.89.090601,SCHULTENS2015100}. Metastability (nonergodicity) can lead to hysteresis. The critical point of collapse of metastability can also be considered a tipping point.


However, to the best of our knowledge, most previous studies used slow-to-start rules~\cite{takayasu19931,benjamin1996cellular,schadschneider1997traffic}, which consider the delay in restarting a blocked particle, and other similar rules to represent metastability. In addition, these studies considered periodic boundary conditions (PBCs), not open boundary conditions (OBCs). 

To date, only few OBC models have succeeded in presenting metastability with relatively complex rules. This is because the stochastic elements $(\alpha, \beta)$ associated with OBCs make it difficult to maintain metastable states. Ref.~\cite{PhysRevLett.91.238302} identified a phase transition between two different long-lived metastable states in the TASEP with Langmuir kinetics for OBCs. Similar phenomena were also observed in Ref.~\cite{PhysRevLett.89.090601}. Ref.~\cite{SCHULTENS2015100} revealed the dependence of the system length on the initial length using a queuing model incorporating excluded volume effects and Langmuir kinetics. In addition, Refs.~\cite{PhysRevE.79.036104,NAKATA20105353} investigated a dilemma game at a bottleneck and succeeded in reproducing a metastable phase for OBCs.




In this paper, we investigate metastability in a TASEP with a branching-merging structure. The main difference between the present model and previous similar models~\cite{PhysRevE.69.066128,Pronina_2005,PhysRevE.77.051108,LIU20094068,Nishi_2011,PhysRevE.87.062116,Chatterjee_2015,PhysRevE.91.062818,PhysRevE.99.052133} is the consideration of conflicts at the merging point. Surprisingly, we observe metastability for both PBCs and OBCs. In particular, two phenomena related to metastability are observed for OBCs: hysteresis induced by metastability and a nonergodic phase. This new phase exhibits dependence on initial conditions under the same input and output probabilities. It should be emphasized that the metastability observed in the present model is unique  in that it arises simply from the presence of a branching-merging structure, which is a typical structure in network models, and a conflict at the merging point.




\section{MODEL}
\begin{figure}[htbp]
\begin{center}
\includegraphics[width=8cm,clip]{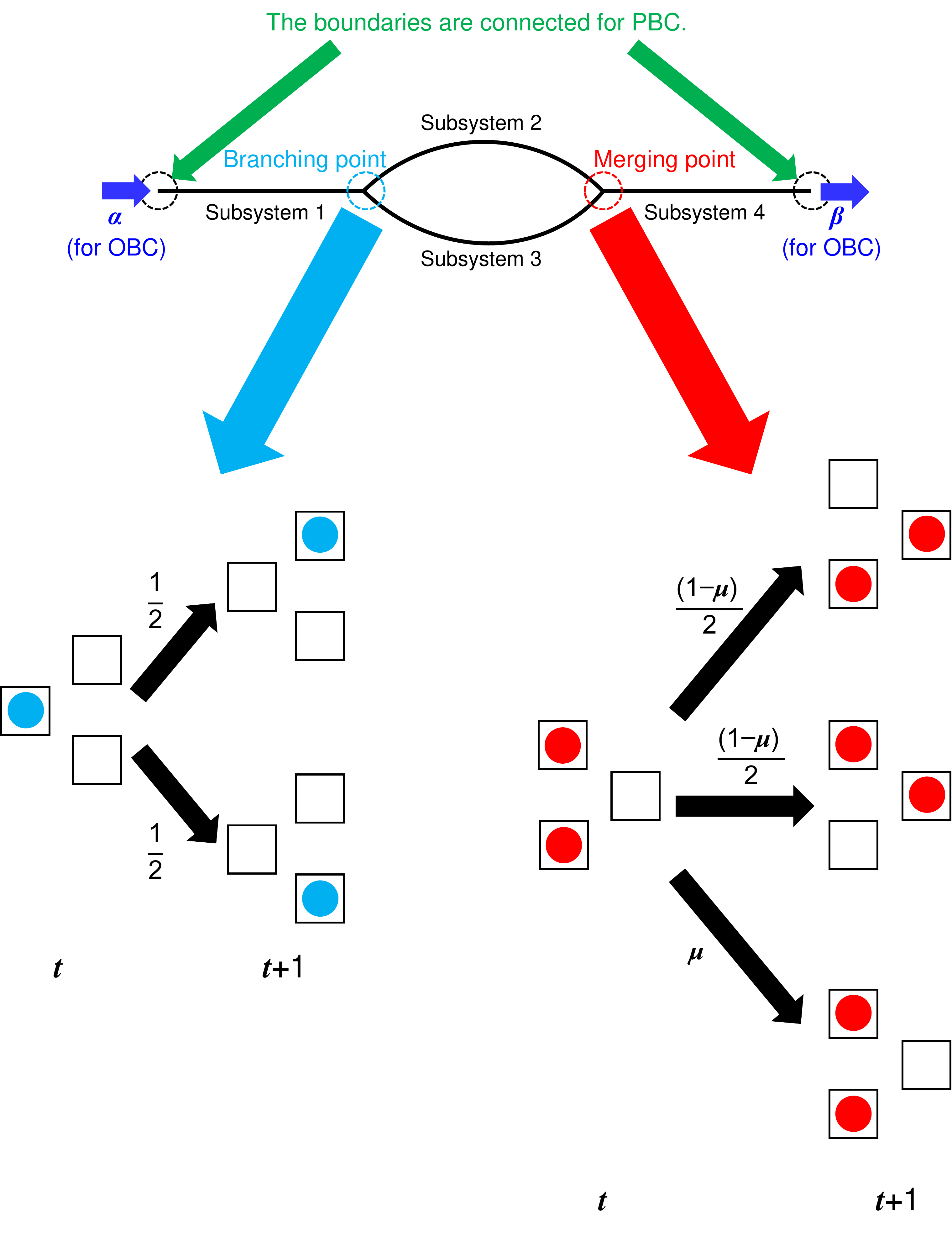}
\caption{(Color Online) Schematic illustration of the TASEP-based simple network. For periodic boundary conditions (PBCs), the right and left boundaries are connected. In contrast, for open boundary conditions (OBCs), particles enter (leave) the system from the left (right) boundary with probability $\alpha$ ($\beta$). Updating schemes from time $t$ to $t+1$ at the branching (merging) point are illustrated in the lower left (right) panel. Only situations that include stochastic elements are depicted.}
\label{fig:model}
\end{center}
\end{figure}

We study a TASEP-based simple network, as illustrated in Fig. \ref{fig:model}. The system has four parts: a head subsystem (Subsystem 1), middle subsystems (Subsystems 2 and 3), and a tail subsystem (Subsystem 4). Each subsystem consists of $L_{\rm h}$, $L_{\rm m}$, and $L_{\rm t}$ sites, where each site can be either empty or occupied by a single particle. The model employs parallel updating with discrete time, signifying that all particles are updated simultaneously. In the bulk region, particles move to the right-neighboring site if that site is empty. If the site is already occupied, the particles remain at their present site. In the case of OBCs, particles enter the system from the left boundary with probability $\alpha$ and exit from the right boundary with probability $\beta$.

At the branching point, which marks the boundary between Subsystem 1 and Subsystem 2/3, a particle randomly selects a subsystem to enter if both of the first sites in the middle subsystems are empty. If either of these sites are occupied, the particle proceeds to the empty site. It remains at its present site if both of these sites are occupied. 

At the merging point, which is the boundary between Subsystem 2/3 and Subsystem 4, a conflict is considered using friction parameter $\mu$~\cite{kirchner2003simulation,PhysRevE.67.056122,yanagisawa2009introduction}. Specifically, when two particles exist at the merging point, their movement is prohibited with probability $\mu$; that is, the particles remain at their site. Therefore, the conflict is resolved with probability $1-\mu$, allowing one of the particles, which is randomly chosen, to move to the next site. We note that particles behave according to the standard TASEP rules in all other situations.

In the following, we focus on the fundamental case where $L_{\rm h}=L_{\rm m}=L_{\rm t}=L$.
To obtain the steady-state values, we evolve the system for $10^6$ time steps and calculate the averages of $10^6$ time steps in each simulation, unless otherwise specified.

\section{METASTABILITY FOR PERIODIC BOUNDARY CONDITIONS}
We first investigate the fundamental diagram for PBCs. The diagram can be divided into three regimes; free-flow (FF), merge-induced (MI), and jam-flow (JF) regimes. The flow $Q$ is calculated as the average number of moving particles in one time step divided by $3L$, regarding Subsystems 2 and 3 as one subsystem with the continuity equation.

Based on the theoretical analysis in Appendix \ref{sec:fundamental}, the flow in the thermodynamic limit can be summarized as
\begin{eqnarray}
Q=\left\{ \begin{array}{ll}
\frac{4}{3}\rho & {\rm for} \ 0\leq\rho\leq\rho_2, \\
\frac{1-\mu}{2-\mu} & {\rm for} \ \rho_1\leq\rho\leq\rho_3, \\
\frac{4}{3}(1-\rho) & \rm{for} \ \rho_3\leq\rho\leq1,
\end{array} \right.
\label{eq:Q_PBC}
\end{eqnarray}
where $\rho_1=\frac{3(1-\mu)}{4(2-\mu)}$, $\rho_2=\frac{3}{8}$, and $\rho_3=1-\frac{3(1-\mu)}{4(2-\mu)}$. 
\begin{figure}[htbp]
\begin{center}
\includegraphics[width=7cm,clip]{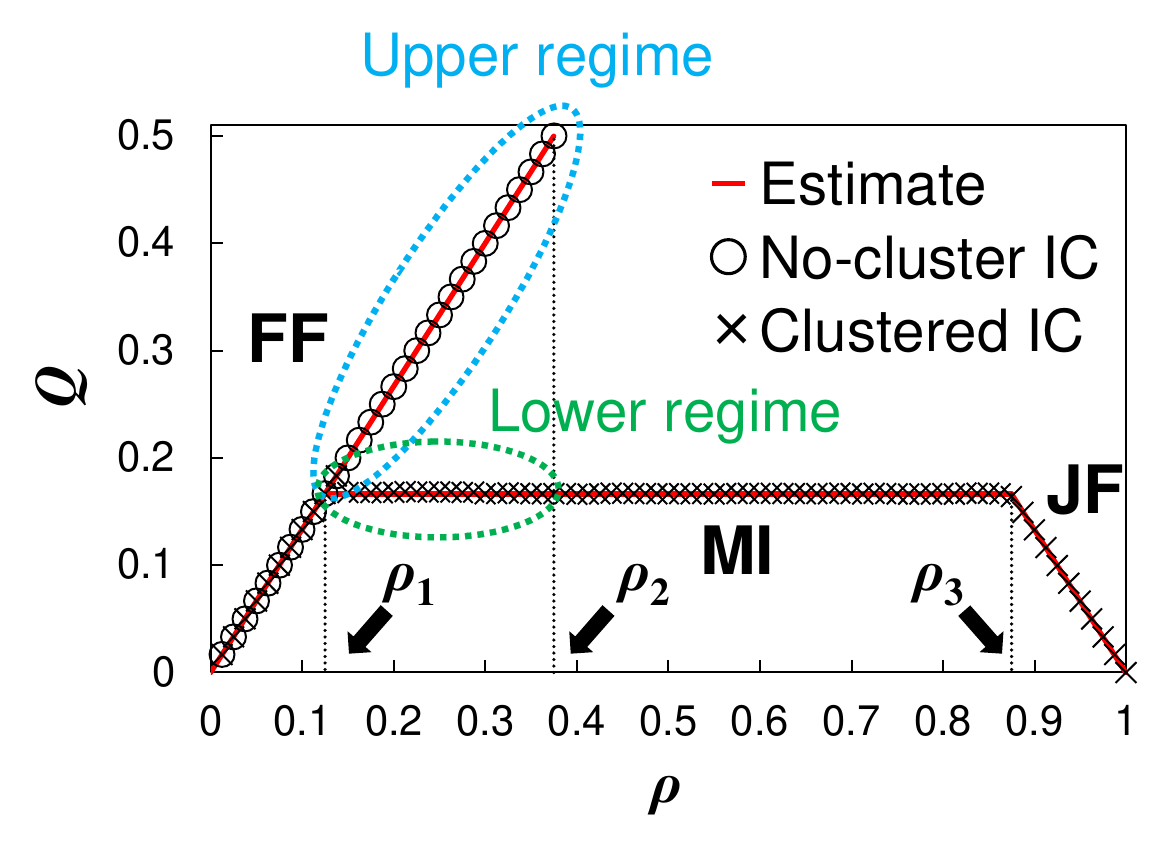}
\caption{(Color Online) Fundamental diagram with $\mu=0.8$. Circles represent the simulation results for a non-clustered initial condition (IC), while crosses represent the simulation results for a clustered IC. Red solid lines represent the theoretical lines for $L\to\infty$, that is, Eq. (\ref{eq:Q_PBC}). All simulations are conducted with $L=2500$.}
\label{fig:FD}
\end{center}
\end{figure}

Figure \ref{fig:FD} presents the fundamental diagram of the system, comparing the simulation results and the theoretical results (i.e., Eq. (\ref{eq:Q_PBC})). The simulation results exhibit excellent agreement with the theoretical results. However, we note that for finite systems, a large cluster is likely to dissolve due to fluctuations (see the vicinity of $\rho=\rho_1$ in Fig. \ref{fig:FD}), as observed in the slow-to-start model~\cite{PhysRevLett.86.2498}.

Surprisingly, for $\rho_1<\rho\leq\rho_2$, two regimes can exist depending on the initial conditions (for further details, see Appendix \ref{sec:initial}). The upper regime is a metastable regime. With a non-clustered initial condition, where all the particles have at least one empty site ahead, the upper regime can be obtained because conflicts do not occur. In contrast, with a clustered initial condition, where all the particles are positioned continuously behind the merging point, the lower regime can be obtained because the particles accumulate at the merging point. Similar phenomena have been reported in previous studies~\cite{PhysRevE.55.5597,barlovic1998metastable,PhysRevLett.86.2498,PhysRevE.66.046113,PhysRevE.68.026135,Nishinari_2004,PhysRevE.72.035102,nishimura2006metastable,Sakai_2006,Kanai_2006,moussa2007metastable,HU2007397,zhu2007modified,Nishinari_2010, MIURA2020125152, PhysRevE.79.036104,NAKATA20105353}.

With a small disturbance in the initial condition, when the system presents the metastable FF regime at $\rho\in(\rho_1, \rho_2]$ (i.e., the upper regime), the system can transition to the MI regime (i.e., the lower regime). Therefore, $\forall\rho\in(\rho_1, \rho_2]$ can be regarded as a critical density for collapse of metastability for PBCs.

We summarize the results for various $\mu$ and confirm the same phenomena in Appendix \ref{sec:FDapp}.


\section{METASTABILITY FOR OPEN BOUNDARY CONDITIONS}
For OBCs, the system has three potentially rate-limiting points: the left boundary, the merging point, and the right boundary. Therefore, the system exhibits a phase determined by these three points. 

\begin{figure}[htbp]
\begin{center}
\includegraphics[width=9cm,clip]{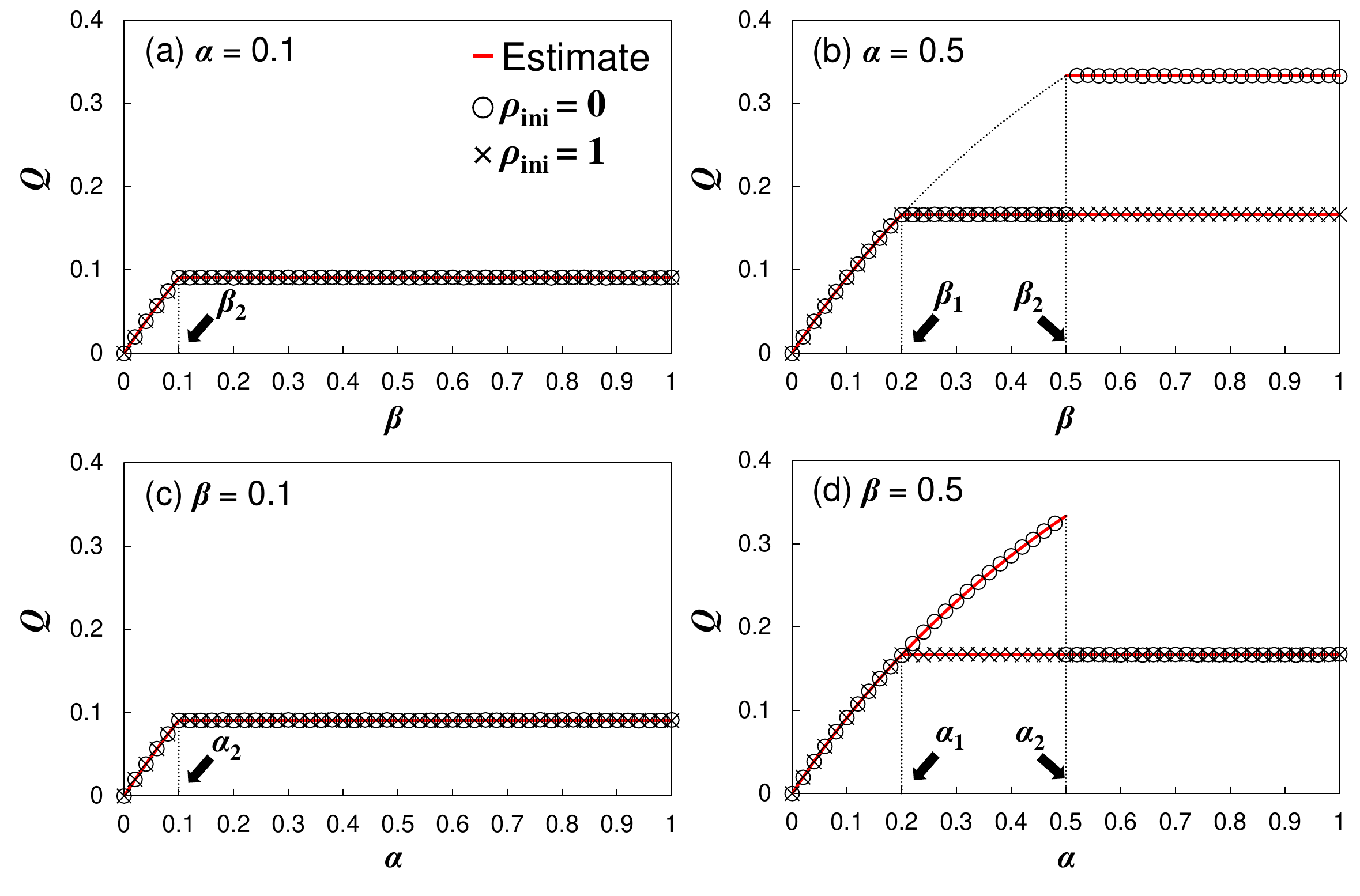}
\caption{(Color Online) Simulation results (circles/crosses) and theoretical estimates (curves) of $Q$ for (a) $\alpha=0.1$, (b) $\alpha=0.5$, (c) $\beta=0.1$, and (d) $\beta=0.5$, where $\mu=0.8$. Circles represent the results of $\rho_{\rm ini}=0$ (i.e., all sites are empty), while crosses represent the results of $\rho_{\rm ini}=1$ (i.e., all sites are occupied). The values $(\beta_1, \beta_2, \alpha_1, \alpha_2)$ are described in the text. All simulations are conducted with $L=500$.}
\label{fig:Qfixed}
\end{center}
\end{figure}

We first consider the case with fixed $\alpha>1-\mu(=\alpha_1)$, corresponding to Fig. \ref{fig:Qfixed} (b). With a small $\beta$, the system is governed by the right boundary; therefore, the flow $Q_{\rm R}$~\cite{de1999exact} is given by
\begin{equation}
Q_{\rm R}=\frac{\beta}{1+\beta}.
\label{eq:QHD}
\end{equation}
When $\beta$ exceeds a certain value $\beta_1$ and $\beta<\alpha$, particles accumulate behind the merging point, indicating that the system is governed by the merging point. Thus, the flow $Q_{\rm M}$ can be expressed as
\begin{equation}
Q_{\rm M}=\frac{1-\mu}{2-\mu},
\label{eq:Qmerge}
\end{equation}
which corresponds to the middle equation in Eq. (\ref{eq:Q_PBC}) with the same discussion for PBCs. Accordingly, $\beta_1$ can be calculated as
\begin{equation}
Q_{\rm R}(\beta_1)=Q_{\rm M}\Leftrightarrow \beta_1=1-\mu.
\end{equation}
In the case of $\beta>\alpha(=\beta_2)$, surprisingly, the flow can exhibit two different values. For relatively low initial densities, conflicts can be neatly avoided, and the system is ultimately governed by the left boundary. Therefore, the flow $Q_{\rm L}$~\cite{de1999exact} is given by
\begin{equation}
Q_{\rm L}=\frac{\alpha}{1+\alpha}.
\label{eq:QLD}
\end{equation}
In contrast with relatively high initial densities, which cause the accumulation of particles behind the merging point, the system is ultimately governed by the merging point and the flow is given as Eq. (\ref{eq:Qmerge}).

When $\alpha$ is fixed below $\alpha_1$, corresponding to Fig. \ref{fig:Qfixed} (a), the phase governed by the merging point does not appear; therefore, the flow for $\beta>\alpha$ is given by Eq. (\ref{eq:QLD}).

We then consider the case with a fixed $\beta>\beta_1$, corresponding to Fig. \ref{fig:Qfixed} (d). With a small $\alpha$, the flow is given by Eq. (\ref{eq:QLD}). When $\alpha$ exceeds a certain value $\alpha_1$ and $\alpha<\beta$, the flow can exhibit two different values, Eq. (\ref{eq:Qmerge}) or (\ref{eq:QLD}), depending on the initial conditions. Finally, when $\alpha>\beta(=\alpha_2)$, the flow is given by Eq. (\ref{eq:QHD}). When $\beta$ is fixed below $\beta_1$, corresponding to Fig. \ref{fig:Qfixed} (c), the phase governed by the merging point does not appear; therefore, the flow for $\alpha>\beta$ is given by Eq. (\ref{eq:QHD}).

Figure \ref{fig:Qfixed} compares the simulation results and our estimates of the flow for various ($\alpha, \beta$) with the initial global density $\rho_{\rm ini}\in\{0,1\}$. The figure indicates that the simulation results are in excellent agreement with our estimates.

In addition, hysteresis can be observed by measuring the space-averaged flow $\bar{Q}$ when changing $\beta$ ($\alpha$) with fixed $\alpha$ ($\beta$). Specifically, we increase $\beta$ ($\alpha$) by $10^{-3}$ and calculate $\bar{Q}$ every $10^4$ time steps. Then, we decrease $\beta$ ($\alpha$) by $10^{-3}$. The detailed calculation scheme is discussed in Appendix \ref{sec:calhys}. Figure \ref{fig:hys} presents the hysteresis plots. 
In Fig. \ref{fig:hys}(a), $\bar{Q}$ changes continuously when $\beta$ increases from 0 to 1; however, a drastic change in $\bar{Q}$ is observed at $\beta\approx\beta_2$ when $\beta$ decreases from 1 to 0. In contrast, $\bar{Q}$ changes continuously when $\alpha$ decreases from 1 to 0; however, a drastic change is observed at $\alpha\approx\alpha_2$ when $\alpha$ increases from 0 to 1, as illustrated in Fig. \ref{fig:hys}(b). Therefore, the boundary condition $\alpha=\beta(>1-\mu)$ corresponds to a critical condition for metastability collapse. We summarize the results for various $\mu$ and confirm the same phenomena in Appendix \ref{sec:hysapp}.

\begin{figure}[htbp]
\begin{center}
\includegraphics[width=9cm,clip]{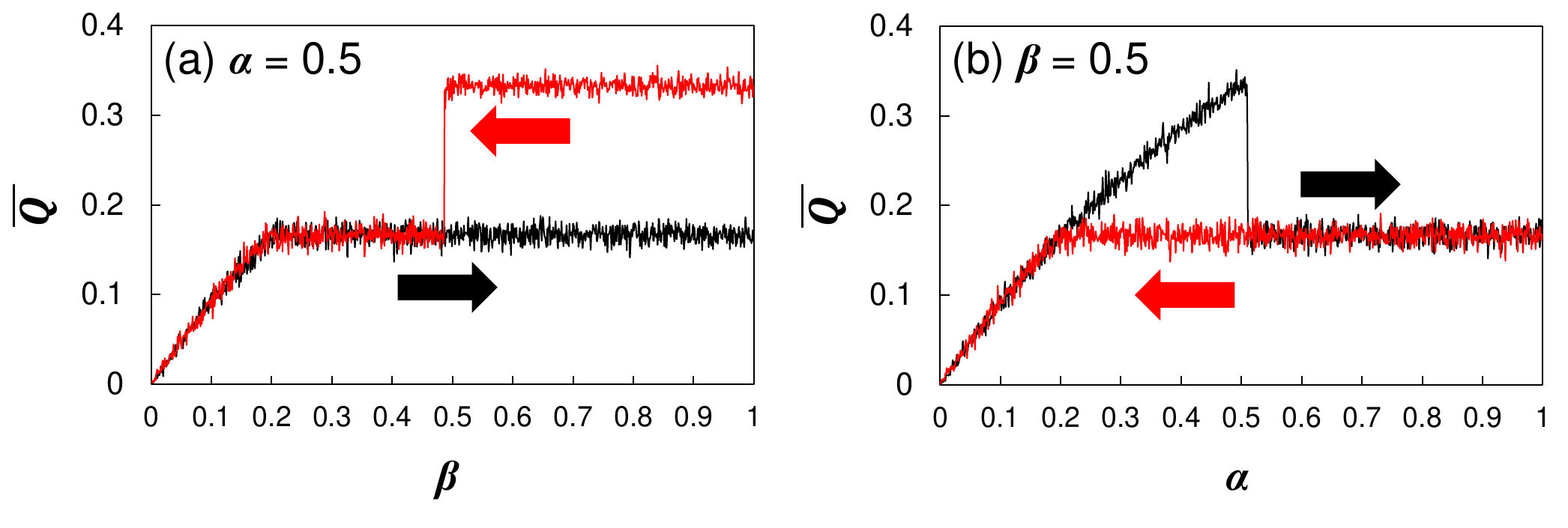}
\caption{(Color Online) Hysteresis plots for (a) $\alpha=0.5$ and (b) $\beta=0.5$, where $\mu=0.8$. The black lines start from $\beta=0$ for (a) and $\alpha=0$ for (b), while the red lines start form $\beta=1$ for (a) and $\alpha=1$ for (b), respectively. All simulations are conducted with $L=500$. }
\label{fig:hys}
\end{center}
\end{figure}

\begin{figure}[htbp]
\begin{center}
\includegraphics[width=9cm,clip]{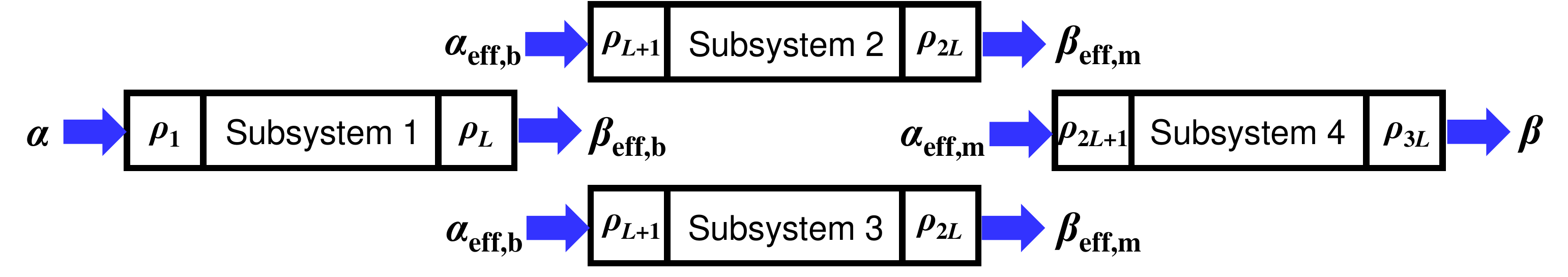}
\caption{(Color Online) Effective entrance/exit probabilities and steady-state densities at the junctions for each Subsystem. The entrance probabilities of Subsystem 1, 2, 3, and 4 are represented as $\alpha$, $\alpha_{\rm eff,b}$, $\alpha_{\rm eff,b}$, and $\alpha_{\rm eff,m}$, respectively, while the exit probabilities of Subsystem 1, 2, 3, and 4 are represented as $\beta_{\rm eff,b}$, $\beta_{\rm eff,m}$, $\beta_{\rm eff,m}$, and $\beta$, respectively. The steady-state densities at sites $1$, $L$, $L+1$, $2L$, $2L+1$, and $3L$ are represented as $\rho_{1}$, $\rho_L$, $\rho_{L+1}$, $\rho_{2L}$, $\rho_{2L+1}$, and $\rho_{3L}$, respectively.}
\label{fig:model2}
\end{center}
\end{figure}

Next, we investigate the phase diagram using a simple approximation similar to those used in previous studies~\cite{Pronina_2005,PhysRevE.77.051108,LIU20094068,Chatterjee_2015}. Figure \ref{fig:model2} presents a schematic of each subsystem. One notable difference between this study and previous studies is the use of different expressions of the effective probabilities at the merging point $\alpha_{\rm eff, m}$ on the phase in Subsystem 4. Specifically, $\alpha_{\rm eff, m}$ can be expressed as
\begin{equation}
\alpha_{\rm eff, m}=2\rho_{2L}
\label{eq:branch2}
\end{equation}
when Subsystems 2 and 3 are in the LD phase, where a conflict rarely occurs, and as
\begin{equation}
\alpha_{\rm eff, m}=1-\mu
\label{eq:branch3}
\end{equation}
when Subsystems 2 and 3 are in the HD phase, where a conflict generally occurs. 

Based on the simple approximations of and additional discussions for the phase boundaries, seven phases ((LD, LD, LD), (HD, HD, LD), (HD, HD, HD), (MC, LD, MC), (SW, SW, SW), (SW$_1$, SW$_1$, SW$_2$) and (HD, HD, SW)) are obtained (see details in Appendix \ref{sec:simple} and \ref{sec:PB}).
Figure \ref{fig:PD} presents the phase diagram of the system (see also Appendix \ref{sec:phase}). Surprisingly, the region where $1-\mu<\alpha<1$ and $\alpha<\beta$, or $\alpha=\beta=1$ exhibits a nonergodic phase. For $1-\mu<\alpha<1$ and $\alpha<\beta$, the system yields either of (LD, LD, LD) with low-initial-density conditions or (HD, HD, LD) with high-initial-density conditions. In contrast, for $\alpha=\beta=1$, the phase transitions from (LD, LD, LD) to (MC, LD, MC). This phenomenon can be interpreted as ergodicity breaking. The density profiles are also discussed in Appendix \ref{sec:profile}.

\begin{figure}[htbp]
\begin{center}
\includegraphics[width=9cm,clip]{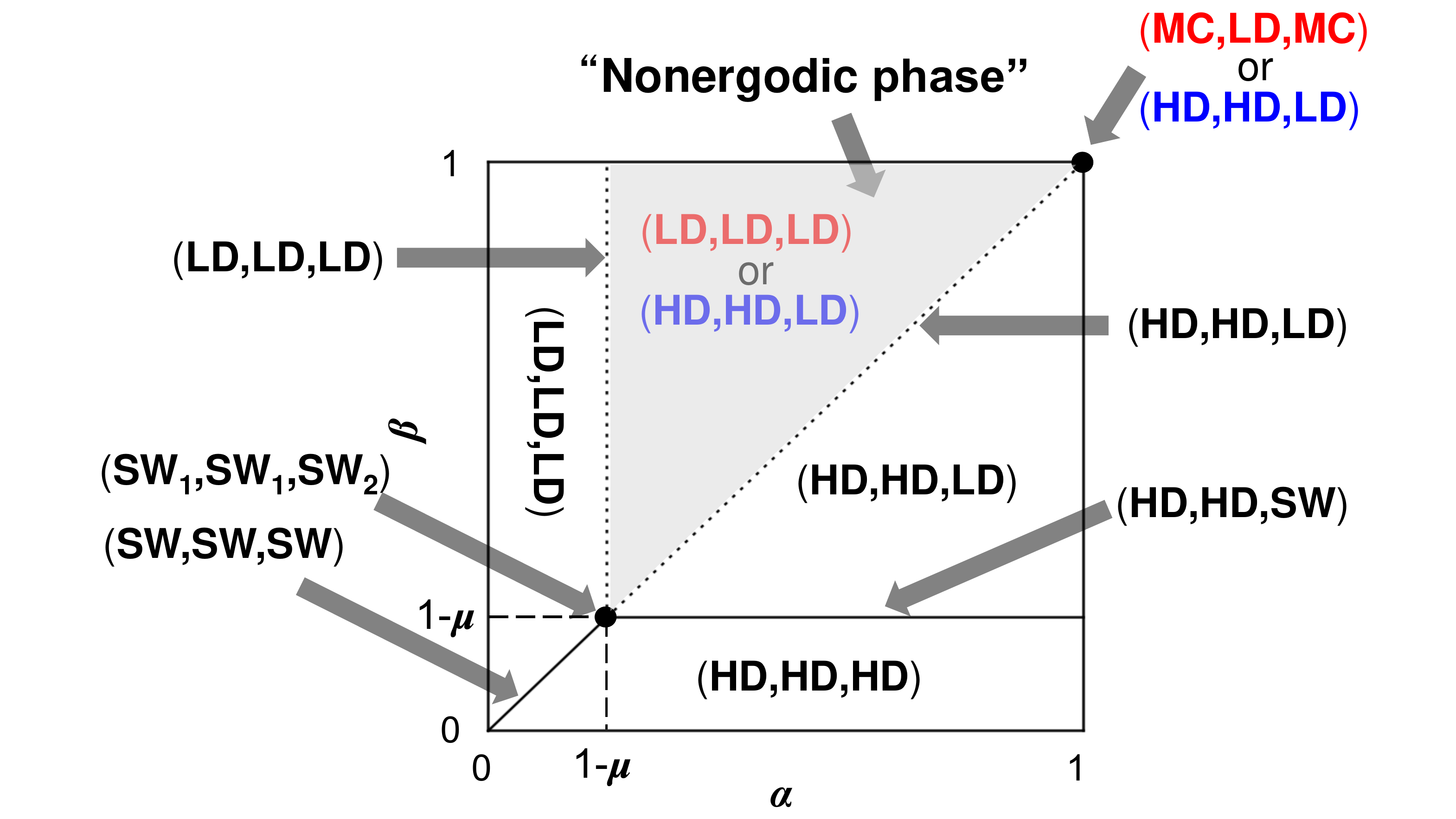}
\caption{(Color Online) Phase diagram of the system. The red (blue) phase is obtained with low-initial-density (high-initial-density) conditions. For the dotted line corresponding to $\alpha=1-\mu$ and $\beta>1-\mu$, the system presents the (LD, LD, LD), whereas for the dotted line corresponding to $\alpha=\beta<1$, the system exhibits the (HD, HD, LD). The gray zone represents the nonergodic phase.}
\label{fig:PD}
\end{center}
\end{figure}

\section{Critical initial densities in the nonergodic phase}
\label{sec:tipping}
As discussed above, the phase of the system can vary depending on initial densities $\rho_{\rm ini}$. In this section, we discuss the critical initial density $\rho_{\rm ini, cr}$ in the nonergodic phase. The steady-state transitions to (HD, HD, LD) with $\rho_{\rm ini}<\rho_{\rm ini, cr}$; in contrast, it transitions to (LD, LD, LD) or (MC, LD, MC) with $\rho_{\rm ini}>\rho_{\rm ini, cr}$.

Starting from $\rho_{\rm ini}=\rho_{\rm ini, cr}$, $L\rho_{\rm ini, cr}$ particles initially exist in each subsystem on average. In the worst-case scenario, all particles in Subsystems 1, 2, and 3 are involved in a conflict at the merging point and therefore enter Subsystem 4 every $\frac{1}{Q_{\rm M}}$ time steps. In this case, $\rho_{\rm ini, cr}$ can be estimated by ensuring that the first entering particle arrives at the merging point just as the conflicts of the initial existing particles are almost resolved. We therefore can formulate this condition as
\begin{equation}
\frac{3L\rho_{\rm ini, cr}}{Q_{\rm M}}\approx\frac{2L}{1}\Leftrightarrow\rho_{\rm ini, cr}\approx\frac{2(1-\mu)}{3(2-\mu)}.
\label{eq:tipping}
\end{equation}

\begin{figure}[htbp]
\begin{center}
\includegraphics[width=7cm,clip]{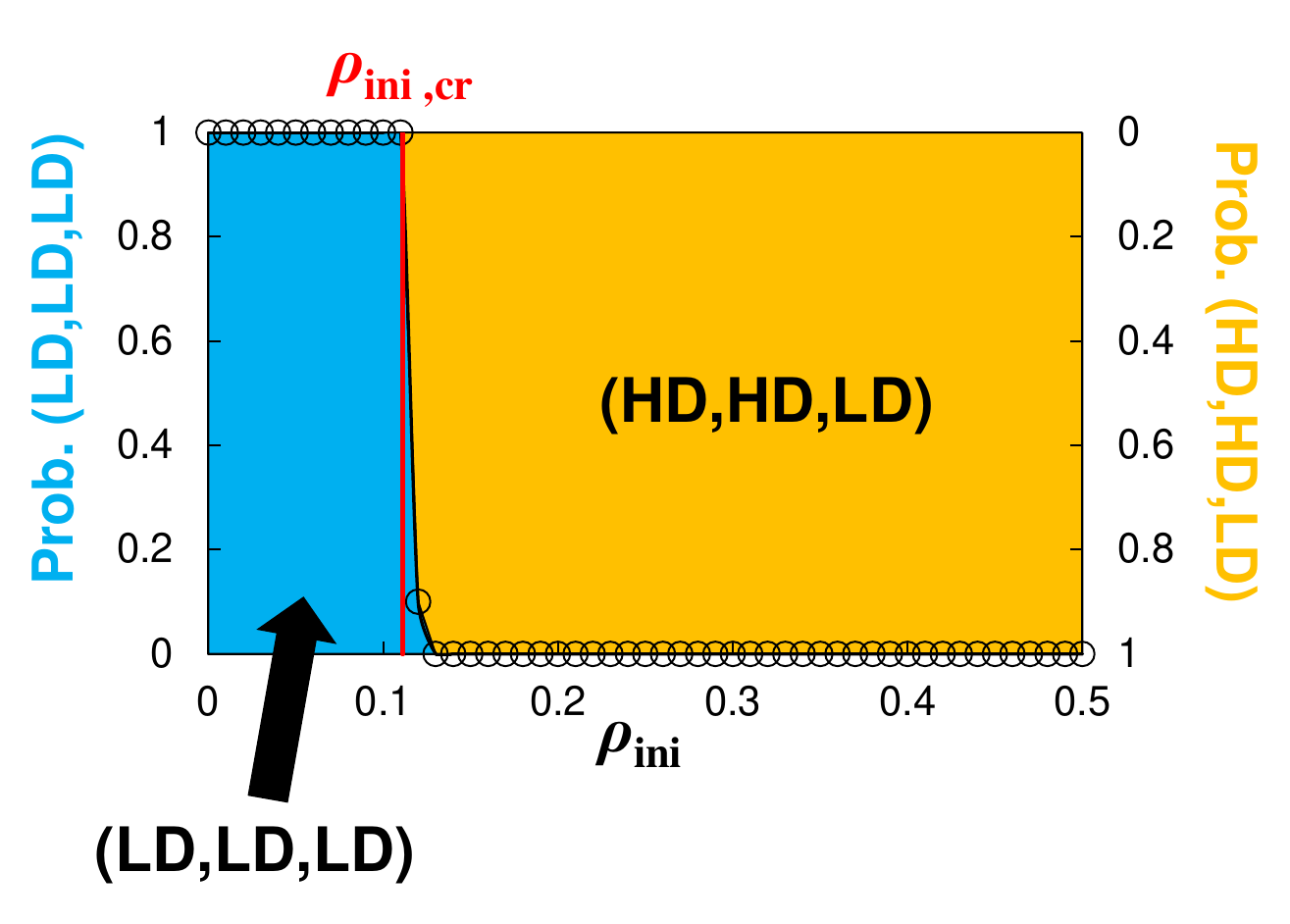}
\caption{(Color Online) Probability of two steady-state phases (LD, LD, LD) (blue) or (HD, HD, LD) (orange) as a function of $\rho_{\rm ini}$, with the red line representing the estimated $\rho_{\rm ini}=\rho_{\rm ini, cr}$ (i.e., Eq. (\ref{eq:tipping})) for $(\mu, \alpha,\beta)=(0.8, 0.85, 0.9)$, with which the system always satisfies $\alpha>1-\mu$ and $\alpha<\beta$. We calculate the probabilities for each $\rho_{\rm ini}$ in increments of 0.01 (black circles) by 10-time simulations with random configurations. We determine the steady-state phase based on the flow; for example, if the simulated flow is near the estimated flow of (LD, LD, LD), the steady-state phase is determined as (LD, LD, LD). All simulations are conducted with $L=2500$.}
\label{fig:tipping}
\end{center}
\end{figure}

Figure \ref{fig:tipping} presents the probabilities of the two steady-state phases (LD, LD, LD) or (HD, HD, LD), as a function of $\rho_{\rm ini}$, with a line representing the estimated $\rho_{\rm ini}=\rho_{\rm ini, cr}$. We observe that the probability of the (LD, LD, LD) phase sharply decreases around $\rho_{\rm ini}=\rho_{\rm ini, cr}$. Despite being a coarse approximation, Eq. (\ref{eq:tipping}) is an excellent estimate of $\rho_{\rm ini}=\rho_{\rm ini, cr}$, corresponding to the critical initial density in the noergodic phase for OBCs. We note that unlike for PBCs, the initial density, not the initial configuration, plays a primary role in determining the steady-state for OBCs. The area of (LD, LD, LD) becomes a little wider than the theory because no conflicts can be realized depending on the initial configuration, especially for low-initial-density conditions.

We summarize the results for various $\mu$ and confirm the same phenomena in Appendix \ref{sec:tippingapp}.

\section{Discussion}
In this paper, we present a TASEP-based simple network with conflicts at a merging point that exhibits metastability (i.e., nonergodicity) for both PBCs and OBCs. Metastability induces hysteresis for OBCs. By using simple expansions of the TASEP, namely, by considering a branching-merging structure, which is a fundamental element of network models, and a conflict rule at the merging point, we can obtain a nonergodic phase, where the initial condition determines the steady-state for OBCs. This phenomenon has not been reported in previous studies. Moreover, we successfully identify the critical conditions (i.e., tipping points) where a small disturbance causes the collapse of metastability. These critical conditions include (i) the critical initial density for PBCs, (ii) the critical boundary conditions for OBCs, and (iii) the critical initial density for OBCs. 

We would like to discuss the robustness of observed metastability, which depends on (i) deterministic hopping probabilities, (ii) the same length of the two branches in the network, and (iii) initial conditions. As for (i), metastability can be observed in the TASEP-based models and other related models where the number of stochastic elements are relatively small. It is true that the restriction of hopping probabilities as $p=1$ is certainly a strong approximation compared to the previous investigations. Instead, we observe metastability not only for PBCs but also for OBCs, where the stochastic elements--input probabilities and output probabilities--make it difficult to maintain metastability. This is one of the strengths of this paper. As for (ii), metastability in this paper requires that the two branches have the same length, or at least, the length of one branch is a multiple that of the other. We admit that this requirement is relatively strong and one limitation of the present investigation. As for (iii), initial conditions critically influence on whether the system yields to metastability. This fact holds true for many of the previous investigations~\cite{PhysRevE.55.5597,barlovic1998metastable,PhysRevLett.86.2498,PhysRevE.66.046113,PhysRevE.68.026135,moussa2007metastable,HU2007397,zhu2007modified}, which considered two different initial conditions--homogeneous and megajam states for PBCs. Homogeneous states are, in a sense, artificial, because there is a very slim chance of obtaining the states if initial conditions are randomly chosen. In this paper, however, the steady state for OBC depends on initial densities, rather than initial configurations, which we can consider more robust than the previous ones.

For complex network systems in various fields, it is important to identify the critical points to be able to anticipate undesired or positive changes. In recent studies~\cite{neri2011totally, neri2013exclusion, neri2013modeling, shen2020totally, wang2021physical}, TASEP networks have been proposed as models for such complex systems. The rules identified in this study can be applied to these models to identify metastability. Despite the simplicity of the present model, we believe that our findings can provide new insights into the metastability (nonergodicity) of network systems.


\section*{ACKNOWLEDGEMENT}
This work was partially supported by JST-Mirai Program Grant No. JPMJMI20D1, Japan, and JSPS KAKENHI Grant Nos. JP21H01570 and JP21H01352.

\appendix

\section{Fundamental diagrams}
\label{sec:fundamental}
Here, we derive Eq. (\ref{eq:Q_PBC}).
In the free-flow regime, the flow is determined by the number of particles. Considering the definition of flow, the flow can be calculated as
\begin{equation}
Q=\frac{4L\rho}{3L}=\frac{4}{3}\rho,
\end{equation}
where $\rho$ is the global density.

In the merge-induced regime, the flow is determined by the merging point. The average number of time steps in which either of the two particles involved in a conflict must exit the merging point is given as $\frac{1}{1-\mu}$. Therefore, the average density of Subsystems 4 and 1 is reduced to $\frac{1}{1+\frac{1}{1-\mu}}$, resulting in the following flow:
\begin{equation}
Q=\frac{1}{1+\frac{1}{1-\mu}}=\frac{1-\mu}{2-\mu}.
\label{eq:mergeQ}
\end{equation}
In this case, the average density of Subsystems 2 and 3 halves, becoming $\frac{1}{2\left(1+\frac{1}{1-\mu}\right)}$. As a result, the global density $\rho$ becomes
\begin{equation}
\rho_1=\frac{2L\left(\frac{1}{1+\frac{1}{1-\mu}}\right)+2L\left[\frac{1}{2\left(1+\frac{1}{1-\mu}\right)}\right]}{4L}=\frac{3(1-\mu)}{4(2-\mu)},
\end{equation}
which is the transitional density from the free-flow regime to the merge-induced regime.

To ensure that there is no conflict at the merging point with the maximum number of particles, the average gap between particles is reduced to $1$ in Subsystems 4 and 1, indicating that the average density of Subsystems 4 and 1 is $\frac{1}{2}$. Therefore, the average density of Subsystems 2 and 3 halves, becoming $\frac{1}{4}$. Eventually, $\rho_2$, the maximum density at which the free-flow regime can exist, is given by
\begin{equation}
\rho_2=\frac{2L\times\frac{1}{2}+2L\times\frac{1}{4}}{4L}=\frac{3}{8}.
\end{equation}

In the jam-flow regime, the flow is determined by the number of empty sites. Considering the definition of flow and particle-hole symmetry, the flow can be calculated as
\begin{equation}
Q=\frac{4L(1-\rho)}{3L}=\frac{4}{3}(1-\rho),
\end{equation}
and the value $\rho_3$, the transitional density from the merge-induced regime to the jam-flow regime, is represented as $\rho_3=1-\rho_1$.

Based on the above discussion, we obtain Eq. (\ref{eq:Q_PBC}).

\section{Initial conditions for PBCs}
\label{sec:initial}

For PBCs, we consider two types of initial conditions; non-clustered and clustered initial conditions. The difference between the two conditions lies in the configuration rather than the density. Therefore, they can be defined under the same density, only at relatively low densities. 

In a non-clustered initial condition, more than or equal to one-site interval is maintained between the two adjacent particles. In the simulations, we put all the particles in either of Subsystem 1, 2, and 4. For $\rho\leq\frac{1}{8}$, all the particles are positioned in Subsystem 1 with one-site interval. For $\frac{1}{8}<\rho\leq\frac{1}{4}$, the particles which cannot be accommodated in Subsystem 1 are positioned in Subsystem 2 with one-site interval. For $\frac{1}{4}<\rho\leq\frac{3}{8}$, the particles which cannot be accommodated in Subsystem 1 or 2 are positioned in Subsystem 4 with one-site interval. We note that for $\rho>\frac{3}{8}$ a non-clustered initial condition cannot be realized.

On the other hand, in a clustered condition, all the particles are positioned behind the merging point to intentionally generate conflicts. In the simulations, half of the particles are placed as one cluster in both of Subsystems 2 and 3 for $0<\rho\leq\frac{1}{2}$. For $\rho>\frac{1}{2}$, the particles which cannot be accommodated in Subsystem 2 or 3 are positioned in Subsystems 1 and 4 with no interval, starting from the right edge.

\begin{figure}[htbp]
\begin{center}
\includegraphics[width=8cm,clip]{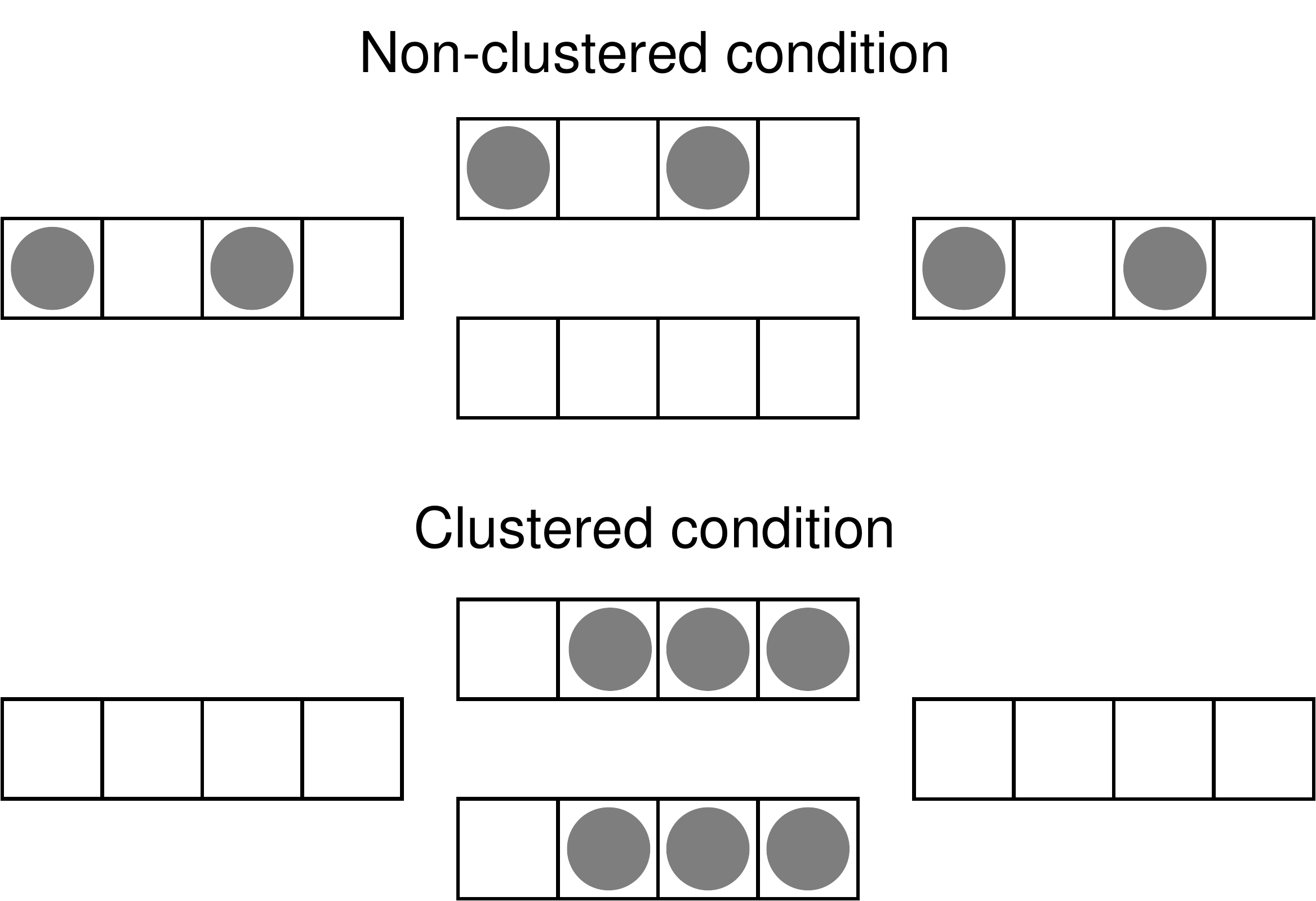}
\caption{(Color Online) Two initial conditions for PBCs when $L=4$ and $\rho=\frac{3}{8}$. Note that the two conditions can be defined when $\rho\leq\frac{3}{8}$; otherwise, only the clustered condition can be defined.}
\label{fig:initial_periodic}
\end{center}
\end{figure}

Figure \ref{fig:initial_periodic} compares the examples of the two initial conditions.

\section{Fundamental diagrams with various $\mu$}
\label{sec:FDapp}
\begin{figure}[htbp]
\begin{center}
\includegraphics[width=9cm,clip]{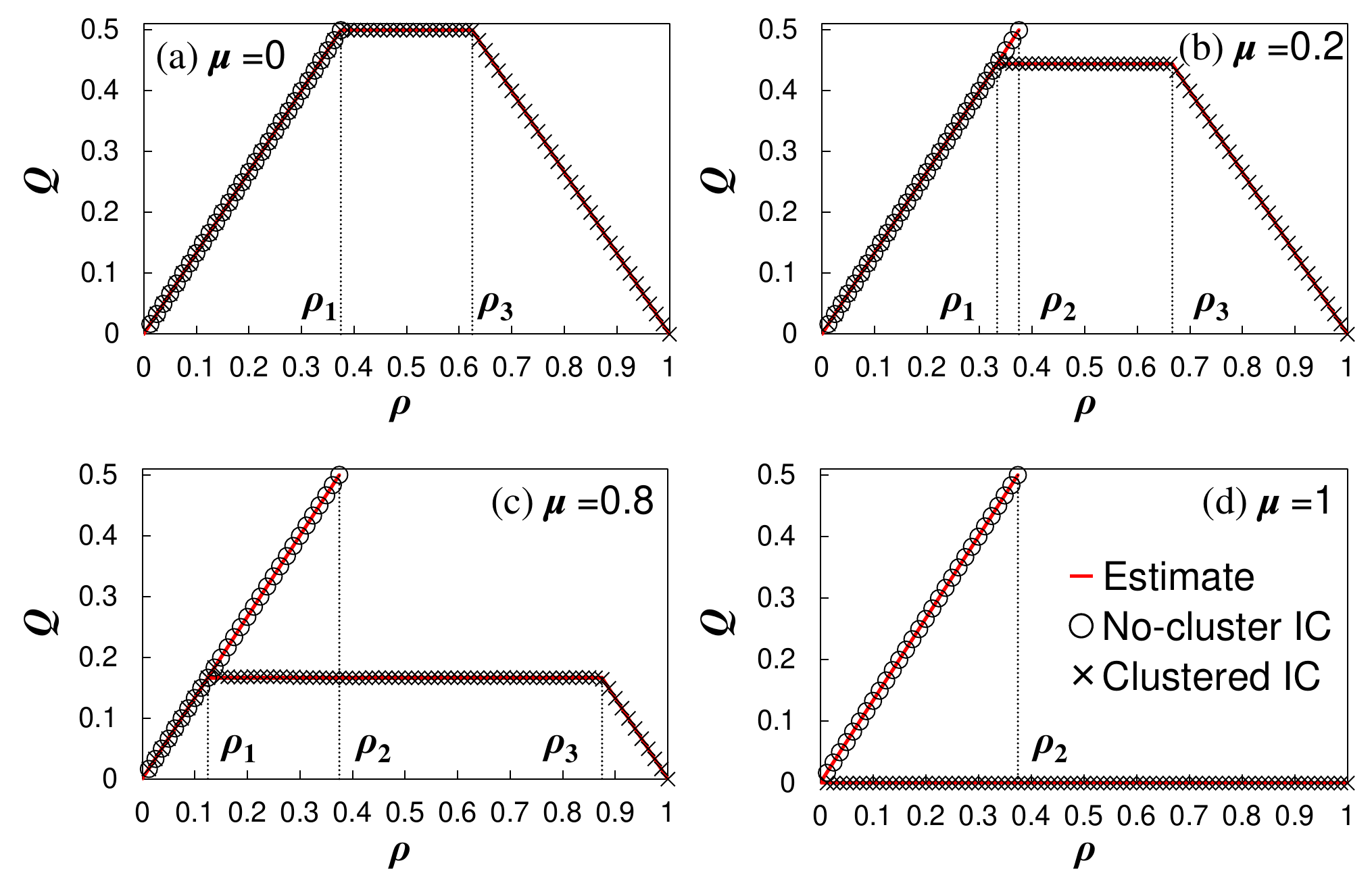}
\caption{(Color Online) Fundamental diagram with (a) $\mu=0$, (b) $\mu=0.2$, (c) $\mu=0.8$, and (d) $\mu=1$. We note that the panel (c) is identical to Fig. \ref{fig:FD}. Other explanations are the same as those of Fig. \ref{fig:FD}.}
\label{fig:FDapp}
\end{center}
\end{figure}

Figure \ref{fig:FDapp} presents the fundamental diagram of the system with various $\mu\in\{0,0.2,0.8,1\}$. Except for the case of $\mu=0$, we confirm the same phenomena as Fig. \ref{fig:FD}. For the case of $\mu=0$, no conflict at the merging point, and therefore, the metastable FF regime vanishes.

\section{Hysteresis plots with various $\mu$}
\label{sec:hysapp}

\begin{figure}[htbp]
\begin{center}
\includegraphics[width=9cm,clip]{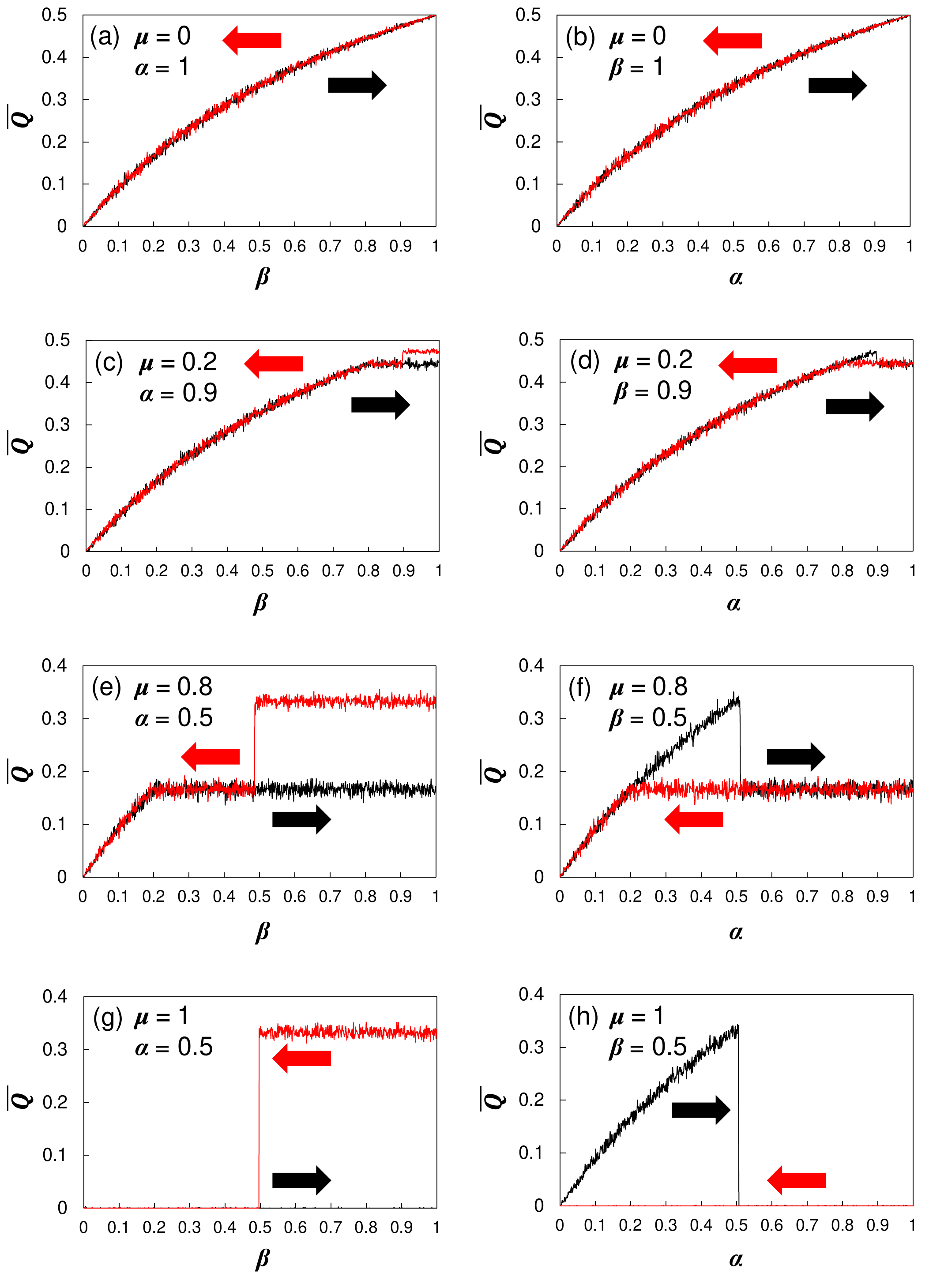}
\caption{(Color Online) Hysteresis plots for(a) $(\mu, \alpha)=(0, 1)$, (b) $(\mu,\beta)=(0, 1)$, (c) $(\mu, \alpha)=(0.2, 0.9)$, (d) $(\mu, \beta)=(0.2, 0.9)$, (e) $(\mu, \alpha)=(0.8,0.5)$, (f) $(\mu, \beta)=(0.8,0.5)$, (g) $(\mu, \alpha)=(1, 0.5)$, and (h) $(\mu, \beta)=(1, 0.5)$. We note that the panels (e) and (f) are identical to Fig. \ref{fig:hys}. The black lines in the panel (g) and (h), entirely or partially, overlap the horizontal axis. Other explanations are the same as those of Fig. \ref{fig:hys}.}
\label{fig:Hysteresisapp}
\end{center}
\end{figure}

Figure \ref{fig:Hysteresisapp} presents the hysteresis plots for various $\mu\in\{0,0.2,0.8,1\}$. Except for the case of $\mu=0$, we confirm the same phenomena as Fig. \ref{fig:hys}.

\section{Calculation scheme of Fig. \ref{fig:hys}}
\label{sec:calhys}

\begin{figure}[htbp]
\begin{center}
\includegraphics[width=8cm,clip]{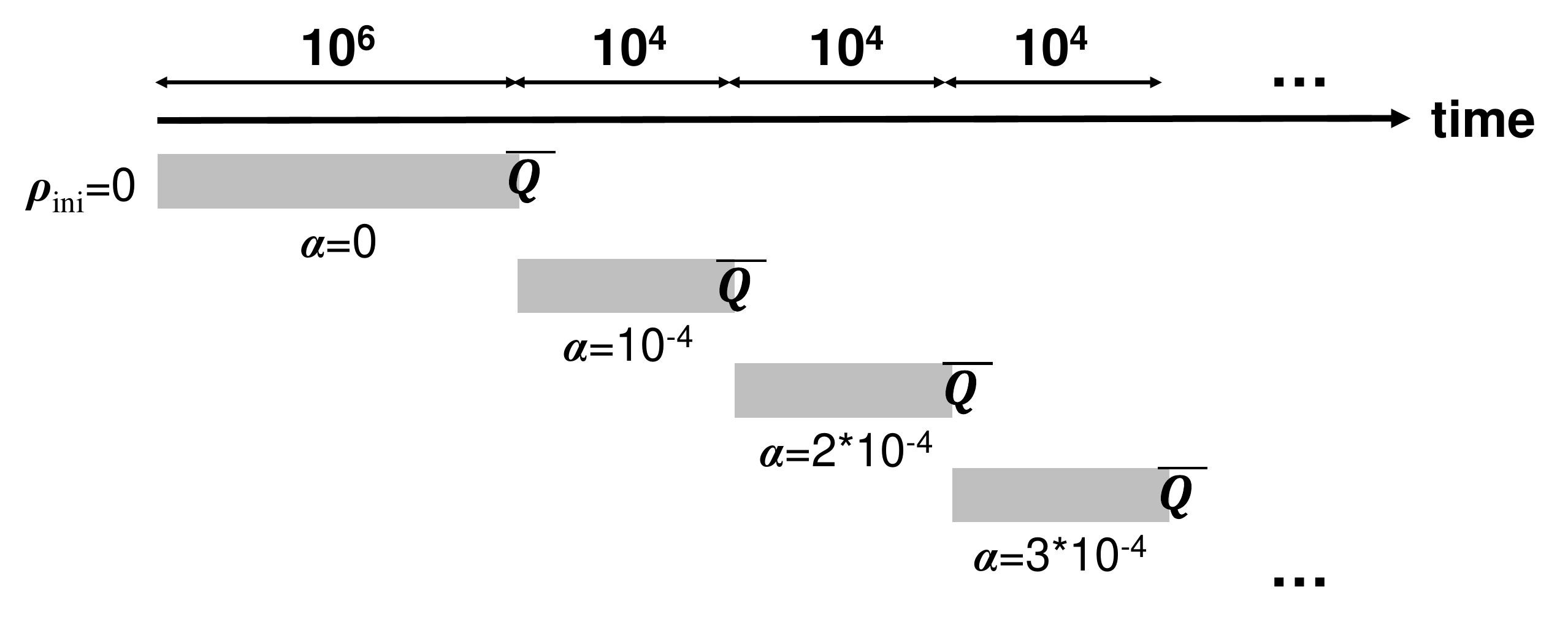}
\caption{(Color Online) Schematic of the calculation scheme with fixed $\beta$, starting from $\alpha$. The space-averaged flow $\bar{Q}$ is calculated every $10^4$ time steps.}
\label{fig:calculation}
\end{center}
\end{figure}

We here explain the details of calculation scheme of Fig. \ref{fig:hys}. The specific scheme with a fixed $\beta$ is as follows.
\begin{enumerate}
\item The simulation starts from $\rho_{\rm ini}=0$ with $(\alpha, \beta)=(0,0.5)$ ($(\alpha, \beta)=(1,0.5)$).
\item The system is evolved for $10^6$ time steps.
\item We increase (decrease) $\alpha$ by $10^{-3}$ every $10^4$ time steps; i.e., the system is evolved for $10^4$ with a certain set of $(\alpha, \beta)$. At the same time, we calculate snapshoted values of $\bar{Q}$. We note that we do not restart the simulation i.e., all the particles in the system remain at the same sites, when $\alpha$ is changed.
\end{enumerate}
This scheme can be illustrated as Fig. \ref{fig:calculation}. The case with a fixed $\alpha$ is absolutely identical.

\section{Simple approximation for phase diagrams}
\label{sec:simple}
We first recall the phase diagram of the one-lane $L$-site totally asymmetric simple exclusion process with no branches or junctions for parallel updating~\cite{de1999exact}. When the entrance governs the system (i.e., $\alpha<\beta\leq1$), a low-density (LD) phase is observed with
\begin{equation}
Q=\frac{\alpha}{1+\alpha}, \ \rho_{\rm bulk}=\frac{\alpha}{1+\alpha}, \ \rho_{1}=\frac{\alpha}{1+\alpha}, \ \rho_{L}=\frac{\alpha}{\beta(1+\alpha)},
\label{eq:JL}
\end{equation}
where $Q$ is the flow of the system, $\rho_{\rm bulk}$ is the bulk density, $\rho_{1}$ is the density of the first site, and $\rho_{L}$ is the density of the $L$th site.

When the exit governs the system (i.e., $\beta<\alpha\leq1$), a high-density (HD) phase is observed with
\begin{equation}
Q=\frac{\beta}{1+\beta}, \ \rho_{\rm bulk}=\frac{1}{1+\beta}, \ \rho_{1}=1-\frac{\beta}{\alpha(1+\beta)}, \ \rho_{L}=\frac{1}{1+\beta}.
\label{eq:JH}
\end{equation}

When $\alpha=\beta<1$, a shock-wave (SW) phase, which is also referred to as a transition line or coexistence line, is observed with
\begin{equation}
Q=\frac{\alpha}{1+\alpha}, \ \rho_i=\frac{\alpha}{1+\alpha}+\frac{1-\alpha}{1+\alpha}\frac{i}{L},
\label{eq:JS}
\end{equation}
where $\rho_i$ is the density of the $i$th ($1\leq i\leq L$) site.

When $\alpha=\beta=1$, a maximal current (MC) phase is observed with
\begin{equation}
Q=\frac{1}{2}, \ \rho_{\rm bulk}=\frac{1}{2}.
\label{eq:JM}
\end{equation}

Based on the above discussion, we investigate the phase diagram of this system. With the notion of flow conservation, we have
\begin{equation}
Q_1=Q_2+Q_3=Q_4\leq\frac{1}{2}, \ Q_2=Q_3,
\label{eq:J}
\end{equation}
where $Q_j \ (j=1,2,3,4)$ is the flow of subsystem $j$.

The effective probabilities at the branching point $\alpha_{\rm eff, b}$ and $\beta_{\rm eff, b}$ can be expressed as
\begin{equation}
\alpha_{\rm eff, b}=\rho_L, \ \beta_{\rm eff, b}=2(1-\rho_{L+1}),
\label{eq:branch1}
\end{equation}
whereas the effective probabilities at the merging point $\alpha_{\rm eff, m}$ and $\beta_{\rm eff, m}$ can be expressed as
\begin{equation}
\alpha_{\rm eff, m}=2\rho_{2L}, \ \beta_{\rm eff, m}=1-\rho_{2L+1},
\label{eq:branch2}
\end{equation}
when Subsystems 2 and 3 are in the LD phase, where a conflict rarely occurs. These expressions are the same as those reported in previous studies~\cite{Pronina_2005,PhysRevE.77.051108,LIU20094068,Chatterjee_2015}. However, when Subsystems 2 and 3 are in the HD phase, a conflict almost always occurs, and the expression of $\alpha_{\rm eff, m}$ becomes
\begin{equation}
\alpha_{\rm eff, m}=1-\mu.
\label{eq:branch3}
\end{equation}

Because the three stationary phases (LD, HD, and MC) can be found in each lane and Subsystems 2 and 3 have identical phases due to a symmetry, the number of possible phase combinations of the system is equal to $27(=3^3)$. However, it is evident that 17 of these phases cannot exist. Nine of them cannot exist because $J_2\leq\frac{1}{4}$ according to Eq. (\ref{eq:J}), indicating that an MC phase cannot be present in Subsystems 2 and 3. In addition, eight of the 17 phases cannot exist because the MC phase cannot occur in Subsystem 1 or 4 alone since $J_1=J_4$. Therefore, there are 10 valid phase combinations  as follows: (LD, LD, LD), (LD, LD, HD), (LD, HD, LD), (LD, HD, HD), (HD ,LD, LD), (HD, LD, HD), (HD, HD, LD), (HD, HD, HD), (MC, LD, MC), and (MC, HD, MC).
In the following discussion, we examine the requirements and density profile for each phase using a simple approximation.


\begin{enumerate}
\item (LD, LD, LD) phase

The following conditions must be satisfied:
\begin{equation}
\alpha<\beta_{\rm eff, b}, \ \alpha_{\rm eff, b}<\beta_{\rm eff, m}, \ \alpha_{\rm eff, m}<\beta.
\label{eq:LDLDLD}
\end{equation}
Using Eqs. (\ref{eq:JL}) and (\ref{eq:J}), we obtain
\begin{equation}
\frac{\alpha}{1+\alpha}=\frac{2\alpha_{\rm eff, b}}{1+\alpha_{\rm eff, b}}=\frac{\alpha_{\rm eff, m}}{1+\alpha_{\rm eff, m}},
\label{eq:LDLDLD_alpha}
\end{equation} 
resulting in
\begin{equation}
\alpha_{\rm eff, b}=\frac{\alpha}{2+\alpha}, \ \alpha_{\rm eff, m}=\alpha.
\label{eq:LDLDLD_alpha_d}
\end{equation}
From Eqs. (\ref{eq:JL}), (\ref{eq:branch1}), and (\ref{eq:branch2}), we obtain
\begin{eqnarray}
&&\rho_L=\frac{\alpha}{\beta_{\rm eff,b}(1+\alpha)}, \ \rho_{L+1}=\frac{\alpha_{\rm eff,b}}{1+\alpha_{\rm eff,b}},\\
&&\rho_{2L}=\frac{\alpha_{\rm eff,b}}{\beta_{\rm eff,m}(1+\alpha_{\rm eff,b})}, \ \rho_{2L+1}=\frac{\alpha_{\rm eff,m}}{1+\alpha_{\rm eff,m}},\\
&&\beta_{\rm eff,b}=\frac{2+\alpha}{1+\alpha}(>\alpha), \ \beta_{\rm eff,m}=\frac{1}{1+\alpha}(>\alpha_{\rm eff, b}).
\label{eq:LDLDLD_beta}
\end{eqnarray}
Based on Eqs. (\ref{eq:LDLDLD_alpha_d}) and (\ref{eq:LDLDLD_beta}), Eq. (\ref{eq:LDLDLD}) can be simplified to
\begin{equation}
\alpha<\beta.
\end{equation}
We note that $\beta_{\rm eff,b}>1$ despite the fact that $\beta_{\rm eff,b}$ is a probability. Therefore, hereafter we consider $\beta_{\rm eff,b}=1$ for practical purposes.

\item (LD, LD, HD) phase

The following conditions must be satisfied:
\begin{equation}
\alpha<\beta_{\rm eff, b}, \ \alpha_{\rm eff, b}<\beta_{\rm eff, m}, \ \alpha_{\rm eff, m}>\beta.
\label{eq:LDLDHD}
\end{equation}
Using Eqs. (\ref{eq:JL}), (\ref{eq:JH}), and (\ref{eq:J}), we obtain
\begin{equation}
\frac{\alpha}{1+\alpha}=\frac{2\alpha_{\rm eff, b}}{1+\alpha_{\rm eff, b}}=\frac{\beta}{1+\beta},
\label{eq:LDLDHD_alpha}
\end{equation} 
resulting in
\begin{equation}
\alpha=\beta, \ \alpha_{\rm eff, b}=\frac{\alpha}{2+\alpha}.
\end{equation}
From Eqs. (\ref{eq:JL}), (\ref{eq:JH}), (\ref{eq:branch1}), and (\ref{eq:branch2}), we obtain
\begin{eqnarray}
&&\rho_L=\frac{\alpha}{\beta_{\rm eff,b}(1+\alpha)}, \ \rho_{L+1}=\frac{\alpha_{\rm eff,b}}{1+\alpha_{\rm eff,b}},\\
&&\rho_{2L}=\frac{\alpha_{\rm eff,b}}{\beta_{\rm eff,m}(1+\alpha)}, \ \rho_{2L+1}=1-\frac{\beta}{\alpha_{\rm eff,m}(1+\beta)},\\
&&\beta_{\rm eff,b}=\frac{2+\alpha}{1+\alpha}(>\alpha).
\label{eq:LDLDHD_beta}
\end{eqnarray}
We note that $\alpha_{\rm eff, m}$ and $\beta_{\rm eff, m}$ cannot be determined from the above relationships. 

\item (LD, HD, LD) phase

The following conditions must be satisfied:
\begin{equation}
\alpha<\beta_{\rm eff, b}, \ \alpha_{\rm eff, b}>\beta_{\rm eff, m}, \ \alpha_{\rm eff, m}<\beta.
\label{eq:LDHDLD}
\end{equation}
Using Eqs. (\ref{eq:JL}), (\ref{eq:JH}), and (\ref{eq:J}), we obtain
\begin{equation}
\frac{\alpha}{1+\alpha}=\frac{2\beta_{\rm eff, m}}{1+\beta_{\rm eff, m}}=\frac{\alpha_{\rm eff, m}}{1+\alpha_{\rm eff, m}},
\label{eq:LDHDLD_alpha}
\end{equation} 
resulting in
\begin{equation}
\alpha_{\rm eff, m}=\alpha, \ \beta_{\rm eff, m}=\frac{\alpha}{2+\alpha}.
\end{equation}
From Eqs. (\ref{eq:JL}), (\ref{eq:JH}), (\ref{eq:branch1}), and (\ref{eq:branch3}), we obtain
\begin{eqnarray}
&&\rho_L=\frac{\alpha}{\beta_{\rm eff,b}(1+\alpha)},\\
&&\rho_{L+1}=1-\frac{\beta_{\rm eff,m}}{\alpha_{\rm eff,b}(1+\beta_{\rm eff,m})},\\
&&\rho_{2L}=\frac{1}{1+\beta_{\rm eff,m}}, \ \rho_{2L+1}=\frac{\alpha_{\rm eff,m}}{(1+\alpha_{\rm eff,m})},\\
&&\alpha_{\rm eff,m}=\alpha=1-\mu.
\label{eq:LDHDLD_beta}
\end{eqnarray}
We note that $\alpha_{\rm eff, b}$ and $\beta_{\rm eff, b}$ cannot be determined from the above relationships. 

\item (LD, HD, HD) phase

The following conditions must be satisfied:
\begin{equation}
\alpha<\beta_{\rm eff, b}, \ \alpha_{\rm eff, b}>\beta_{\rm eff, m}, \ \alpha_{\rm eff, m}>\beta.
\label{eq:LDHDHD}
\end{equation}
Using Eqs. (\ref{eq:JL}), (\ref{eq:JH}), and (\ref{eq:J}), we obtain
\begin{equation}
\frac{\alpha}{1+\alpha}=\frac{2\beta_{\rm eff, m}}{1+\beta_{\rm eff, m}}=\frac{\beta}{1+\beta},
\label{eq:LDHDHD_alpha}
\end{equation} 
resulting in
\begin{equation}
\alpha=\beta, \ \beta_{\rm eff, m}=\frac{\alpha}{2+\alpha}.
\end{equation}
From Eqs. (\ref{eq:JL}), (\ref{eq:JH}), (\ref{eq:branch1}), and (\ref{eq:branch3}), we obtain
\begin{eqnarray}
&&\rho_L=\frac{\alpha}{\beta_{\rm eff,b}(1+\alpha)},\\
&&\rho_{L+1}=1-\frac{\beta_{\rm eff,m}}{\alpha_{\rm eff,b}(1+\beta_{\rm eff,m})},\\
&&\rho_{2L}=\frac{1}{1+\beta_{\rm eff,m}}, \ \rho_{2L+1}=1-\frac{\beta}{\alpha_{\rm eff,m}(1+\beta)},\\
&&\alpha_{\rm eff,m}=1-\mu.
\label{eq:LDHDHD_beta}
\end{eqnarray}
We note that $\alpha_{\rm eff, b}$ and $\beta_{\rm eff, b}$ cannot be determined from the above relationships. 

\item (HD, LD, LD) phase

The following conditions must be satisfied:
\begin{equation}
\alpha>\beta_{\rm eff, b}, \ \alpha_{\rm eff, b}<\beta_{\rm eff, m}, \ \alpha_{\rm eff, m}<\beta.
\label{eq:HDLDLD}
\end{equation}
Using Eqs. (\ref{eq:JL}), (\ref{eq:JH}), and (\ref{eq:J}), we obtain
\begin{equation}
\frac{\beta_{\rm eff, b}}{1+\beta_{\rm eff, b}}=\frac{2\alpha_{\rm eff, b}}{1+\alpha_{\rm eff, b}}=\frac{\alpha_{\rm eff, m}}{1+\alpha_{\rm eff, m}},
\label{eq:HDLDLD_alpha}
\end{equation} 
resulting in
\begin{equation}
\alpha_{\rm eff, b}=\frac{\beta_{\rm eff, b}}{2+\beta_{\rm eff, b}}, \ \alpha_{\rm eff, m}=\beta_{\rm eff, b}.
\label{eq:HDLDLD_e}
\end{equation}
From Eqs. (\ref{eq:JL}), (\ref{eq:branch1}), and (\ref{eq:HDLDLD_e}), we obtain
\begin{eqnarray}
&&\rho_{L+1}=\frac{\alpha_{\rm eff, b}}{1+\alpha_{\rm eff, b}}, \ \beta_{\rm eff, b}=\sqrt{2},
\end{eqnarray}
which never satisfies $\alpha>\beta_{\rm eff, b}$; therefore, this phase cannot exist.

\item (HD, LD, HD) phase

The following conditions must be satisfied:
\begin{equation}
\alpha>\beta_{\rm eff, b}, \ \alpha_{\rm eff, b}<\beta_{\rm eff, m}, \ \alpha_{\rm eff, m}>\beta.
\label{eq:HDLDHD}
\end{equation}
Using Eqs. (\ref{eq:JL}), (\ref{eq:JH}), and (\ref{eq:J}), we obtain
\begin{equation}
\frac{\beta_{\rm eff, b}}{1+\beta_{\rm eff, b}}=\frac{2\alpha_{\rm eff, b}}{1+\alpha_{\rm eff, b}}=\frac{\beta}{1+\beta},
\label{eq:HDLDHD_J}
\end{equation} 
resulting in
\begin{equation}
\alpha_{\rm eff, b}=\frac{\beta}{2+\beta}, \ \beta_{\rm eff, b}=\beta.
\label{eq:HDLDHD_e}
\end{equation}
From Eqs. (\ref{eq:JL}), (\ref{eq:branch1}), and (\ref{eq:HDLDHD_e}), we obtain
\begin{eqnarray}
&&\rho_{L}=\frac{1}{1+\beta_{\rm eff, b}}, \ \alpha_{\rm eff, b}=\frac{1}{1+\beta_{\rm eff, b}}=\frac{\beta}{1+\beta}.
\label{eq:HDLDHD_e2}
\end{eqnarray}
Eq. (\ref{eq:HDLDHD_e}) contradicts Eq. (\ref{eq:HDLDHD_e2}); therefore, this phase cannot exist.

\item (HD, HD, LD) phase

The following conditions must be satisfied:
\begin{equation}
\alpha>\beta_{\rm eff, b}, \ \alpha_{\rm eff, b}>\beta_{\rm eff, m}, \ \alpha_{\rm eff, m}<\beta.
\label{eq:HDHDLD}
\end{equation}
Using Eqs. (\ref{eq:JL}), (\ref{eq:JH}), and (\ref{eq:J}), we obtain
\begin{equation}
\frac{\beta_{\rm eff, b}}{1+\beta_{\rm eff, b}}=\frac{2\beta_{\rm eff, m}}{1+\beta_{\rm eff, m}}=\frac{\alpha_{\rm eff, m}}{1+\alpha_{\rm eff, m}},
\label{eq:HDHDLD_J}
\end{equation} 
resulting in
\begin{equation}
\alpha_{\rm eff, m}=\beta_{\rm eff, b}, \ \beta_{\rm eff, m}=\frac{\beta_{\rm eff, b}}{2+\beta_{\rm eff, b}}.
\label{eq:HDHDLD_e}
\end{equation}
From Eqs. (\ref{eq:JL}), (\ref{eq:JH}), (\ref{eq:branch3}) and (\ref{eq:HDHDLD_e}), we obtain
\begin{eqnarray}
&&\rho_L=\frac{1}{1+\beta_{\rm eff,b}}, \ \rho_{L+1}=1-\frac{\beta_{\rm eff,m}}{\alpha_{\rm eff,b}(1+\beta_{\rm eff,m})},\\
&&\rho_{2L}=\frac{1}{1+\beta_{\rm eff,m}}, \ \rho_{3L}=\frac{\alpha_{\rm eff,m}}{\beta(1+\alpha_{\rm eff,m})},\\
&&\alpha_{\rm eff, m}=\beta_{\rm eff, b}=1-\mu,\label{eq:HDHDLD_e2}\\
&&\alpha_{\rm eff, b}=\frac{1}{2-\mu}, \ \beta_{\rm eff, m}=\frac{1-\mu}{3-\mu}(<\alpha_{\rm eff, b}).\label{eq:HDHDLD_e3}
\end{eqnarray}
Based on Eqs. (\ref{eq:HDHDLD_e2}), and (\ref{eq:HDHDLD_e3}), Eq. (\ref{eq:HDHDLD}) can be simplified to
\begin{equation}
\alpha>1-\mu, \ \beta>1-\mu.
\end{equation}

\item (HD, HD, HD) phase

The following conditions must be satisfied:
\begin{equation}
\alpha>\beta_{\rm eff, b}, \ \alpha_{\rm eff, b}>\beta_{\rm eff, m}, \ \alpha_{\rm eff, m}>\beta.
\label{eq:HDHDHD}
\end{equation}
Using Eqs. (\ref{eq:JL}), (\ref{eq:JH}), and (\ref{eq:J}), we obtain
\begin{equation}
\frac{\beta_{\rm eff, b}}{1+\beta_{\rm eff, b}}=\frac{2\beta_{\rm eff, m}}{1+\beta_{\rm eff, m}}=\frac{\beta}{1+\beta},
\label{eq:HDHDHD_J}
\end{equation} 
resulting in
\begin{equation}
\beta_{\rm eff, b}=\beta, \ \beta_{\rm eff, m}=\frac{\beta}{2+\beta}.
\label{eq:HDHDHD_e}
\end{equation}
From Eqs. (\ref{eq:JL}), (\ref{eq:JH}), (\ref{eq:branch3}) and (\ref{eq:HDHDLD_e}), we get
\begin{eqnarray}
&&\rho_L=\frac{1}{1+\beta_{\rm eff,b}}, \ \rho_{L+1}=1-\frac{\beta_{\rm eff,m}}{\alpha_{\rm eff,b}(1+\beta_{\rm eff,m})},\\
&&\rho_{2L}=\frac{1}{1+\beta_{\rm eff,m}}, \ \rho_{2L+1}=1-\frac{1}{\alpha_{\rm eff, m}(1+\beta)},\\
&&\alpha_{\rm eff, m}=1-\mu, \ \alpha_{\rm eff, b}=\frac{1}{1+\beta}(>\beta_{\rm eff, m}).
\label{eq:HDHDHD_e2}
\end{eqnarray}
Based on Eqs. (\ref{eq:HDHDHD_e}) and (\ref{eq:HDHDHD_e2}), Eq. (\ref{eq:HDHDHD}) can be simplified to
\begin{equation}
\alpha>\beta, \ \beta<1-\mu.
\end{equation}

\item (MC, LD, MC) phase

The following conditions must be satisfied:
\begin{equation}
\alpha=\beta_{\rm eff, b}=\alpha_{\rm eff, m}=\beta=1, \ \alpha_{\rm eff, b}<\beta_{\rm eff, m}.
\label{eq:MCLDMC}
\end{equation}
Using Eqs. (\ref{eq:JL}), (\ref{eq:JM}), and (\ref{eq:J}), we obtain
\begin{equation}
\frac{2\alpha_{\rm eff, b}}{1+\alpha_{\rm eff, b}}=\frac{1}{2}\Leftrightarrow \alpha_{\rm eff, b}=\frac{1}{3}.
\label{eq:MCLDMC_J}
\end{equation}
From Eqs. (\ref{eq:JL}), (\ref{eq:JM}), (\ref{eq:branch2}) and (\ref{eq:MCLDMC_J}), we obtain
\begin{eqnarray}
&&\rho_L=\alpha_{\rm eff, b}, \ \rho_{L+1}=\frac{\alpha_{\rm eff,b}}{1+\alpha_{\rm eff,b}},\\
&&\rho_{2L}=\frac{\alpha_{\rm eff,m}}{2}, \ \rho_{2L+1}=1-\beta_{\rm eff,m},\\
&&\beta_{\rm eff,m}=\frac{1}{2}>\alpha_{\rm eff, b}.
\label{eq:MCLDMC_beta}
\end{eqnarray}

\item (MC, HD, MC) phase

The following conditions must be satisfied:
\begin{equation}
\alpha=\beta_{\rm eff, b}=\alpha_{\rm eff, m}=\beta=1, \ \alpha_{\rm eff, b}>\beta_{\rm eff, m},
\label{eq:MCHDMC}
\end{equation}
which contradicts Eq. (\ref{eq:branch3}); therefore, this phase cannot exist.
\end{enumerate}

Based on the above analysis, it can be seen that the system can exhibit seven possible phases, specifically, (LD, LD, LD), (LD, LD, HD), (LD, HD, LD), (LD, HD, HD), (HD, HD, LD), (HD, HD, HD), and (MC, LD, MC). Surprisingly, the region in which $\alpha>\beta, \alpha\geq1-\mu$, and $\beta\geq1-\mu$, or $\alpha=\beta=1$, can exhibit two possible phases, which we refer to as nonergodic phases.


\section{Phase boundaries}
\label{sec:PB}
This section investigates the phase boundaries, specifically, (i) $\alpha=1-\mu<\beta$, (ii) $\alpha=\beta>1-\mu$, (iii) $\beta=1-\mu<\alpha$, (iv) $\alpha=\beta<1-\mu$, and (v) $\alpha=\beta=1-\mu$. This investigation is performed because the phase and the density profile for the boundaries cannot be determined from the simple approximation in Appendix \ref{sec:simple}. We note that all the simulations below with $L=500$.

\begin{enumerate}
\item $\alpha=1-\mu<\beta$

With low-initial-density conditions, the system clearly exhibits the (LD, LD, LD) phase from an early stage in the simulation. 

In contrast, with high-initial-density conditions, the steps to reach a steady state are somewhat complex; specifically, 
\begin{enumerate}
\item Conflicts occur due to congestion at the merging point, leading to $\alpha_{\rm eff, m}=1-\mu<\beta$. This results in the LD phase in Subsystem 4. 
\item A SW arises at the right boundary and moves throughout Subsystems 1, 2, and 3 because the maximal input flow equals the flow at the merging point, that is,
\begin{equation}
\frac{\alpha}{1+\alpha}=\frac{2\beta_{\rm eff,m}}{1+\beta_{\rm eff,m}}\left(=\frac{2\times\frac{1-\mu}{3-\mu}}{1+\frac{1-\mu}{3-\mu}}\right),
\end{equation}
leading to a temporary SW phase in Subsystems 1, 2, and 3.
\item Once the SW reaches the merging point, the value of $\beta_{\rm eff,m}$ changes from $\beta_{\rm eff,m}=\frac{1-\mu}{3-\mu}$ to $\beta_{\rm eff,m}=\frac{1}{1+\alpha}$, leading to the disappearance of the SW.
\item Subsystems 1, 2, and 3 exhibit an LD phase because the maximal input flow is less than the flow at the merging point, that is,
\begin{equation}
\frac{\alpha}{1+\alpha}<\frac{2\beta_{\rm eff,m}}{1+\beta_{\rm eff,m}}\left(=\frac{2\times\frac{1}{1+\alpha}}{1+\frac{1}{1+\alpha}}\right).
\end{equation}
Therefore, the system yields the (LD, LD, LD) phase. 
\end{enumerate}
Figure \ref{fig:st0205} presents space-time plots for $(\alpha, \beta, \mu)=(0.2, 0.5, 0.8)$. In this figure and subsequent figures, the space-time plots only illustrate the states of Subsystems 1, 2, and 4 because the states of Subsystem 3 are almost identical to those of Subsystem 2. Figure \ref{fig:st0205} confirms the above explanations. Refer to Sec. \ref{sec:profile} about the density profile.
\begin{figure}[htbp]
\begin{center}
\includegraphics[width=10cm,clip]{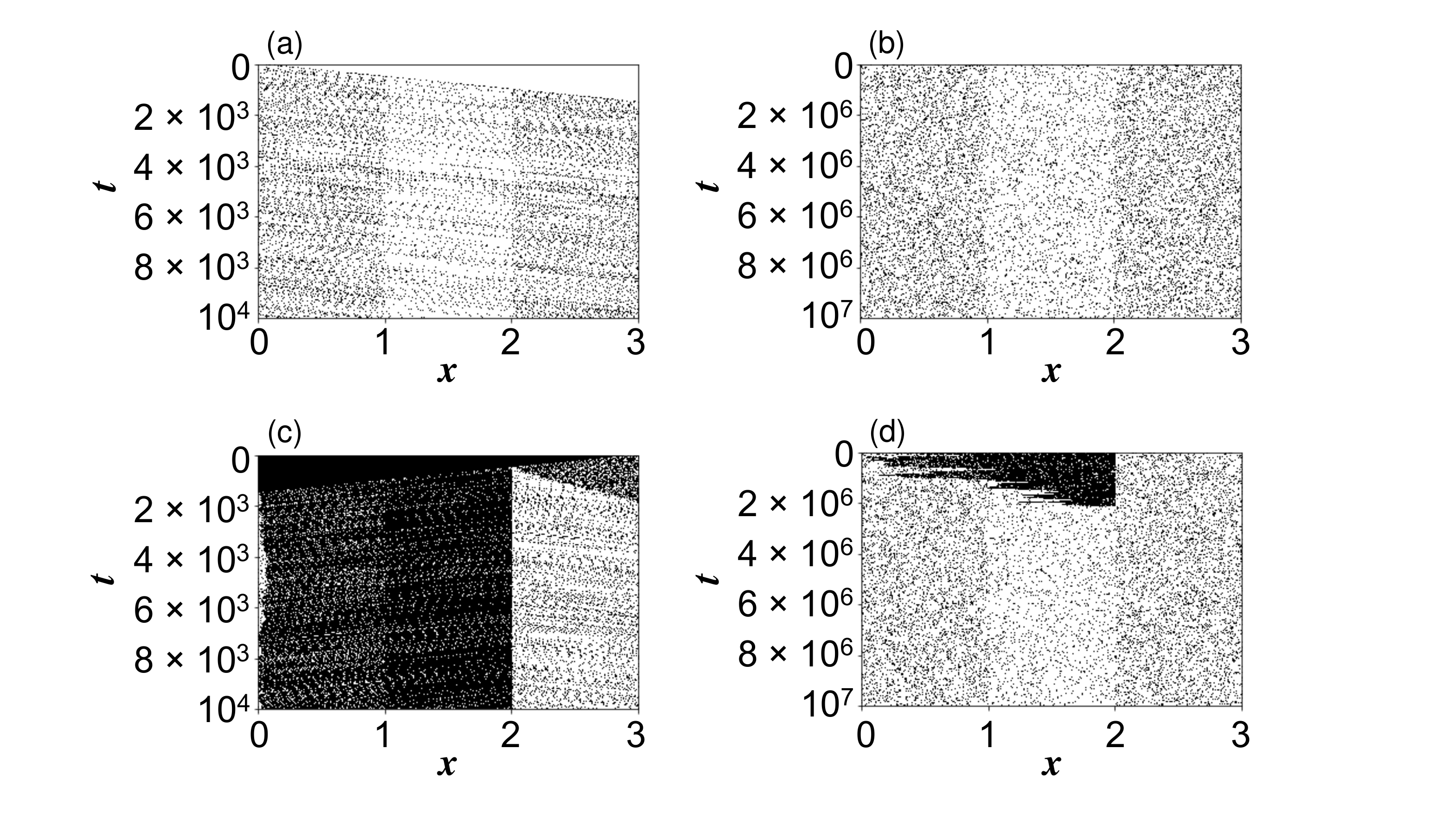}
\caption{(Color Online) Space-time plots for $(\alpha, \beta, \mu)=(0.2, 0.5, 0.8)$. The value $t$ represents the simulation time. The low-initial-density condition ($\rho_{\rm ini}=0$) is adopted for (a) and (b), while the high-initial-density condition ($\rho_{\rm ini}=1$) for (c) and (d). Snapshots with $t\in[0,10^4]$ are plotted for (a) and (c), while snapshots with $t\in[0,10^7]$ are plotted every $10^3$ time steps for (b) and (d).}
\label{fig:st0205}
\end{center}
\end{figure}

\item $\alpha=\beta>1-\mu$

With low-initial-density conditions, the steps to a steady state are as follows:
\begin{enumerate}
\item A SW arises at the right boundary because $\alpha=\beta$.
\item Once the SW reaches the merging point, conflicts occur at the merging point and the value of $\alpha_{\rm eff, m}$ changes from $\alpha_{\rm eff, m}=\alpha$ to $\alpha_{\rm eff, m}=1-\mu$. This leads to the disappearance of the SW.
\item Subsystems 1, 2, and 3 exhibit the HD phase because the maximal input flow exceeds the flow at the merging point, that is,
\begin{equation}
\frac{\alpha}{1+\alpha}>\frac{2\beta_{\rm eff,m}}{1+\beta_{\rm eff,m}}\left(=\frac{2\times\frac{1-\mu}{3-\mu}}{1+\frac{1-\mu}{3-\mu}}\right).
\end{equation}
In contrast, Subsystem 4 exhibits an LD phase because $\alpha_{\rm eff, m}<\beta$. Therefore, the system exhibits the (HD, HD, LD) phase.
\end{enumerate}
With high-initial-density conditions, the steps are as follows:
\begin{enumerate}
\item Conflicts occur due to congestion at the merging point, leading to $\alpha_{\rm eff, m}=1-\mu<\beta$. This results in the LD phase in Subsystem 4. 
\item Finally, Subsystems 1, 2, and 3 present the HD phase, because the maximal input flow exceeds the flow at the merging point, i.e.,
\begin{equation}
\frac{\alpha}{1+\alpha}>\frac{2\beta_{\rm eff,m}}{1+\beta_{\rm eff,m}}\left(=\frac{2\times\frac{1-\mu}{3-\mu}}{1+\frac{1-\mu}{3-\mu}}\right).
\end{equation}
Therefore, the system yields the (HD, HD, LD) phase.
\end{enumerate}
Figure \ref{fig:st0505} presents space-time plots for $(\alpha, \beta, \mu)=(0.5, 0.5, 0.8)$, which confirm the above explanations. Refer to Sec. \ref{sec:profile} about the density profile.
\begin{figure}[htbp]
\begin{center}
\includegraphics[width=10cm,clip]{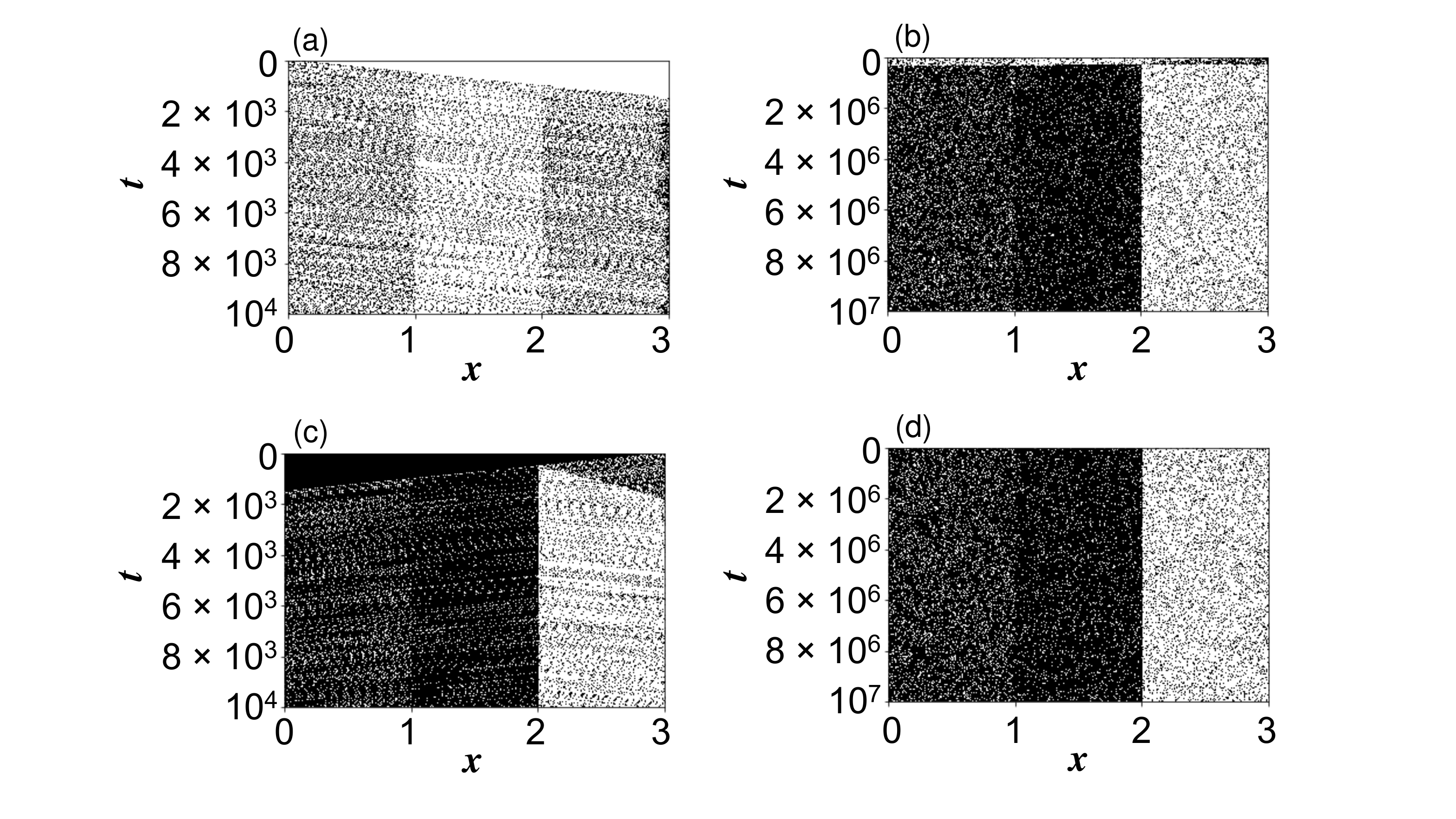}
\caption{(Color Online) Space-time plots for $(\alpha, \beta, \mu)=(0.5, 0.5, 0.8)$. Snapshots with $t\in[0,10^4]$ are plotted for (a) and (c), while snapshots with $t\in[0,10^6]$ are plotted every $10^3$ time steps for (b) and (d). The other conditions are the same as in Fig. \ref{fig:st0205}.}
\label{fig:st0505}
\end{center}
\end{figure}

\item $\beta=1-\mu<\alpha$

With low-initial-density conditions, the steps to a steady state are as follows:
\begin{enumerate}
\item Particles accumulate at the right boundary because the input flow exceeds the output flow, that is,
\begin{equation}
\frac{\alpha}{1+\alpha}>\frac{\beta}{1+\beta}.
\end{equation}
\item Conflicts occur due to congestion at the merging point, leading to $\alpha_{\rm eff, m}=1-\mu=\beta$. This results in the SW phase in Subsystem 4.
\item Subsystems 1, 2, and 3 exhibit the HD phase because the input flow exceeds the flow at the merging point, that is,
\begin{equation}
\frac{\alpha}{1+\alpha}>\frac{2\beta_{\rm eff,m}}{1+\beta_{\rm eff,m}}\left(=\frac{2\times\frac{1-\mu}{3-\mu}}{1+\frac{1-\mu}{3-\mu}}\right).
\end{equation}
Therefore, the system yields the (HD, HD, SW) phase.
\end{enumerate}
We note that with high-initial-density conditions the steps start from (b).

Figure \ref{fig:st0502} presents space-time plots for $(\alpha, \beta, \mu)=(0.5, 0.2, 0.8)$, which confirm the above explanations.
\begin{figure}[htbp]
\begin{center}
\includegraphics[width=10cm,clip]{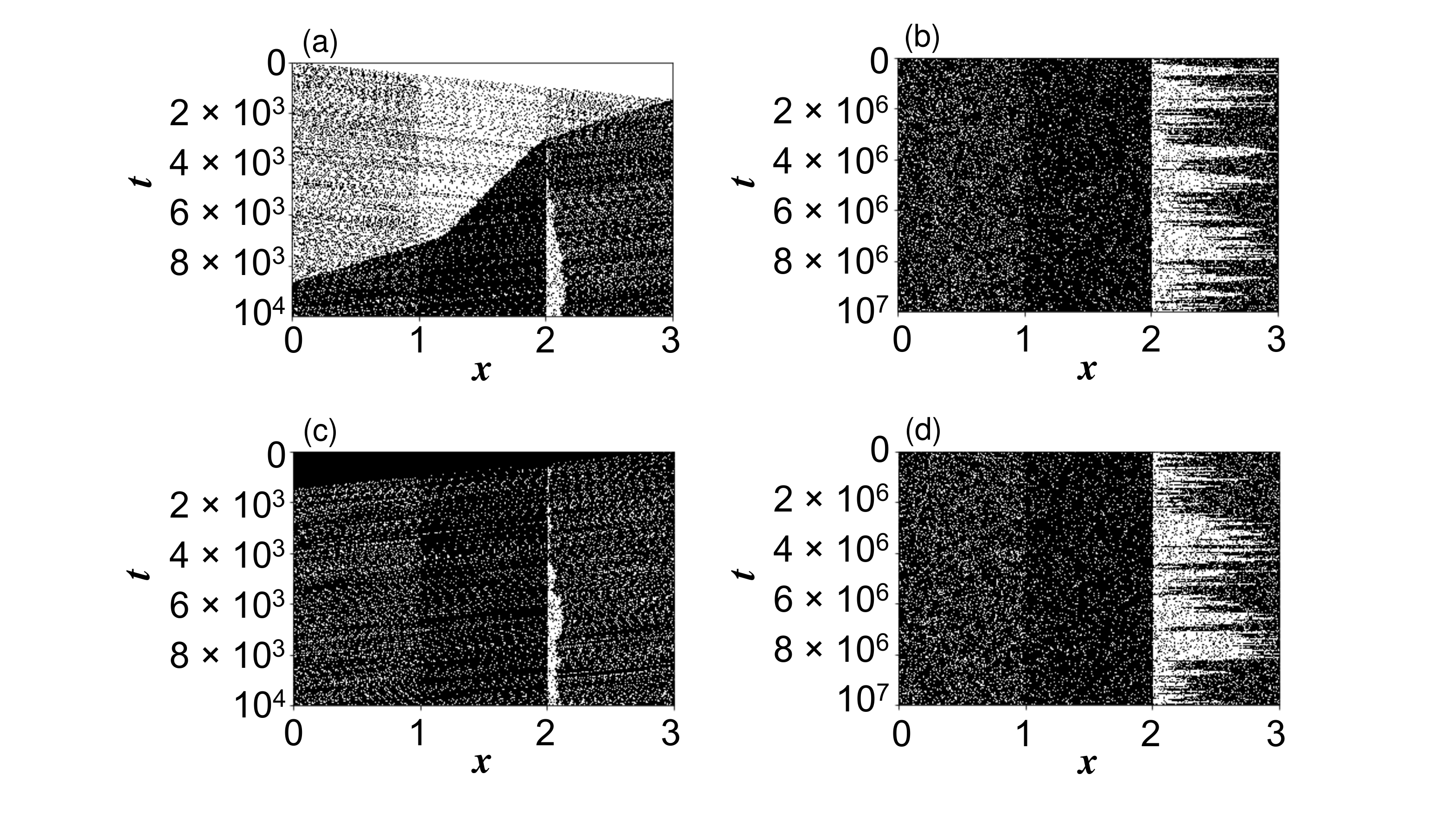}
\caption{(Color Online) Space-time plots for $(\alpha, \beta, \mu)=(0.5, 0.2, 0.8)$. Snapshots with $t\in[0,10^4]$ are plotted for (a) and (c), while snapshots with $t\in[0,10^6]$ are plotted every $10^3$ time steps for (b) and (d). The other conditions are the same as in Fig. \ref{fig:st0205}.}
\label{fig:st0502}
\end{center}
\end{figure}

The density profile for Subsystems 1, 2, and 3 is the same as that for the (HD, HD, HD) phase, whereas that for Subsystem 4 ($2<x\leq3$) can be represented as follows~\cite{de1999exact}:
\begin{equation}
\rho(x)=\frac{1-\mu}{2-\mu}+\frac{\mu}{2-\mu}(x-2).
\end{equation}

\item $\alpha=\beta<1-\mu$

A SW arises at the right (left) boundary with low-initial-density (high-initial-density) conditions because $\alpha=\beta$. The SW does not disappear, and another shock wave arises because the maximal input flow cannot exceed the flow at the merging point. Therefore, the shock wave moves throughout the system, and the system yields the (SW, SW, SW) phase.

Figure \ref{fig:st0101} presents space-time plots for $(\alpha, \beta, \mu)=(0.1, 0.1, 0.8)$, which confirm the above explanations.
\begin{figure}[htbp]
\begin{center}
\includegraphics[width=10cm,clip]{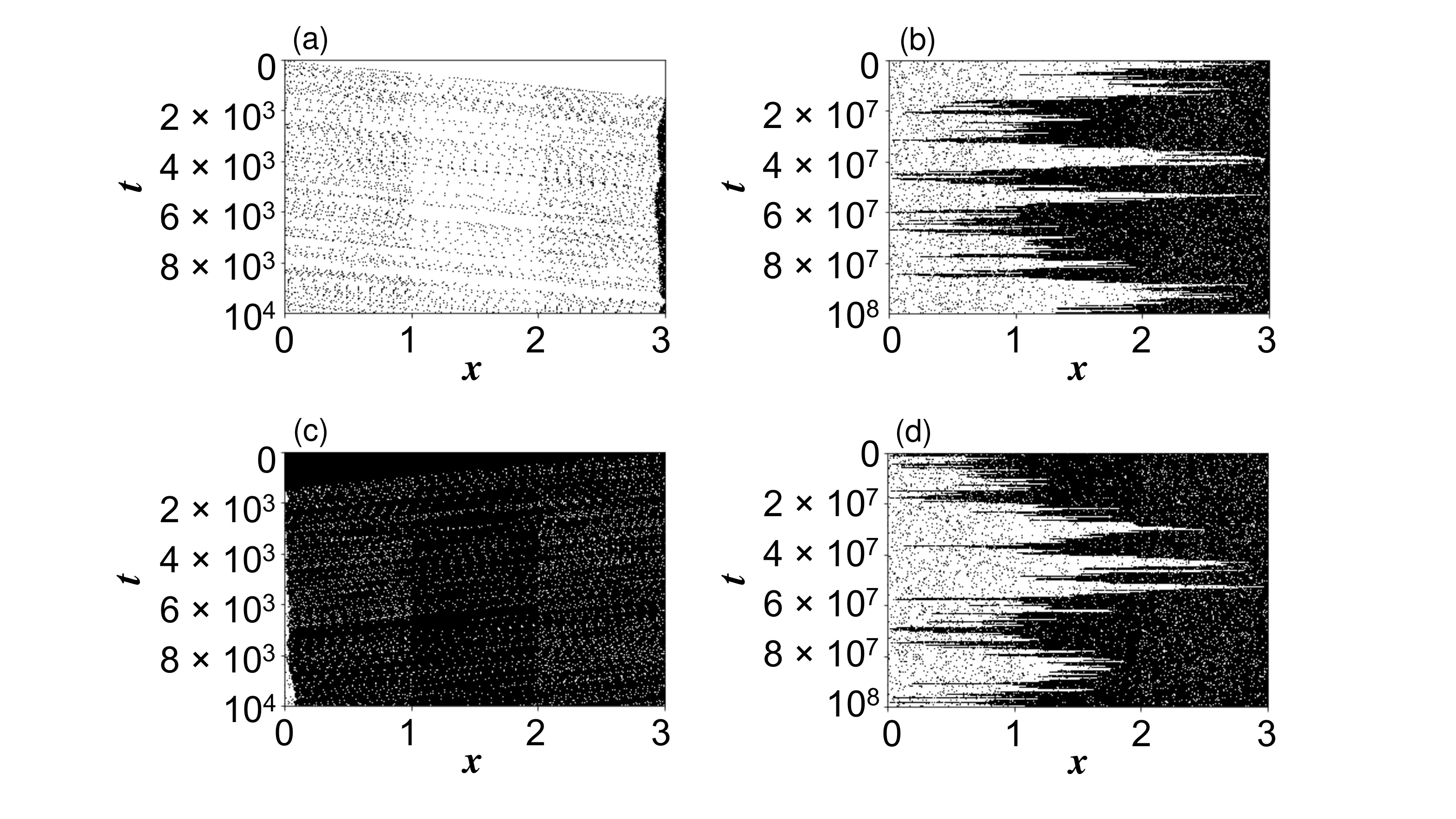}
\caption{(Color Online) Space-time plots for $(\alpha, \beta, \mu)=(0.1, 0.1, 0.8)$. Snapshots with $t\in[0,10^4]$ are plotted for (a) and (c), while snapshots with $t\in[0,10^8]$ are plotted every $10^4$ time steps for (b) and (d). The other conditions are the same as in Fig. \ref{fig:st0205}.}
\label{fig:st0101}
\end{center}
\end{figure}

To obtain the density profile, domain wall theory~\cite{kolomeisky1998phase,Pronina_2005,PhysRevE.77.051108} is used. First, the domain wall in each subsystem have a random walk though a speed
\begin{eqnarray}
v_{\rm DW}=\frac{Q_+-Q_-}{\rho_+^j-\rho_-^j},
\label{eq:DW}
\end{eqnarray}
where $+$ $(-)$ represents the phase to the right (left) of the domain wall, $j$ represents the subsystem number, and the right direction of an axis is defined as the positive direction.

We define the position of the domain wall as $x=\frac{i}{L}$, where $i$ is the site number. The domain wall moves with velocity $v_{1}$ in Subsystem 1 when $0<x\leq1$, with velocity $v_{2}$ in Subsystem 2 and 3 when $1<x\leq2$, and with velocity $v_{4}$ in Subsystem 4 when $2<x\leq3$, as illustrated in Fig. \ref{fig:DW}. The velocity is given as
\begin{eqnarray}
v_{j}=\frac{Q_j}{\rho_+^j-\rho_-^j}, \ {\rm for} \ j=1, 2, 4,
\label{eq:vj}
\end{eqnarray}
where
\begin{eqnarray}
&&\rho_-^1=\frac{\alpha}{1+\alpha}, \  \rho_+^1=\frac{1}{1+\alpha}, \label{eq:rho1}\\
&&\rho_-^2=\frac{\alpha}{2(1+\alpha)},  \ \rho_+^2=\frac{2+\alpha}{2(1+\alpha)}, \label{eq:rho2}\\
&&\rho_-^4=\frac{\beta}{1+\beta}, \  \rho_+^4=\frac{1}{1+\beta}, \label{eq:rho4}\\
&&Q_1=\frac{\alpha}{1+\alpha}, \  Q_2=\frac{\alpha}{2(1+\alpha)}, \ Q_4=\frac{\beta}{1+\beta}.\label{eq:rhoq}
\end{eqnarray}
We note that this simple approximation determines whether the system exhibits the (LD, LD, HD) or (LD, HD, HD) phase, and the values of $(\rho_-^1, \rho_+^1, \rho_-^2, \rho_+^2, \rho_-^4, \rho_+^4, Q_1, Q_2, Q_4)$ are determined under the assumption.
\begin{figure}[htbp]
\begin{center}
\includegraphics[width=8cm,clip]{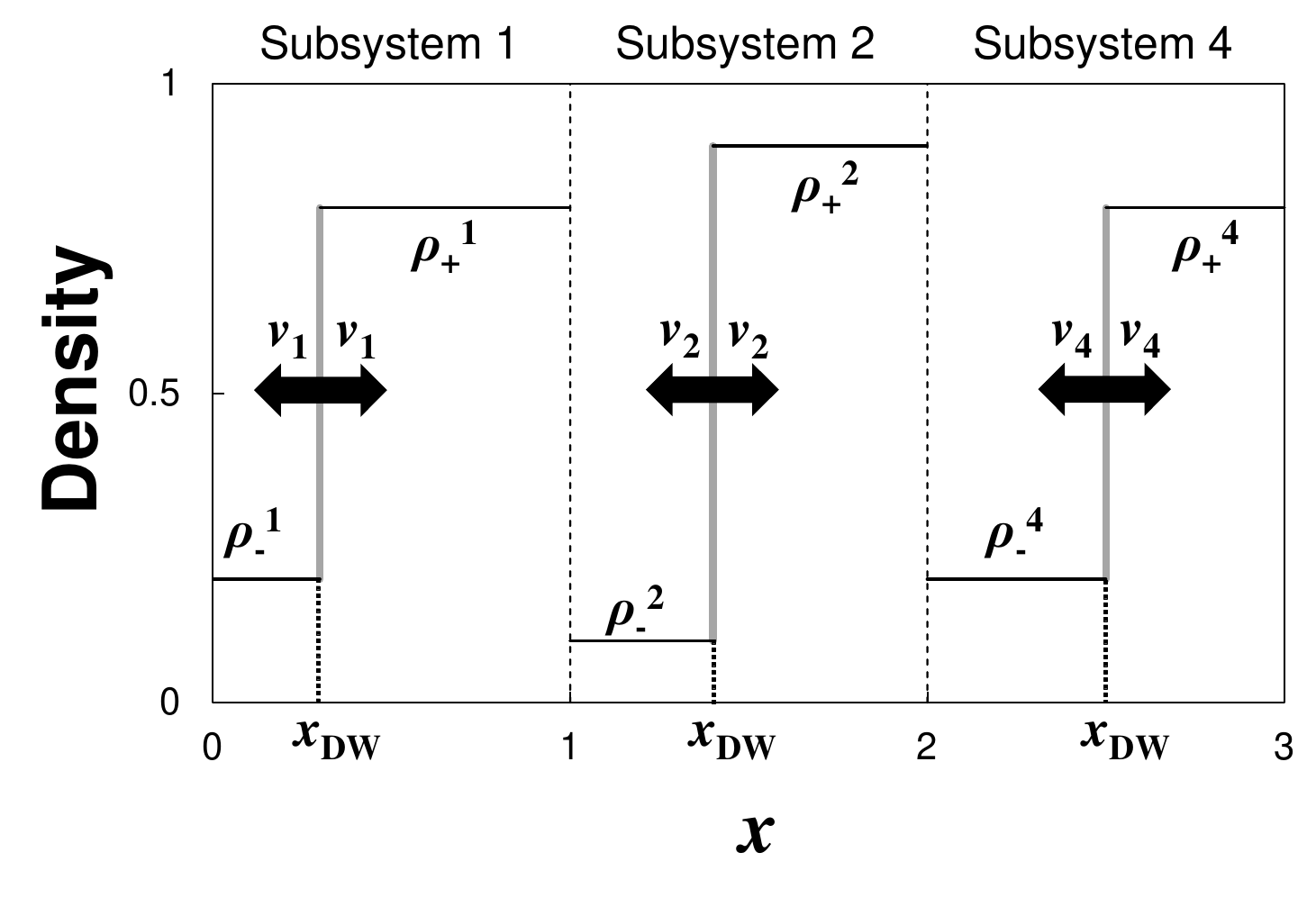}
\caption{(Color Online) Schematic of domain wall dynamics for $\alpha=\beta<1-\mu$. The domain wall moves with velocity $v_{1}$, $v_{1}$, and $v_{4}$ in Subsystems 1, 2, and 4, respectively. We note that only one domain wall can exist in the system at the same time.}
\label{fig:DW}
\end{center}
\end{figure}

Using Eqs. (\ref{eq:vj})--(\ref{eq:rhoq}), $v_1$, $v_2$, and $v_4$ can be expressed as
\begin{equation}
v_1=\frac{\alpha}{1-\alpha}, \ v_2=\frac{\alpha}{2}, \ v_4=\frac{\beta}{1-\beta}.
\label{eq:v124}
\end{equation}
The probability of a domain wall in a site in Subsystems 1, 2, and 4 is equal to $\frac{q_1}{L}$, $\frac{q_2}{L}$, and $\frac{q_4}{L}$, respectively, where $q_1$, $q_2$, and $q_4$ represent the probabilities of the domain wall in Subsystems 1, 2, and 4, respectively. As a result, at the branching/merging point, we have
\begin{equation}
\frac{v_1q_1}{L}=\frac{v_2q_2}{L}=\frac{v_4q_4}{L}.
\label{eq:v124/L}
\end{equation}
In addition, $q_1$, $q_2$, and $q_4$ satisfy the normalization condition
\begin{equation}
q_1+q_2+q_4=1.
\label{eq:normal}
\end{equation}
Using Eqs. (\ref{eq:v124})--(\ref{eq:normal}), we have
\begin{eqnarray}
&&q_1=\frac{\beta(1-\alpha)}{\alpha+3\beta-2\alpha\beta}, \label{eq:q1}\\
&&q_2=\frac{2\beta}{\alpha+3\beta-2\alpha\beta},\label{eq:q2}\\
&&q_4=\frac{\alpha(1-\beta)}{\alpha+3\beta-2\alpha\beta}.\label{eq:q4}
\end{eqnarray}
Therefore, the probabilities of domain walls in a certain region are expressed as
\begin{equation}
  {\rm Prob}(x_{\rm DW}<x)=
  \begin{cases}
    q_1x \ (0<x\leq1), \\
    q_1+q_2(x-1) \ (1<x\leq2), \\
    q_1+q_2+q_4(x-2) \ (2<x\leq3).
  \end{cases}
\end{equation}
Thus, we obtain the density at $x$ in the system $\rho(x)$ as follows:
\begin{widetext}
\begin{equation}
  \rho(x)=
  \begin{cases}
    \rho_-^1(1-q_1x)+\rho_+^1q_1x \ \ \ (0<x\leq1), \\
    \rho_-^2[1-q_1-q_2(x-1)]+\rho_+^2[q_1+q_2(x-1)] \ \ \ (1<x\leq2), \\
    \rho_-^4[1-q_1-q_2-q_3(x-2)]+\rho_+^4[q_1+q_2+q_3(x-2)] \ \ \ (2<x\leq3).
  \end{cases}
\label{eq:rhox}
\end{equation}
\end{widetext}
From Eqs. (\ref{eq:rho1})--(\ref{eq:rho4}), (\ref{eq:q1})--(\ref{eq:q4}), (\ref{eq:rhox}), and $\alpha=\beta$, we have
\begin{widetext}
\begin{equation}
  \rho(x)=
  \begin{cases}
    \displaystyle\frac{\alpha}{1+\alpha}+\frac{(1-\alpha)^2}{(1+\alpha)(4-2\alpha)}x \ \ \ (0<x\leq1). \\
\\
     \displaystyle\frac{1+\alpha-\alpha^2}{(1+\alpha)(4-2\alpha)}+\frac{1}{(1+\alpha)(2-\alpha)}(x-1) \ \ \ (1<x\leq2). \\
\\
    \displaystyle\frac{3-\alpha^2}{(1+\alpha)(4-2\alpha)}+\frac{(1-\alpha)^2}{(1+\alpha)(4-2\alpha)}(x-2) \ \ \ (2<x\leq3).
  \end{cases}
\end{equation}
\end{widetext}

\item $\alpha=\beta=1-\mu$

With low-initial-density conditions, the steps to a steady state are as follows:
\begin{enumerate}
\item A SW occurs at the right boundary because $\alpha=\beta$.

\item Once the SW reaches the merging point, conflicts occur at the merging point and the value of $\alpha_{\rm eff, m}$ changes from $\alpha_{\rm eff, m}=\alpha$ to $\alpha_{\rm eff, m}=1-\mu$. 

\item Another SW arises in Subsystems 1, 2, and 3 because the maximal input flow equals the flow at the merging point, that is,
\begin{equation}
\frac{\alpha}{1+\alpha}=\frac{2\beta_{\rm eff,m}}{1+\beta_{\rm eff,m}}\left(=\frac{2\times\frac{1-\mu}{3-\mu}}{1+\frac{1-\mu}{3-\mu}}\right).
\end{equation}
In contrast, the SW in Subsystem 4 is still present because the flow at the merging point equals to the maximal output flow, that is,
\begin{equation}
\frac{2\beta_{\rm eff,m}}{1+\beta_{\rm eff,m}}\left(=\frac{2\times\frac{1-\mu}{3-\mu}}{1+\frac{1-\mu}{3-\mu}}\right)=\frac{\beta}{1+\beta}.
\end{equation}
Therefore, the system yields the (SW$_1$, SW$_1$, SW$_2$) phase. We note that SW$_1$ and SW$_2$ are clearly separated because different SWs determine the phases.
\end{enumerate}

In contrast, with high-initial-density conditions, the steps to a steady state are as follows:
\begin{enumerate}
\item Conflicts occur due to congestion at the merging point, leading to $\alpha_{\rm eff, m}=1-\mu=\beta$.

\item A SW arises at the left boundary, while another SW arises at the merging point because the maximal input flow, the flow at the merging point, and the maximal output flow all have the same value, that is,
\begin{equation}
\frac{\alpha}{1+\alpha}=\frac{2\beta_{\rm eff,m}}{1+\beta_{\rm eff,m}}\left(=\frac{2\times\frac{1-\mu}{3-\mu}}{1+\frac{1-\mu}{3-\mu}}\right)=\frac{\beta}{1+\beta},
\end{equation}
leading to the (SW$_1$, SW$_1$, SW$_2$) phase.
\end{enumerate}
Figure \ref{fig:st0202} presents space-time plots for $(\alpha, \beta, \mu)=(0.2, 0.2, 0.8)$, which confirms the above explanations.
\begin{figure}[htbp]
\begin{center}
\includegraphics[width=10cm,clip]{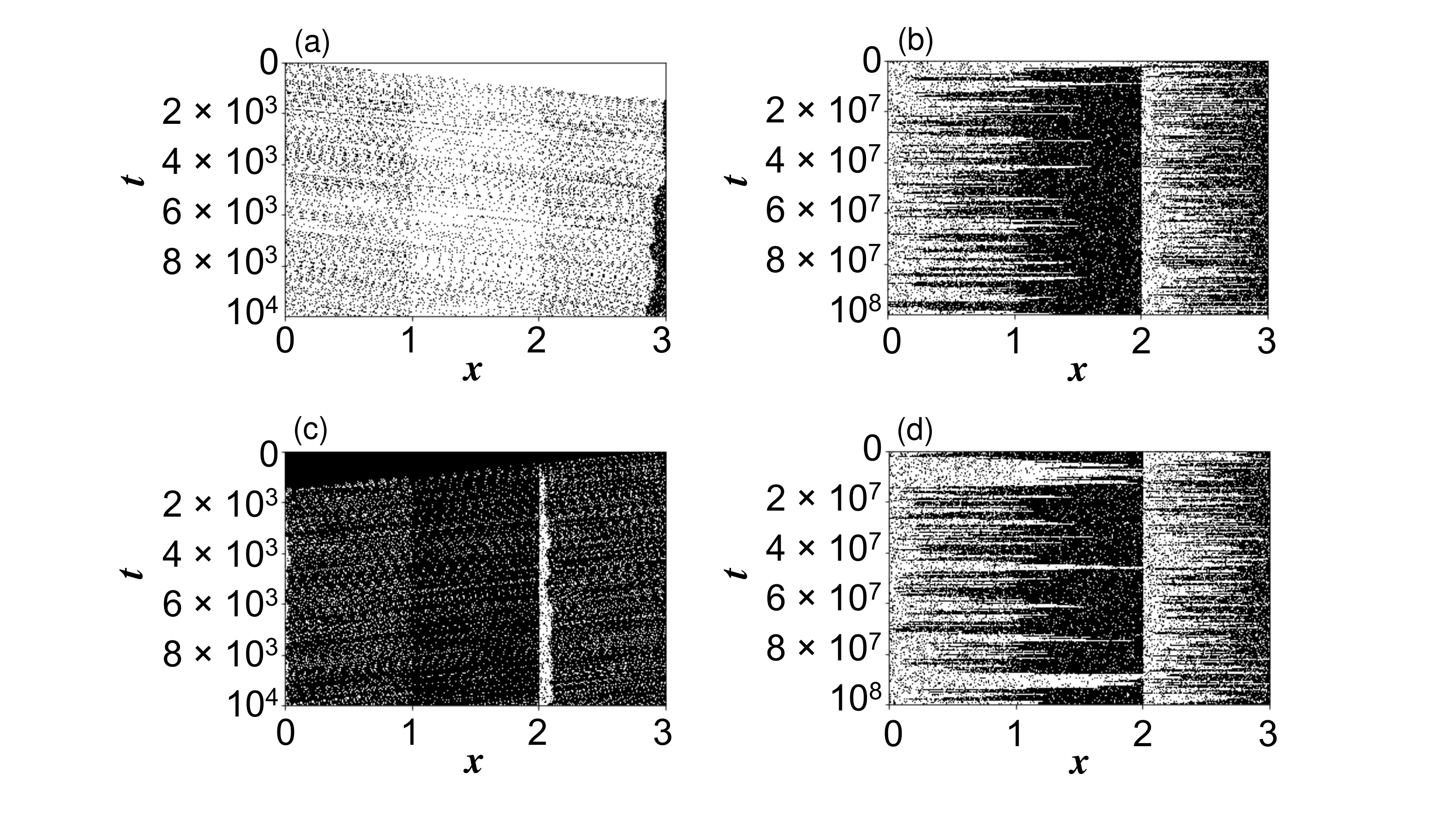}
\caption{(Color Online) Space-time plots for $(\alpha, \beta, \mu)=(0.2, 0.2, 0.8)$. Snapshots with $t\in[0,10^4]$ are plotted for (a) and (c), while the snapshots with $t\in[0,10^8]$ are plotted every $10^4$ time steps for (b) and (d). The other conditions are the same as in Fig. \ref{fig:st0205}.}
\label{fig:st0202}
\end{center}
\end{figure}

Next, we use domain wall theory to obtain the density profile. Unlike in the case of $\alpha=\beta<1-\mu$, two domain wall exist in the system; one moves through Subsystem 1 and 2, while the other moves through Subsystem 4. 

First, the density profile for Subsystem 4 ($2<x\leq3$) can be represented as follows~\cite{de1999exact}:
\begin{equation}
\rho(x)=\frac{1-\mu}{2-\mu}+\frac{\mu}{2-\mu}(x-2).
\end{equation}

Then, we obtain the density profile for Subsystems 1 and 2. The same values are given for $(\rho_-^1, \rho_+^1, \rho_-^2, \rho_+^2, Q_1, Q_2, v_1, v_2)$ as in the case of $\alpha=\beta<1-\mu$.

At the branching point, we have
\begin{equation}
\frac{v_1q_1}{L}=\frac{v_2q_2}{L}.
\label{eq:v124/L_2}
\end{equation}
Here, $q_1$ and $q_2$ satisfy the normalization condition
\begin{equation}
q_1+q_2=1.
\label{eq:normal_2}
\end{equation}
From Eqs. (\ref{eq:v124}), (\ref{eq:v124/L_2}) and (\ref{eq:normal_2}), we have
\begin{eqnarray}
&&q_1=\frac{1-\alpha}{3-\alpha}, \label{eq:q1_2}\\
&&q_2=\frac{2}{3-\alpha}.\label{eq:q2_2}
\end{eqnarray}
From Eqs. (\ref{eq:rhox}), (\ref{eq:q1_2}), and (\ref{eq:q2_2}), we have
\begin{equation}
  \rho(x)=
  \begin{cases}
    \displaystyle\frac{\alpha}{1+\alpha}+\frac{(1-\alpha)^2}{(1+\alpha)(3-\alpha)}x \ \ \ (0<x\leq1), \\
\\
     \displaystyle\frac{2-\alpha}{2(3-\alpha)}+\frac{2}{(1+\alpha)(3-\alpha)}(x-1) \ \ \ (1<x\leq2). 
  \end{cases}
\end{equation}
\end{enumerate}

\section{Phase diagrams}
\label{sec:phase}
\begin{figure}[htbp]
\begin{center}
\includegraphics[width=8cm,clip]{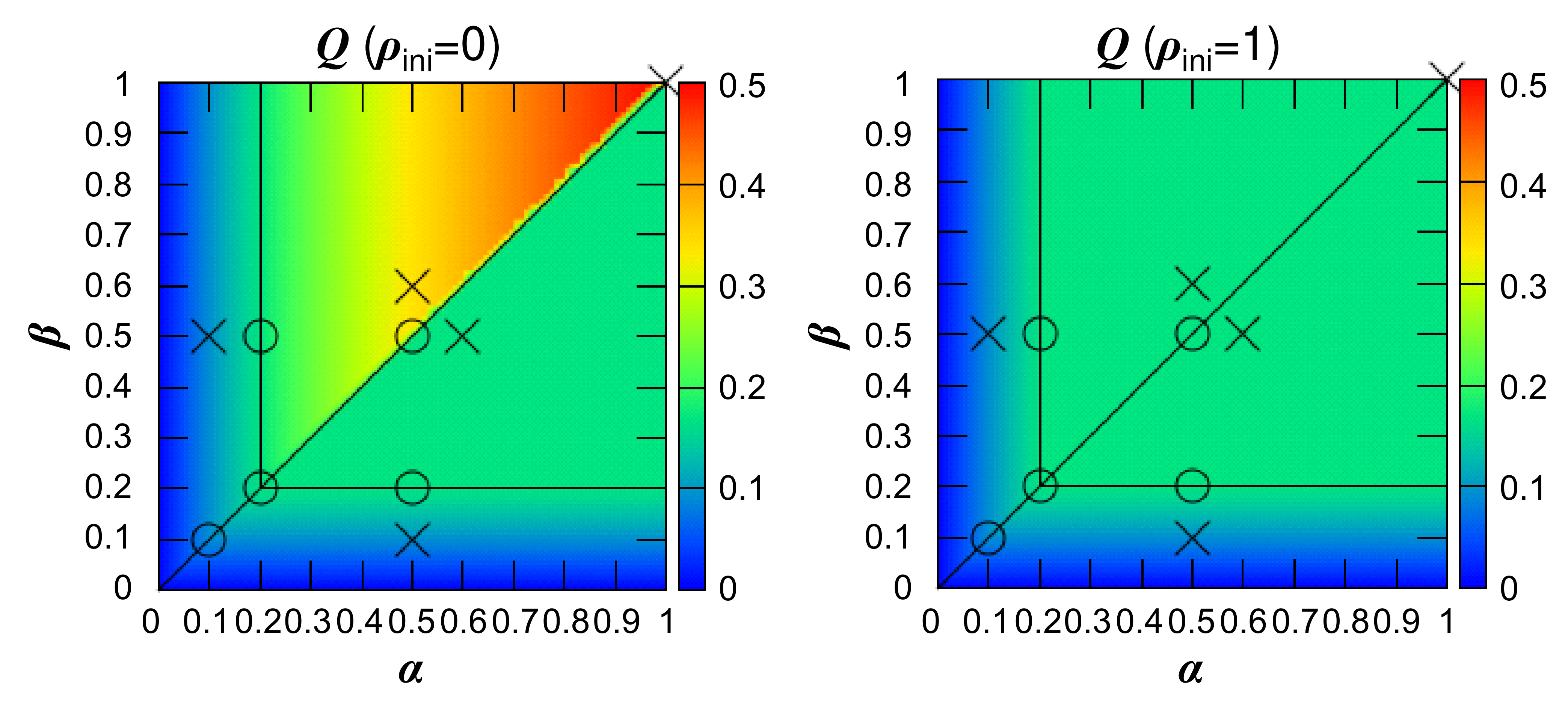}
\caption{(Color Online) (a) Simulation values of $Q$ for $\rho_{\rm ini}=0$. (b) Simulation values of $Q$ for $\rho_{\rm ini}=1$. The five black crosses in each panel represent ($\alpha, \beta$)=(0.1, 0.5), (0.5, 0.1), (0.5, 0.6), (0.6, 0.5), and (1,1), while the five black circles represent ($\alpha, \beta$)=(0.2, 0.5), (0.5, 0.5), (0.5, 0.2), (0.1, 0.1), and (0.2,0.2), which are used in Sec. \ref{sec:profile}.}
\label{fig:phase2}
\end{center}
\end{figure}

\begin{figure}[htbp]
\begin{center}
\includegraphics[width=10cm,clip]{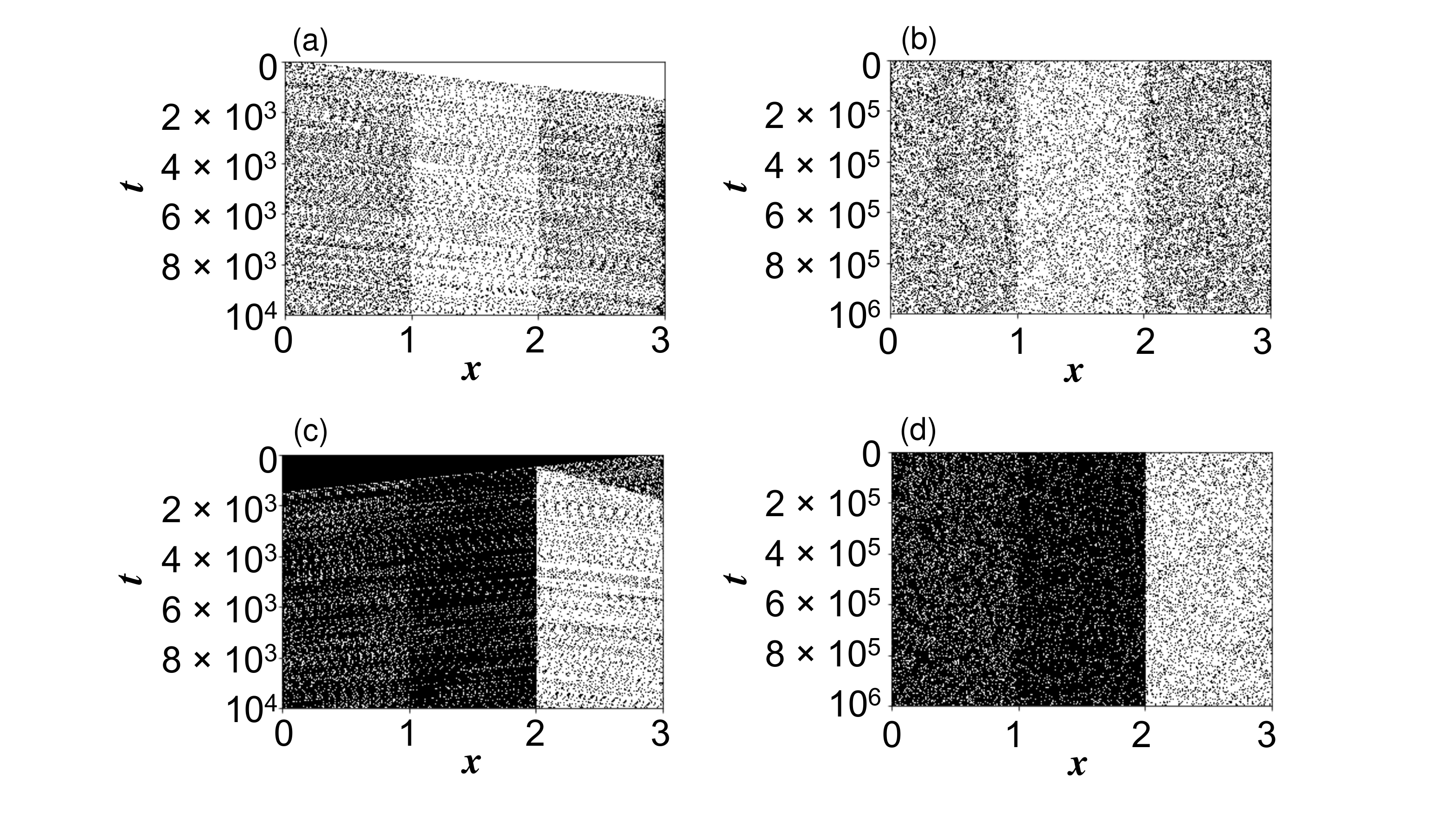}
\caption{(Color Online) Space-time plots for $(\alpha, \beta, \mu)=(0.5, 0.6, 0.8)$. Snapshots with $t\in[0,10^4]$ are plotted for (a) and (c), while the snapshots with $t\in[0,10^6]$ are plotted every $10^2$ time steps for (b) and (d). The other conditions are the same as in Fig. \ref{fig:st0205}.}
\label{fig:st0506}
\end{center}
\end{figure}

\begin{figure}[htbp]
\begin{center}
\includegraphics[width=10cm,clip]{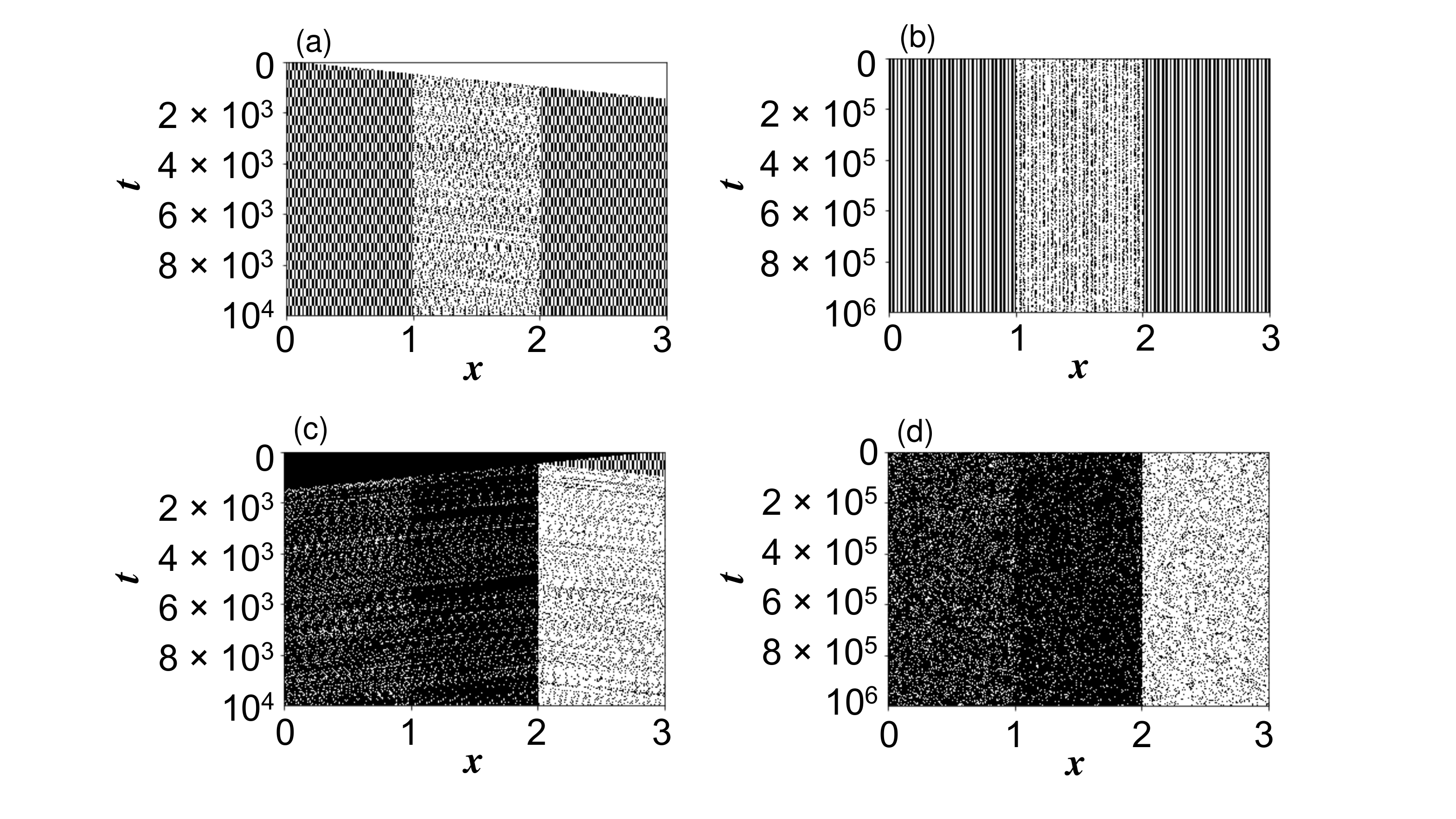}
\caption{(Color Online) Space-time plots for $(\alpha, \beta, \mu)=(1,1,0.8)$. Snapshots with $t\in[0,10^4]$ are plotted for (a) and (c), while the snapshots with $t\in[0,10^6]$ are plotted every $10^2$ time steps for (b) and (d). The other conditions are the same as in Fig. \ref{fig:st0205}.}
\label{fig:st11}
\end{center}
\end{figure}

Based on the above discussion, we describe the phase diagram presented Fig. \ref{fig:PD}. To validate our theoretical approximation, we perform simulations with two kinds of initial conditions: $\rho_{\rm ini}=0$ (low-initial-density condition) and $\rho_{\rm ini}=1$ (high-initial-density condition), as illustrated in Fig. \ref{fig:phase2}. As expected, a nonergodic phase, in which the phase depends on the initial conditions, is observed. The dependency of the phase on the initial conditions can be confirmed by Figs. \ref{fig:st0506} and \ref{fig:st11}, which present space-time plots for $(\alpha, \beta, \mu)=(0.5, 0.6, 0.8)$ and for $(\alpha, \beta, \mu)=(1, 1, 0.8)$, respectively.

\section{Density profile}
\label{sec:profile}
This section compares the density profiles obtained from the theoretical analysis and the simulations. 

\begin{figure}[htbp]
\begin{center}
\includegraphics[width=7.5cm,clip]{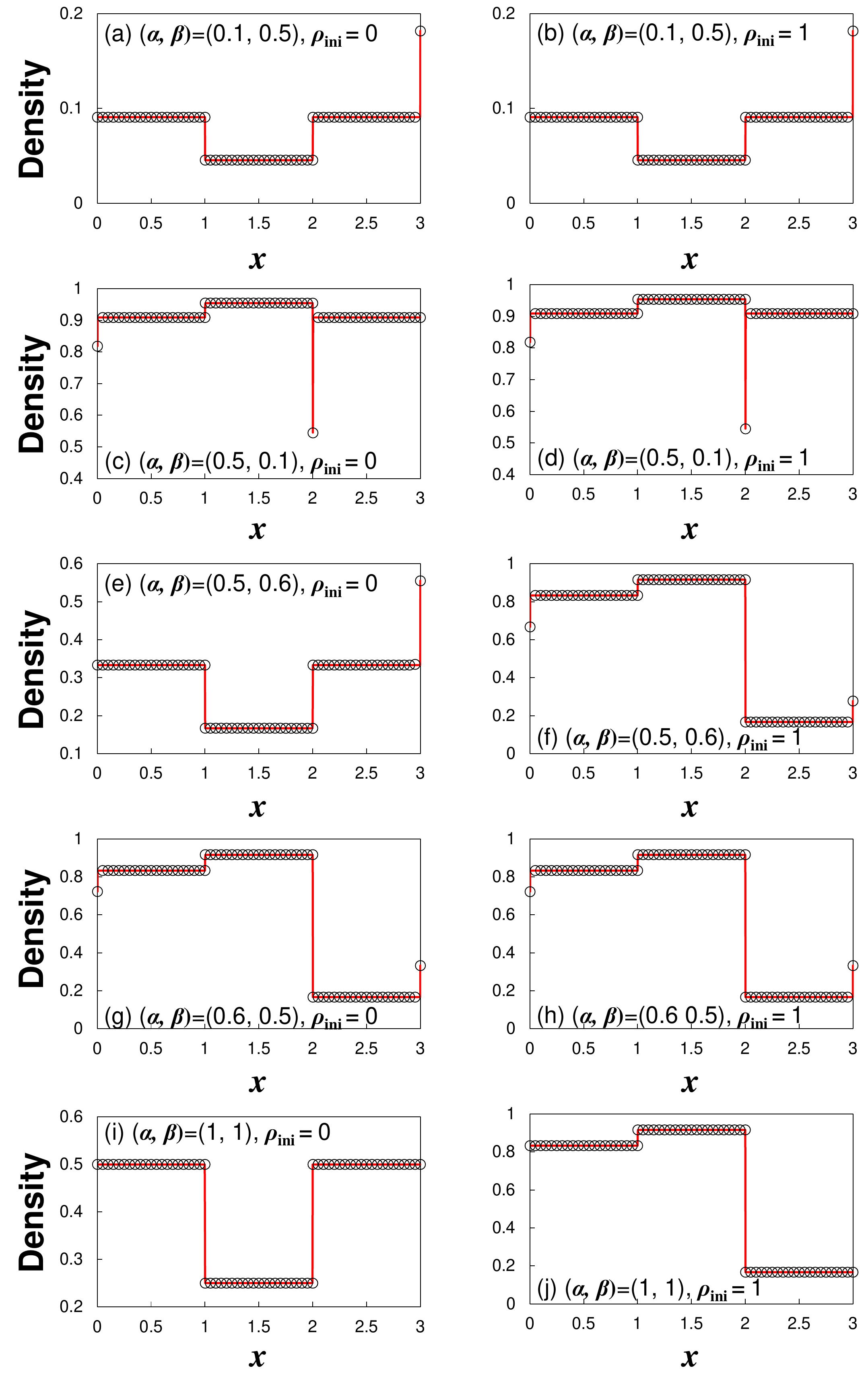}
\caption{(Color Online) Density profiles with $\mu=0.8$ for each phase. Circles represent the simulation results, while red solid lines represent estimates of the theoretical results	. Other parameters are displayed in each panel.}
\label{fig:density1}
\end{center}
\end{figure}

\begin{figure}[htbp]
\begin{center}
\includegraphics[width=7.5cm,clip]{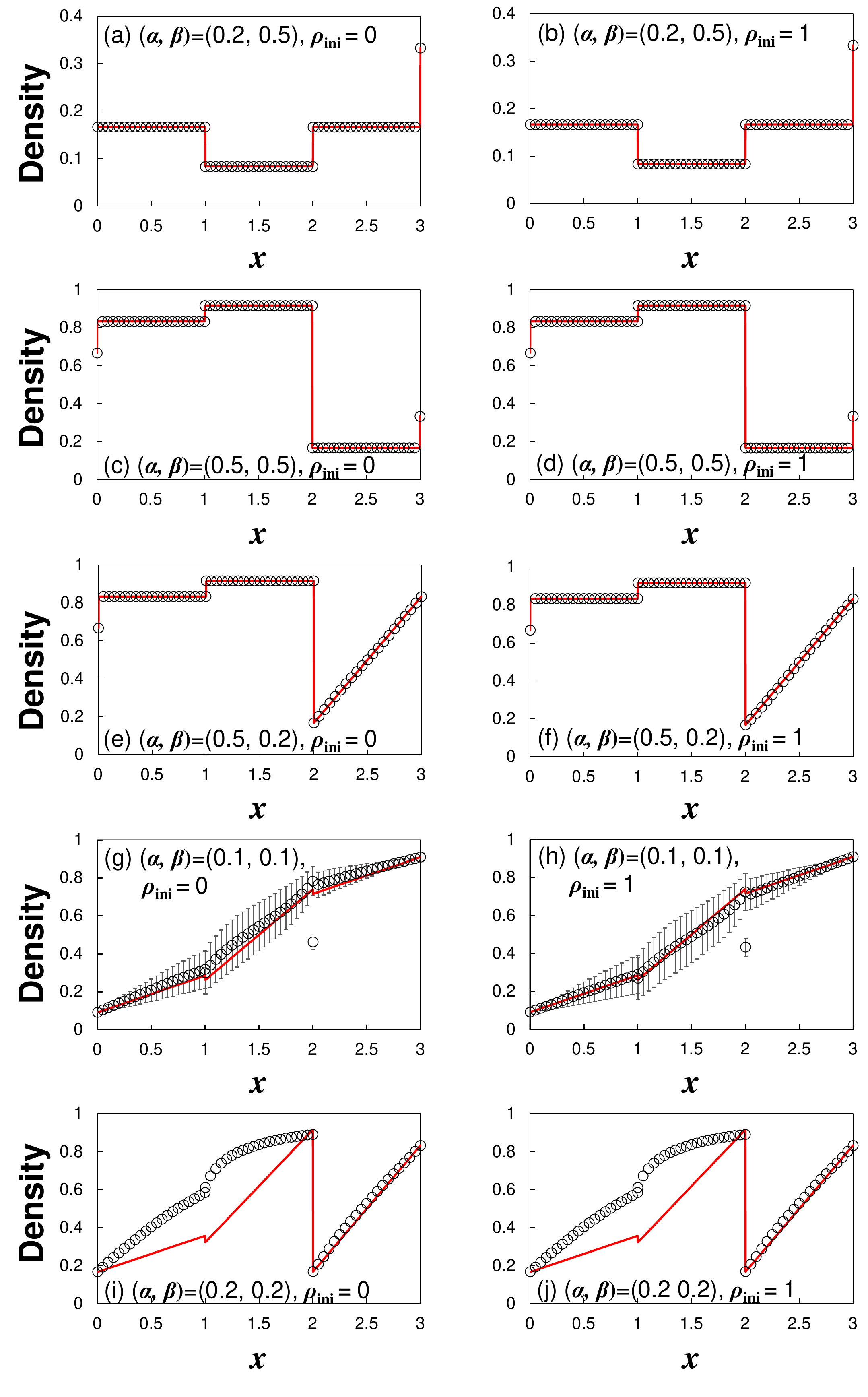}
\caption{(Color Online) Density profiles with $\mu=0.8$ for each phase boundary. Circles represent the simulation results, while red solid lines represent the theoretical results. Other parameters are displayed in each panel. To obtain the densities, we evolve the system for $10^8$ time steps and calculate the averages of $10^8$ time steps for (a), (b), (e), (f), (i), and (j). In contrast, we use the average of 10 trial of the densities for $10^8$ time steps after evolving the system for $10^8$ time steps with one standard error for (g) and (h) because their standard errors are relatively large.}
\label{fig:density2}
\end{center}
\end{figure}

Figure \ref{fig:density1} presents density profiles for the black crosses in Fig. \ref{fig:phase2}, which exist in each phase, for the two initial conditions, comparing the simulation and theoretical results. We note that only the density profiles of Subsystems 1, 2, and 4 are presented because the profiles of Subsystems 2 and 3 are equivalent due to their symmetry (the same holds for Fig. \ref{fig:density2}). The simulation results are in excellent agreement with the theoretical results.

Figure \ref{fig:density2} presents density profiles for the black circles in Fig. \ref{fig:phase2}, which exist on each phase boundary, for the two initial conditions, comparing the simulation results and theoretical results. The simulation results are generally in excellent agreement with the results of the simple approximations, except for $(\alpha,\beta)=(0.1,0.1), (0.2,0.2)$. For $(\alpha,\beta)=(0.1,0.1)$, the deviation of densities for each trial is relatively large because a SW moves through the system; however, the average generally agrees with the theoretical results. 

The above discussion demonstrates the validity of our theoretical analysis.

\section{Probability of two steady-state phases with various $\mu$}
\label{sec:tippingapp}

\begin{figure}[htbp]
\begin{center}
\includegraphics[width=9cm,clip]{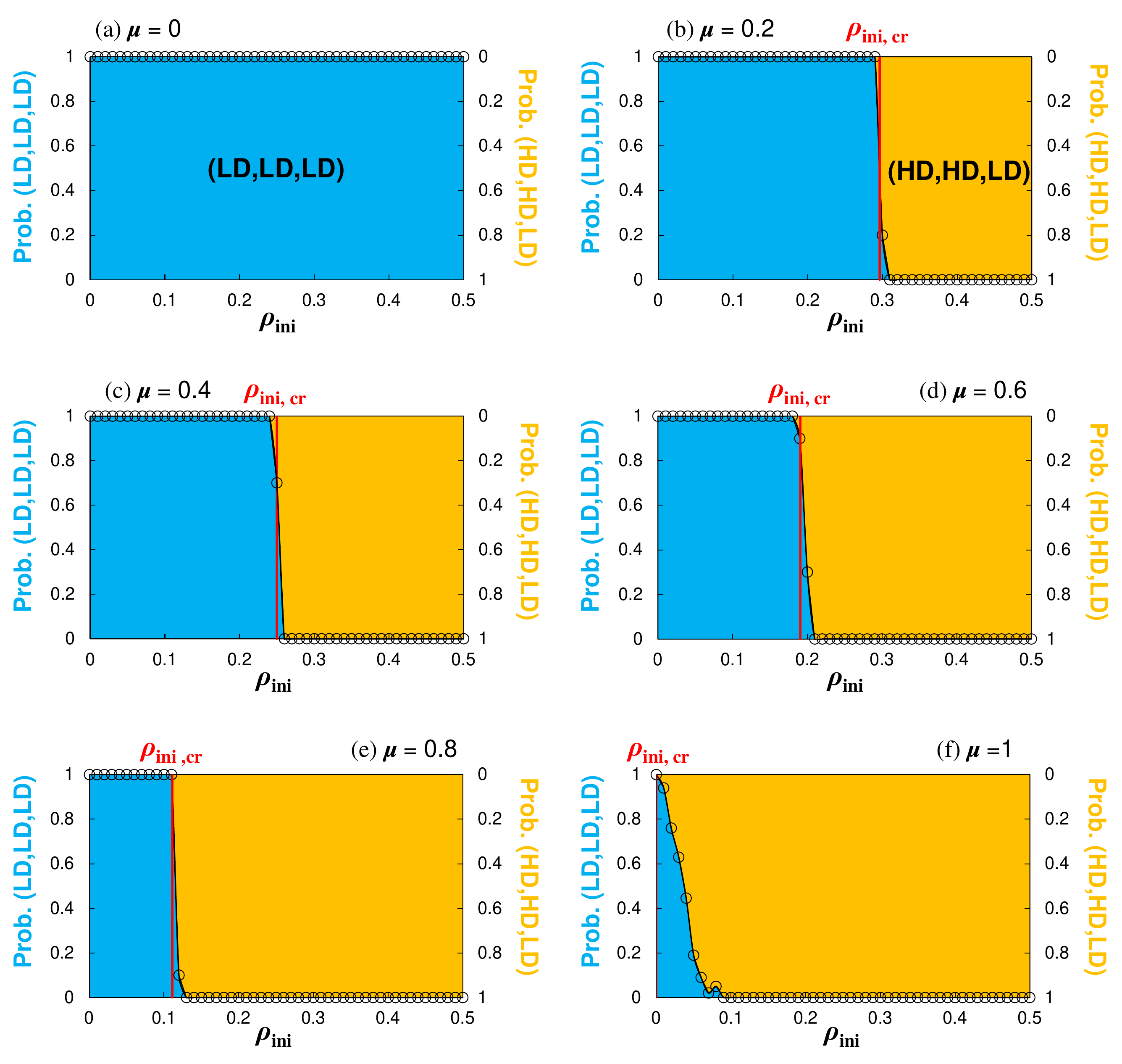}
\caption{(Color Online) Probability of two steady-state phases (LD, LD, LD) (blue) or (HD, HD, LD) (orange) as a function of $\rho_{\rm ini}$, with the red line representing the estimated $\rho_{\rm ini}=\rho_{\rm ini, cr}$ (i.e., Eq. (\ref{eq:tipping})) for (a) $\mu=0$, (b) $\mu=0.2$, (c) $\mu=0.4$, (d) $\mu=0.6$, (e) $\mu=0.8$, (f) $\mu=1$. For all the cases, we set $(\alpha,\beta)=(0.85,0.9)$, with which the system always satisfies $\alpha>1-\mu$ and $\alpha<\beta$, i.e., the system is in the nonergodic phase, except for the case of $\mu=0$. We calculate the probabilities for each $\rho_{\rm ini}$ in increments of 0.01 (black circles) by 100-time (for $\mu=1$) and 10-time (for others) simulations with random configurations. We note that the panel (e) is identical to Fig. \ref{fig:tipping}. Other explanations are the same as those of Fig. \ref{fig:tipping}.}
\label{fig:tippingapp}
\end{center}
\end{figure}

Figure \ref{fig:tippingapp} presents the hysteresis plots for various $\mu\in\{0,0.2,0.4,0.6,0.8,1\}$. Except for the case of $\mu=0$, we confirm the same phenomena as Fig. \ref{fig:tipping}. We stress that Eq. (\ref{eq:tipping}) does not hold true for the case of $\mu=0$ (no collision), because $Q$ never equals to $Q=Q_{\rm M}$, and therefore, the red line cannot be depicted for all $(\alpha,\beta)$. For the case of $\mu=1$, the theoretical value of $\rho_{\rm ini, cr}$ becomes 0; however, various initial configurations, with which collisions can be avoided, exist in low-initial-density conditions, resulting in an existence of the (LD,LD,LD) phase for small $\rho_{\rm ini, cr}$.

\clearpage


\begin{thebibliography}{49}%
\makeatletter
\providecommand \@ifxundefined [1]{%
 \@ifx{#1\undefined}
}%
\providecommand \@ifnum [1]{%
 \ifnum #1\expandafter \@firstoftwo
 \else \expandafter \@secondoftwo
 \fi
}%
\providecommand \@ifx [1]{%
 \ifx #1\expandafter \@firstoftwo
 \else \expandafter \@secondoftwo
 \fi
}%
\providecommand \natexlab [1]{#1}%
\providecommand \enquote  [1]{``#1''}%
\providecommand \bibnamefont  [1]{#1}%
\providecommand \bibfnamefont [1]{#1}%
\providecommand \citenamefont [1]{#1}%
\providecommand \href@noop [0]{\@secondoftwo}%
\providecommand \href [0]{\begingroup \@sanitize@url \@href}%
\providecommand \@href[1]{\@@startlink{#1}\@@href}%
\providecommand \@@href[1]{\endgroup#1\@@endlink}%
\providecommand \@sanitize@url [0]{\catcode `\\12\catcode `\$12\catcode
  `\&12\catcode `\#12\catcode `\^12\catcode `\_12\catcode `\%12\relax}%
\providecommand \@@startlink[1]{}%
\providecommand \@@endlink[0]{}%
\providecommand \url  [0]{\begingroup\@sanitize@url \@url }%
\providecommand \@url [1]{\endgroup\@href {#1}{\urlprefix }}%
\providecommand \urlprefix  [0]{URL }%
\providecommand \Eprint [0]{\href }%
\providecommand \doibase [0]{http://dx.doi.org/}%
\providecommand \selectlanguage [0]{\@gobble}%
\providecommand \bibinfo  [0]{\@secondoftwo}%
\providecommand \bibfield  [0]{\@secondoftwo}%
\providecommand \translation [1]{[#1]}%
\providecommand \BibitemOpen [0]{}%
\providecommand \bibitemStop [0]{}%
\providecommand \bibitemNoStop [0]{.\EOS\space}%
\providecommand \EOS [0]{\spacefactor3000\relax}%
\providecommand \BibitemShut  [1]{\csname bibitem#1\endcsname}%
\let\auto@bib@innerbib\@empty
\bibitem [{\citenamefont {Scheffer}\ \emph {et~al.}(2012)\citenamefont
  {Scheffer}, \citenamefont {Carpenter}, \citenamefont {Lenton}, \citenamefont
  {Bascompte}, \citenamefont {Brock}, \citenamefont {Dakos}, \citenamefont
  {van~de Koppel}, \citenamefont {van~de Leemput}, \citenamefont {Levin},
  \citenamefont {van Nes}, \citenamefont {Pascual},\ and\ \citenamefont
  {Vandermeer}}]{doi:10.1126/science.1225244}%
  \BibitemOpen
  \bibfield  {author} {\bibinfo {author} {\bibfnamefont {M.}~\bibnamefont
  {Scheffer}}, \bibinfo {author} {\bibfnamefont {S.~R.}\ \bibnamefont
  {Carpenter}}, \bibinfo {author} {\bibfnamefont {T.~M.}\ \bibnamefont
  {Lenton}}, \bibinfo {author} {\bibfnamefont {J.}~\bibnamefont {Bascompte}},
  \bibinfo {author} {\bibfnamefont {W.}~\bibnamefont {Brock}}, \bibinfo
  {author} {\bibfnamefont {V.}~\bibnamefont {Dakos}}, \bibinfo {author}
  {\bibfnamefont {J.}~\bibnamefont {van~de Koppel}}, \bibinfo {author}
  {\bibfnamefont {I.~A.}\ \bibnamefont {van~de Leemput}}, \bibinfo {author}
  {\bibfnamefont {S.~A.}\ \bibnamefont {Levin}}, \bibinfo {author}
  {\bibfnamefont {E.~H.}\ \bibnamefont {van Nes}}, \bibinfo {author}
  {\bibfnamefont {M.}~\bibnamefont {Pascual}}, \ and\ \bibinfo {author}
  {\bibfnamefont {J.}~\bibnamefont {Vandermeer}},\ }\href {\doibase
  10.1126/science.1225244} {\bibfield  {journal} {\bibinfo  {journal}
  {Science}\ }\textbf {\bibinfo {volume} {338}},\ \bibinfo {pages} {344}
  (\bibinfo {year} {2012})}\BibitemShut {NoStop}%
\bibitem [{\citenamefont {Zeng}\ \emph {et~al.}(2020)\citenamefont {Zeng},
  \citenamefont {Gao}, \citenamefont {Shekhtman}, \citenamefont {Guo},
  \citenamefont {Lv}, \citenamefont {Wu}, \citenamefont {Liu}, \citenamefont
  {Levy}, \citenamefont {Li}, \citenamefont {Gao}, \citenamefont {Stanley},\
  and\ \citenamefont {Havlin}}]{doi:10.1073/pnas.1907493117}%
  \BibitemOpen
  \bibfield  {author} {\bibinfo {author} {\bibfnamefont {G.}~\bibnamefont
  {Zeng}}, \bibinfo {author} {\bibfnamefont {J.}~\bibnamefont {Gao}}, \bibinfo
  {author} {\bibfnamefont {L.}~\bibnamefont {Shekhtman}}, \bibinfo {author}
  {\bibfnamefont {S.}~\bibnamefont {Guo}}, \bibinfo {author} {\bibfnamefont
  {W.}~\bibnamefont {Lv}}, \bibinfo {author} {\bibfnamefont {J.}~\bibnamefont
  {Wu}}, \bibinfo {author} {\bibfnamefont {H.}~\bibnamefont {Liu}}, \bibinfo
  {author} {\bibfnamefont {O.}~\bibnamefont {Levy}}, \bibinfo {author}
  {\bibfnamefont {D.}~\bibnamefont {Li}}, \bibinfo {author} {\bibfnamefont
  {Z.}~\bibnamefont {Gao}}, \bibinfo {author} {\bibfnamefont {H.~E.}\
  \bibnamefont {Stanley}}, \ and\ \bibinfo {author} {\bibfnamefont
  {S.}~\bibnamefont {Havlin}},\ }\href {\doibase 10.1073/pnas.1907493117}
  {\bibfield  {journal} {\bibinfo  {journal} {Proc. Natl. Acad. Sci. U. S. A.}\
  }\textbf {\bibinfo {volume} {117}},\ \bibinfo {pages} {17528} (\bibinfo
  {year} {2020})}\BibitemShut {NoStop}%
\bibitem [{\citenamefont {Macy}\ \emph {et~al.}(2021)\citenamefont {Macy},
  \citenamefont {Ma}, \citenamefont {Tabin}, \citenamefont {Gao},\ and\
  \citenamefont {Szymanski}}]{macy2021polarization}%
  \BibitemOpen
  \bibfield  {author} {\bibinfo {author} {\bibfnamefont {M.~W.}\ \bibnamefont
  {Macy}}, \bibinfo {author} {\bibfnamefont {M.}~\bibnamefont {Ma}}, \bibinfo
  {author} {\bibfnamefont {D.~R.}\ \bibnamefont {Tabin}}, \bibinfo {author}
  {\bibfnamefont {J.}~\bibnamefont {Gao}}, \ and\ \bibinfo {author}
  {\bibfnamefont {B.~K.}\ \bibnamefont {Szymanski}},\ }\href@noop {} {\bibfield
   {journal} {\bibinfo  {journal} {Proceedings of the National Academy of
  Sciences}\ }\textbf {\bibinfo {volume} {118}},\ \bibinfo {pages}
  {e2102144118} (\bibinfo {year} {2021})}\BibitemShut {NoStop}%
\bibitem [{\citenamefont {Chowdhury}\ \emph {et~al.}(2000)\citenamefont
  {Chowdhury}, \citenamefont {Santen},\ and\ \citenamefont
  {Schadschneider}}]{chowdhury2000statistical}%
  \BibitemOpen
  \bibfield  {author} {\bibinfo {author} {\bibfnamefont {D.}~\bibnamefont
  {Chowdhury}}, \bibinfo {author} {\bibfnamefont {L.}~\bibnamefont {Santen}}, \
  and\ \bibinfo {author} {\bibfnamefont {A.}~\bibnamefont {Schadschneider}},\
  }\href@noop {} {\bibfield  {journal} {\bibinfo  {journal} {Phys. Rep.}\
  }\textbf {\bibinfo {volume} {329}},\ \bibinfo {pages} {199} (\bibinfo {year}
  {2000})}\BibitemShut {NoStop}%
\bibitem [{\citenamefont {Chowdhury}\ \emph {et~al.}(2005)\citenamefont
  {Chowdhury}, \citenamefont {Schadschneider},\ and\ \citenamefont
  {Nishinari}}]{chowdhury2005physics}%
  \BibitemOpen
  \bibfield  {author} {\bibinfo {author} {\bibfnamefont {D.}~\bibnamefont
  {Chowdhury}}, \bibinfo {author} {\bibfnamefont {A.}~\bibnamefont
  {Schadschneider}}, \ and\ \bibinfo {author} {\bibfnamefont {K.}~\bibnamefont
  {Nishinari}},\ }\href@noop {} {\bibfield  {journal} {\bibinfo  {journal}
  {Phys. Life Rev.}\ }\textbf {\bibinfo {volume} {2}},\ \bibinfo {pages} {318}
  (\bibinfo {year} {2005})}\BibitemShut {NoStop}%
\bibitem [{\citenamefont {Schadschneider}\ \emph {et~al.}(2010)\citenamefont
  {Schadschneider}, \citenamefont {Chowdhury},\ and\ \citenamefont
  {Nishinari}}]{schadschneider2010stochastic}%
  \BibitemOpen
  \bibfield  {author} {\bibinfo {author} {\bibfnamefont {A.}~\bibnamefont
  {Schadschneider}}, \bibinfo {author} {\bibfnamefont {D.}~\bibnamefont
  {Chowdhury}}, \ and\ \bibinfo {author} {\bibfnamefont {K.}~\bibnamefont
  {Nishinari}},\ }\href@noop {} {\emph {\bibinfo {title} {Stochastic transport
  in complex systems: from molecules to vehicles}}}\ (\bibinfo  {publisher}
  {Elsevier},\ \bibinfo {year} {2010})\BibitemShut {NoStop}%
\bibitem [{\citenamefont {MacDonald}\ \emph {et~al.}(1968)\citenamefont
  {MacDonald}, \citenamefont {Gibbs},\ and\ \citenamefont
  {Pipkin}}]{macdonald1968kinetics}%
  \BibitemOpen
  \bibfield  {author} {\bibinfo {author} {\bibfnamefont {C.~T.}\ \bibnamefont
  {MacDonald}}, \bibinfo {author} {\bibfnamefont {J.~H.}\ \bibnamefont
  {Gibbs}}, \ and\ \bibinfo {author} {\bibfnamefont {A.~C.}\ \bibnamefont
  {Pipkin}},\ }\href@noop {} {\bibfield  {journal} {\bibinfo  {journal}
  {Biopolymers}\ }\textbf {\bibinfo {volume} {6}},\ \bibinfo {pages} {1}
  (\bibinfo {year} {1968})}\BibitemShut {NoStop}%
\bibitem [{\citenamefont {MacDonald}\ and\ \citenamefont
  {Gibbs}(1969)}]{macdonald1969concerning}%
  \BibitemOpen
  \bibfield  {author} {\bibinfo {author} {\bibfnamefont {C.~T.}\ \bibnamefont
  {MacDonald}}\ and\ \bibinfo {author} {\bibfnamefont {J.~H.}\ \bibnamefont
  {Gibbs}},\ }\href@noop {} {\bibfield  {journal} {\bibinfo  {journal}
  {Biopolymers}\ }\textbf {\bibinfo {volume} {7}},\ \bibinfo {pages} {707}
  (\bibinfo {year} {1969})}\BibitemShut {NoStop}%
\bibitem [{\citenamefont {Krauss}\ \emph {et~al.}(1997)\citenamefont {Krauss},
  \citenamefont {Wagner},\ and\ \citenamefont {Gawron}}]{PhysRevE.55.5597}%
  \BibitemOpen
  \bibfield  {author} {\bibinfo {author} {\bibfnamefont {S.}~\bibnamefont
  {Krauss}}, \bibinfo {author} {\bibfnamefont {P.}~\bibnamefont {Wagner}}, \
  and\ \bibinfo {author} {\bibfnamefont {C.}~\bibnamefont {Gawron}},\ }\href
  {\doibase 10.1103/PhysRevE.55.5597} {\bibfield  {journal} {\bibinfo
  {journal} {Phys. Rev. E}\ }\textbf {\bibinfo {volume} {55}},\ \bibinfo
  {pages} {5597} (\bibinfo {year} {1997})}\BibitemShut {NoStop}%
\bibitem [{\citenamefont {Barlovic}\ \emph {et~al.}(1998)\citenamefont
  {Barlovic}, \citenamefont {Santen}, \citenamefont {Schadschneider},\ and\
  \citenamefont {Schreckenberg}}]{barlovic1998metastable}%
  \BibitemOpen
  \bibfield  {author} {\bibinfo {author} {\bibfnamefont {R.}~\bibnamefont
  {Barlovic}}, \bibinfo {author} {\bibfnamefont {L.}~\bibnamefont {Santen}},
  \bibinfo {author} {\bibfnamefont {A.}~\bibnamefont {Schadschneider}}, \ and\
  \bibinfo {author} {\bibfnamefont {M.}~\bibnamefont {Schreckenberg}},\
  }\href@noop {} {\bibfield  {journal} {\bibinfo  {journal} {Eur. Phys. J. B}\
  }\textbf {\bibinfo {volume} {5}},\ \bibinfo {pages} {793} (\bibinfo {year}
  {1998})}\BibitemShut {NoStop}%
\bibitem [{\citenamefont {Appert}\ and\ \citenamefont
  {Santen}(2001)}]{PhysRevLett.86.2498}%
  \BibitemOpen
  \bibfield  {author} {\bibinfo {author} {\bibfnamefont {C.}~\bibnamefont
  {Appert}}\ and\ \bibinfo {author} {\bibfnamefont {L.}~\bibnamefont
  {Santen}},\ }\href {\doibase 10.1103/PhysRevLett.86.2498} {\bibfield
  {journal} {\bibinfo  {journal} {Phys. Rev. Lett.}\ }\textbf {\bibinfo
  {volume} {86}},\ \bibinfo {pages} {2498} (\bibinfo {year}
  {2001})}\BibitemShut {NoStop}%
\bibitem [{\citenamefont {Barlovic}\ \emph {et~al.}(2002)\citenamefont
  {Barlovic}, \citenamefont {Huisinga}, \citenamefont {Schadschneider},\ and\
  \citenamefont {Schreckenberg}}]{PhysRevE.66.046113}%
  \BibitemOpen
  \bibfield  {author} {\bibinfo {author} {\bibfnamefont {R.}~\bibnamefont
  {Barlovic}}, \bibinfo {author} {\bibfnamefont {T.}~\bibnamefont {Huisinga}},
  \bibinfo {author} {\bibfnamefont {A.}~\bibnamefont {Schadschneider}}, \ and\
  \bibinfo {author} {\bibfnamefont {M.}~\bibnamefont {Schreckenberg}},\ }\href
  {\doibase 10.1103/PhysRevE.66.046113} {\bibfield  {journal} {\bibinfo
  {journal} {Phys. Rev. E}\ }\textbf {\bibinfo {volume} {66}},\ \bibinfo
  {pages} {046113} (\bibinfo {year} {2002})}\BibitemShut {NoStop}%
\bibitem [{\citenamefont {Jiang}\ and\ \citenamefont
  {Wu}(2003)}]{PhysRevE.68.026135}%
  \BibitemOpen
  \bibfield  {author} {\bibinfo {author} {\bibfnamefont {R.}~\bibnamefont
  {Jiang}}\ and\ \bibinfo {author} {\bibfnamefont {Q.-S.}\ \bibnamefont {Wu}},\
  }\href {\doibase 10.1103/PhysRevE.68.026135} {\bibfield  {journal} {\bibinfo
  {journal} {Phys. Rev. E}\ }\textbf {\bibinfo {volume} {68}},\ \bibinfo
  {pages} {026135} (\bibinfo {year} {2003})}\BibitemShut {NoStop}%
\bibitem [{\citenamefont {Nishinari}\ \emph {et~al.}(2004)\citenamefont
  {Nishinari}, \citenamefont {Fukui},\ and\ \citenamefont
  {Schadschneider}}]{Nishinari_2004}%
  \BibitemOpen
  \bibfield  {author} {\bibinfo {author} {\bibfnamefont {K.}~\bibnamefont
  {Nishinari}}, \bibinfo {author} {\bibfnamefont {M.}~\bibnamefont {Fukui}}, \
  and\ \bibinfo {author} {\bibfnamefont {A.}~\bibnamefont {Schadschneider}},\
  }\href {\doibase 10.1088/0305-4470/37/9/003} {\bibfield  {journal} {\bibinfo
  {journal} {J. Phys. A}\ }\textbf {\bibinfo {volume} {37}},\ \bibinfo {pages}
  {3101} (\bibinfo {year} {2004})}\BibitemShut {NoStop}%
\bibitem [{\citenamefont {Kanai}\ \emph {et~al.}(2005)\citenamefont {Kanai},
  \citenamefont {Nishinari},\ and\ \citenamefont
  {Tokihiro}}]{PhysRevE.72.035102}%
  \BibitemOpen
  \bibfield  {author} {\bibinfo {author} {\bibfnamefont {M.}~\bibnamefont
  {Kanai}}, \bibinfo {author} {\bibfnamefont {K.}~\bibnamefont {Nishinari}}, \
  and\ \bibinfo {author} {\bibfnamefont {T.}~\bibnamefont {Tokihiro}},\ }\href
  {\doibase 10.1103/PhysRevE.72.035102} {\bibfield  {journal} {\bibinfo
  {journal} {Phys. Rev. E}\ }\textbf {\bibinfo {volume} {72}},\ \bibinfo
  {pages} {035102} (\bibinfo {year} {2005})}\BibitemShut {NoStop}%
\bibitem [{\citenamefont {Nishimura}\ \emph {et~al.}(2006)\citenamefont
  {Nishimura}, \citenamefont {Cheon},\ and\ \citenamefont
  {{\v{S}}eba}}]{nishimura2006metastable}%
  \BibitemOpen
  \bibfield  {author} {\bibinfo {author} {\bibfnamefont {Y.}~\bibnamefont
  {Nishimura}}, \bibinfo {author} {\bibfnamefont {T.}~\bibnamefont {Cheon}}, \
  and\ \bibinfo {author} {\bibfnamefont {P.}~\bibnamefont {eba}},\
  }\href@noop {} {\bibfield  {journal} {\bibinfo  {journal} {J. Phys. Soc.
  Jpn.}\ }\textbf {\bibinfo {volume} {75}},\ \bibinfo {pages} {014801}
  (\bibinfo {year} {2006})}\BibitemShut {NoStop}%
\bibitem [{\citenamefont {Sakai}\ \emph {et~al.}(2006)\citenamefont {Sakai},
  \citenamefont {Nishinari},\ and\ \citenamefont {Iida}}]{Sakai_2006}%
  \BibitemOpen
  \bibfield  {author} {\bibinfo {author} {\bibfnamefont {S.}~\bibnamefont
  {Sakai}}, \bibinfo {author} {\bibfnamefont {K.}~\bibnamefont {Nishinari}}, \
  and\ \bibinfo {author} {\bibfnamefont {S.}~\bibnamefont {Iida}},\ }\href
  {\doibase 10.1088/0305-4470/39/50/002} {\bibfield  {journal} {\bibinfo
  {journal} {J. Phys. A}\ }\textbf {\bibinfo {volume} {39}},\ \bibinfo {pages}
  {15327} (\bibinfo {year} {2006})}\BibitemShut {NoStop}%
\bibitem [{\citenamefont {Kanai}\ \emph {et~al.}(2006)\citenamefont {Kanai},
  \citenamefont {Nishinari},\ and\ \citenamefont {Tokihiro}}]{Kanai_2006}%
  \BibitemOpen
  \bibfield  {author} {\bibinfo {author} {\bibfnamefont {M.}~\bibnamefont
  {Kanai}}, \bibinfo {author} {\bibfnamefont {K.}~\bibnamefont {Nishinari}}, \
  and\ \bibinfo {author} {\bibfnamefont {T.}~\bibnamefont {Tokihiro}},\ }\href
  {\doibase 10.1088/0305-4470/39/12/004} {\bibfield  {journal} {\bibinfo
  {journal} {J. Phys.}\ }\textbf {\bibinfo {volume} {39}},\ \bibinfo {pages}
  {2921} (\bibinfo {year} {2006})}\BibitemShut {NoStop}%
\bibitem [{\citenamefont {Moussa}(2007)}]{moussa2007metastable}%
  \BibitemOpen
  \bibfield  {author} {\bibinfo {author} {\bibfnamefont {N.}~\bibnamefont
  {Moussa}},\ }\href@noop {} {\bibfield  {journal} {\bibinfo  {journal} {Eur.
  Phys. J. B}\ }\textbf {\bibinfo {volume} {58}},\ \bibinfo {pages} {193}
  (\bibinfo {year} {2007})}\BibitemShut {NoStop}%
\bibitem [{\citenamefont {Hu}\ \emph {et~al.}(2007)\citenamefont {Hu},
  \citenamefont {Gao}, \citenamefont {Wang}, \citenamefont {Lu},\ and\
  \citenamefont {Fu}}]{HU2007397}%
  \BibitemOpen
  \bibfield  {author} {\bibinfo {author} {\bibfnamefont {S.-X.}\ \bibnamefont
  {Hu}}, \bibinfo {author} {\bibfnamefont {K.}~\bibnamefont {Gao}}, \bibinfo
  {author} {\bibfnamefont {B.-H.}\ \bibnamefont {Wang}}, \bibinfo {author}
  {\bibfnamefont {Y.-F.}\ \bibnamefont {Lu}}, \ and\ \bibinfo {author}
  {\bibfnamefont {C.-J.}\ \bibnamefont {Fu}},\ }\href {\doibase
  https://doi.org/10.1016/j.physa.2007.08.010} {\bibfield  {journal} {\bibinfo
  {journal} {Physica A}\ }\textbf {\bibinfo {volume} {386}},\ \bibinfo {pages}
  {397} (\bibinfo {year} {2007})}\BibitemShut {NoStop}%
\bibitem [{\citenamefont {Zhu}\ \emph {et~al.}(2007)\citenamefont {Zhu},
  \citenamefont {Ge}, \citenamefont {Dong},\ and\ \citenamefont
  {Dai}}]{zhu2007modified}%
  \BibitemOpen
  \bibfield  {author} {\bibinfo {author} {\bibfnamefont {H.}~\bibnamefont
  {Zhu}}, \bibinfo {author} {\bibfnamefont {H.}~\bibnamefont {Ge}}, \bibinfo
  {author} {\bibfnamefont {L.}~\bibnamefont {Dong}}, \ and\ \bibinfo {author}
  {\bibfnamefont {S.}~\bibnamefont {Dai}},\ }\href@noop {} {\bibfield
  {journal} {\bibinfo  {journal} {Eur. Phys. J. B}\ }\textbf {\bibinfo {volume}
  {57}},\ \bibinfo {pages} {103} (\bibinfo {year} {2007})}\BibitemShut
  {NoStop}%
\bibitem [{\citenamefont {Nishinari}\ \emph {et~al.}(2010)\citenamefont
  {Nishinari}, \citenamefont {Iwamura}, \citenamefont {Saito},\ and\
  \citenamefont {Watanabe}}]{Nishinari_2010}%
  \BibitemOpen
  \bibfield  {author} {\bibinfo {author} {\bibfnamefont {K.}~\bibnamefont
  {Nishinari}}, \bibinfo {author} {\bibfnamefont {M.}~\bibnamefont {Iwamura}},
  \bibinfo {author} {\bibfnamefont {Y.~U.}\ \bibnamefont {Saito}}, \ and\
  \bibinfo {author} {\bibfnamefont {T.}~\bibnamefont {Watanabe}},\ }\href
  {\doibase 10.1088/1742-6596/221/1/012006} {\bibfield  {journal} {\bibinfo
  {journal} {J. Phys. Conf. Ser.}\ }\textbf {\bibinfo {volume} {221}},\
  \bibinfo {pages} {012006} (\bibinfo {year} {2010})}\BibitemShut {NoStop}%
\bibitem [{\citenamefont {Miura}\ \emph {et~al.}(2020)\citenamefont {Miura},
  \citenamefont {Tomoeda},\ and\ \citenamefont {Nishinari}}]{MIURA2020125152}%
  \BibitemOpen
  \bibfield  {author} {\bibinfo {author} {\bibfnamefont {A.}~\bibnamefont
  {Miura}}, \bibinfo {author} {\bibfnamefont {A.}~\bibnamefont {Tomoeda}}, \
  and\ \bibinfo {author} {\bibfnamefont {K.}~\bibnamefont {Nishinari}},\ }\href
  {\doibase https://doi.org/10.1016/j.physa.2020.125152} {\bibfield  {journal}
  {\bibinfo  {journal} {Phys. A}\
  }\textbf {\bibinfo {volume} {560}},\ \bibinfo {pages} {125152} (\bibinfo
  {year} {2020})}\BibitemShut {NoStop}%
\bibitem [{\citenamefont {Yamauchi}\ \emph {et~al.}(2009)\citenamefont
  {Yamauchi}, \citenamefont {Tanimoto}, \citenamefont {Hagishima},\ and\
  \citenamefont {Sagara}}]{PhysRevE.79.036104}%
  \BibitemOpen
  \bibfield  {author} {\bibinfo {author} {\bibfnamefont {A.}~\bibnamefont
  {Yamauchi}}, \bibinfo {author} {\bibfnamefont {J.}~\bibnamefont {Tanimoto}},
  \bibinfo {author} {\bibfnamefont {A.}~\bibnamefont {Hagishima}}, \ and\
  \bibinfo {author} {\bibfnamefont {H.}~\bibnamefont {Sagara}},\ }\href
  {\doibase 10.1103/PhysRevE.79.036104} {\bibfield  {journal} {\bibinfo
  {journal} {Phys. Rev. E}\ }\textbf {\bibinfo {volume} {79}},\ \bibinfo
  {pages} {036104} (\bibinfo {year} {2009})}\BibitemShut {NoStop}%
\bibitem [{\citenamefont {Nakata}\ \emph {et~al.}(2010)\citenamefont {Nakata},
  \citenamefont {Yamauchi}, \citenamefont {Tanimoto},\ and\ \citenamefont
  {Hagishima}}]{NAKATA20105353}%
  \BibitemOpen
  \bibfield  {author} {\bibinfo {author} {\bibfnamefont {M.}~\bibnamefont
  {Nakata}}, \bibinfo {author} {\bibfnamefont {A.}~\bibnamefont {Yamauchi}},
  \bibinfo {author} {\bibfnamefont {J.}~\bibnamefont {Tanimoto}}, \ and\
  \bibinfo {author} {\bibfnamefont {A.}~\bibnamefont {Hagishima}},\ }\href
  {\doibase https://doi.org/10.1016/j.physa.2010.08.005} {\bibfield  {journal}
  {\bibinfo  {journal} {Physica A}\ }\textbf {\bibinfo {volume} {389}},\
  \bibinfo {pages} {5353} (\bibinfo {year} {2010})}\BibitemShut {NoStop}%
\bibitem [{\citenamefont {R\'akos}\ \emph {et~al.}(2003)\citenamefont
  {R\'akos}, \citenamefont {Paessens},\ and\ \citenamefont
  {Sch\"utz}}]{PhysRevLett.91.238302}%
  \BibitemOpen
  \bibfield  {author} {\bibinfo {author} {\bibfnamefont {A.}~\bibnamefont
  {R\'akos}}, \bibinfo {author} {\bibfnamefont {M.}~\bibnamefont {Paessens}}, \
  and\ \bibinfo {author} {\bibfnamefont {G.~M.}\ \bibnamefont {Sch\"utz}},\
  }\href {\doibase 10.1103/PhysRevLett.91.238302} {\bibfield  {journal}
  {\bibinfo  {journal} {Phys. Rev. Lett.}\ }\textbf {\bibinfo {volume} {91}},\
  \bibinfo {pages} {238302} (\bibinfo {year} {2003})}\BibitemShut {NoStop}%
\bibitem [{\citenamefont {Zielen}\ and\ \citenamefont
  {Schadschneider}(2002)}]{PhysRevLett.89.090601}%
  \BibitemOpen
  \bibfield  {author} {\bibinfo {author} {\bibfnamefont {F.}~\bibnamefont
  {Zielen}}\ and\ \bibinfo {author} {\bibfnamefont {A.}~\bibnamefont
  {Schadschneider}},\ }\href {\doibase 10.1103/PhysRevLett.89.090601}
  {\bibfield  {journal} {\bibinfo  {journal} {Phys. Rev. Lett.}\ }\textbf
  {\bibinfo {volume} {89}},\ \bibinfo {pages} {090601} (\bibinfo {year}
  {2002})}\BibitemShut {NoStop}%
\bibitem [{\citenamefont {Schultens}\ \emph {et~al.}(2015)\citenamefont
  {Schultens}, \citenamefont {Schadschneider},\ and\ \citenamefont
  {Arita}}]{SCHULTENS2015100}%
  \BibitemOpen
  \bibfield  {author} {\bibinfo {author} {\bibfnamefont {C.}~\bibnamefont
  {Schultens}}, \bibinfo {author} {\bibfnamefont {A.}~\bibnamefont
  {Schadschneider}}, \ and\ \bibinfo {author} {\bibfnamefont {C.}~\bibnamefont
  {Arita}},\ }\href {\doibase https://doi.org/10.1016/j.physa.2015.03.068}
  {\bibfield  {journal} {\bibinfo  {journal} {Phys. A}\ }\textbf {\bibinfo
  {volume} {433}},\ \bibinfo {pages} {100} (\bibinfo {year}
  {2015})}\BibitemShut {NoStop}%
\bibitem [{\citenamefont {Takayasu}\ and\ \citenamefont
  {Takayasu}(1993)}]{takayasu19931}%
  \BibitemOpen
  \bibfield  {author} {\bibinfo {author} {\bibfnamefont {M.}~\bibnamefont
  {Takayasu}}\ and\ \bibinfo {author} {\bibfnamefont {H.}~\bibnamefont
  {Takayasu}},\ }\href@noop {} {\bibfield  {journal} {\bibinfo  {journal}
  {Fractals}\ }\textbf {\bibinfo {volume} {1}},\ \bibinfo {pages} {860}
  (\bibinfo {year} {1993})}\BibitemShut {NoStop}%
\bibitem [{\citenamefont {Benjamin}\ \emph {et~al.}(1996)\citenamefont
  {Benjamin}, \citenamefont {Johnson},\ and\ \citenamefont
  {Hui}}]{benjamin1996cellular}%
  \BibitemOpen
  \bibfield  {author} {\bibinfo {author} {\bibfnamefont {S.~C.}\ \bibnamefont
  {Benjamin}}, \bibinfo {author} {\bibfnamefont {N.~F.}\ \bibnamefont
  {Johnson}}, \ and\ \bibinfo {author} {\bibfnamefont {P.}~\bibnamefont
  {Hui}},\ }\href@noop {} {\bibfield  {journal} {\bibinfo  {journal} {J. Phys.
  A}\ }\textbf {\bibinfo {volume} {29}},\ \bibinfo {pages} {3119} (\bibinfo
  {year} {1996})}\BibitemShut {NoStop}%
\bibitem [{\citenamefont {Schadschneider}\ and\ \citenamefont
  {Schreckenberg}(1997)}]{schadschneider1997traffic}%
  \BibitemOpen
  \bibfield  {author} {\bibinfo {author} {\bibfnamefont {A.}~\bibnamefont
  {Schadschneider}}\ and\ \bibinfo {author} {\bibfnamefont {M.}~\bibnamefont
  {Schreckenberg}},\ }\href@noop {} {\bibfield  {journal} {\bibinfo  {journal}
  {Ann. Phys.}\ }\textbf {\bibinfo {volume} {509}},\ \bibinfo {pages} {541}
  (\bibinfo {year} {1997})}\BibitemShut {NoStop}%
\bibitem [{\citenamefont {Brankov}\ \emph {et~al.}(2004)\citenamefont
  {Brankov}, \citenamefont {Pesheva},\ and\ \citenamefont
  {Bunzarova}}]{PhysRevE.69.066128}%
  \BibitemOpen
  \bibfield  {author} {\bibinfo {author} {\bibfnamefont {J.}~\bibnamefont
  {Brankov}}, \bibinfo {author} {\bibfnamefont {N.}~\bibnamefont {Pesheva}}, \
  and\ \bibinfo {author} {\bibfnamefont {N.}~\bibnamefont {Bunzarova}},\ }\href
  {\doibase 10.1103/PhysRevE.69.066128} {\bibfield  {journal} {\bibinfo
  {journal} {Phys. Rev. E}\ }\textbf {\bibinfo {volume} {69}},\ \bibinfo
  {pages} {066128} (\bibinfo {year} {2004})}\BibitemShut {NoStop}%
\bibitem [{\citenamefont {Pronina}\ and\ \citenamefont
  {Kolomeisky}(2005)}]{Pronina_2005}%
  \BibitemOpen
  \bibfield  {author} {\bibinfo {author} {\bibfnamefont {E.}~\bibnamefont
  {Pronina}}\ and\ \bibinfo {author} {\bibfnamefont {A.~B.}\ \bibnamefont
  {Kolomeisky}},\ }\href {\doibase 10.1088/1742-5468/2005/07/p07010} {\bibfield
   {journal} {\bibinfo  {journal} {J. Stat. Mech.}\ }\textbf {\bibinfo {volume}
  {2005}},\ \bibinfo {pages} {P07010} (\bibinfo {year} {2005})}\BibitemShut
  {NoStop}%
\bibitem [{\citenamefont {Wang}\ \emph {et~al.}(2008)\citenamefont {Wang},
  \citenamefont {Liu},\ and\ \citenamefont {Jiang}}]{PhysRevE.77.051108}%
  \BibitemOpen
  \bibfield  {author} {\bibinfo {author} {\bibfnamefont {R.}~\bibnamefont
  {Wang}}, \bibinfo {author} {\bibfnamefont {M.}~\bibnamefont {Liu}}, \ and\
  \bibinfo {author} {\bibfnamefont {R.}~\bibnamefont {Jiang}},\ }\href
  {\doibase 10.1103/PhysRevE.77.051108} {\bibfield  {journal} {\bibinfo
  {journal} {Phys. Rev. E}\ }\textbf {\bibinfo {volume} {77}},\ \bibinfo
  {pages} {051108} (\bibinfo {year} {2008})}\BibitemShut {NoStop}%
\bibitem [{\citenamefont {Liu}\ and\ \citenamefont {Wang}(2009)}]{LIU20094068}%
  \BibitemOpen
  \bibfield  {author} {\bibinfo {author} {\bibfnamefont {M.}~\bibnamefont
  {Liu}}\ and\ \bibinfo {author} {\bibfnamefont {R.}~\bibnamefont {Wang}},\
  }\href {\doibase https://doi.org/10.1016/j.physa.2009.05.048} {\bibfield
  {journal} {\bibinfo  {journal} {Phys. A}\ }\textbf {\bibinfo {volume}
  {388}},\ \bibinfo {pages} {4068} (\bibinfo {year} {2009})}\BibitemShut
  {NoStop}%
\bibitem [{\citenamefont {Nishi}\ \emph {et~al.}(2011)\citenamefont {Nishi},
  \citenamefont {Miki}, \citenamefont {Tomoeda}, \citenamefont {Yanagisawa},\
  and\ \citenamefont {Nishinari}}]{Nishi_2011}%
  \BibitemOpen
  \bibfield  {author} {\bibinfo {author} {\bibfnamefont {R.}~\bibnamefont
  {Nishi}}, \bibinfo {author} {\bibfnamefont {H.}~\bibnamefont {Miki}},
  \bibinfo {author} {\bibfnamefont {A.}~\bibnamefont {Tomoeda}}, \bibinfo
  {author} {\bibfnamefont {D.}~\bibnamefont {Yanagisawa}}, \ and\ \bibinfo
  {author} {\bibfnamefont {K.}~\bibnamefont {Nishinari}},\ }\href {\doibase
  10.1088/1742-5468/2011/05/p05027} {\bibfield  {journal} {\bibinfo  {journal}
  {J. Stat. Mech.}\ }\textbf {\bibinfo {volume} {2011}},\ \bibinfo {pages}
  {P05027} (\bibinfo {year} {2011})}\BibitemShut {NoStop}%
\bibitem [{\citenamefont {Pesheva}\ and\ \citenamefont
  {Brankov}(2013)}]{PhysRevE.87.062116}%
  \BibitemOpen
  \bibfield  {author} {\bibinfo {author} {\bibfnamefont {N.~C.}\ \bibnamefont
  {Pesheva}}\ and\ \bibinfo {author} {\bibfnamefont {J.~G.}\ \bibnamefont
  {Brankov}},\ }\href {\doibase 10.1103/PhysRevE.87.062116} {\bibfield
  {journal} {\bibinfo  {journal} {Phys. Rev. E}\ }\textbf {\bibinfo {volume}
  {87}},\ \bibinfo {pages} {062116} (\bibinfo {year} {2013})}\BibitemShut
  {NoStop}%
\bibitem [{\citenamefont {Chatterjee}\ \emph {et~al.}(2015)\citenamefont
  {Chatterjee}, \citenamefont {Chandra},\ and\ \citenamefont
  {Basu}}]{Chatterjee_2015}%
  \BibitemOpen
  \bibfield  {author} {\bibinfo {author} {\bibfnamefont {R.}~\bibnamefont
  {Chatterjee}}, \bibinfo {author} {\bibfnamefont {A.~K.}\ \bibnamefont
  {Chandra}}, \ and\ \bibinfo {author} {\bibfnamefont {A.}~\bibnamefont
  {Basu}},\ }\href {\doibase 10.1088/1742-5468/2015/01/p01012} {\bibfield
  {journal} {\bibinfo  {journal} {J. Stat. Mech.}\ }\textbf {\bibinfo {volume}
  {2015}},\ \bibinfo {pages} {P01012} (\bibinfo {year} {2015})}\BibitemShut
  {NoStop}%
\bibitem [{\citenamefont {Imai}\ and\ \citenamefont
  {Nishinari}(2015)}]{PhysRevE.91.062818}%
  \BibitemOpen
  \bibfield  {author} {\bibinfo {author} {\bibfnamefont {T.}~\bibnamefont
  {Imai}}\ and\ \bibinfo {author} {\bibfnamefont {K.}~\bibnamefont
  {Nishinari}},\ }\href {\doibase 10.1103/PhysRevE.91.062818} {\bibfield
  {journal} {\bibinfo  {journal} {Phys. Rev. E}\ }\textbf {\bibinfo {volume}
  {91}},\ \bibinfo {pages} {062818} (\bibinfo {year} {2015})}\BibitemShut
  {NoStop}%
\bibitem [{\citenamefont {Zhang}\ \emph {et~al.}(2019)\citenamefont {Zhang},
  \citenamefont {Krapivsky},\ and\ \citenamefont
  {Redner}}]{PhysRevE.99.052133}%
  \BibitemOpen
  \bibfield  {author} {\bibinfo {author} {\bibfnamefont {K.}~\bibnamefont
  {Zhang}}, \bibinfo {author} {\bibfnamefont {P.~L.}\ \bibnamefont
  {Krapivsky}}, \ and\ \bibinfo {author} {\bibfnamefont {S.}~\bibnamefont
  {Redner}},\ }\href {\doibase 10.1103/PhysRevE.99.052133} {\bibfield
  {journal} {\bibinfo  {journal} {Phys. Rev. E}\ }\textbf {\bibinfo {volume}
  {99}},\ \bibinfo {pages} {052133} (\bibinfo {year} {2019})}\BibitemShut
  {NoStop}%
\bibitem [{\citenamefont {Kirchner}\ \emph
  {et~al.}(2003{\natexlab{a}})\citenamefont {Kirchner}, \citenamefont
  {Kl{\"u}pfel}, \citenamefont {Nishinari}, \citenamefont {Schadschneider},\
  and\ \citenamefont {Schreckenberg}}]{kirchner2003simulation}%
  \BibitemOpen
  \bibfield  {author} {\bibinfo {author} {\bibfnamefont {A.}~\bibnamefont
  {Kirchner}}, \bibinfo {author} {\bibfnamefont {H.}~\bibnamefont
  {Kl{\"u}pfel}}, \bibinfo {author} {\bibfnamefont {K.}~\bibnamefont
  {Nishinari}}, \bibinfo {author} {\bibfnamefont {A.}~\bibnamefont
  {Schadschneider}}, \ and\ \bibinfo {author} {\bibfnamefont {M.}~\bibnamefont
  {Schreckenberg}},\ }\href@noop {} {\bibfield  {journal} {\bibinfo  {journal}
  {Physica A}\ }\textbf {\bibinfo {volume} {324}},\ \bibinfo {pages} {689}
  (\bibinfo {year} {2003}{\natexlab{a}})}\BibitemShut {NoStop}%
\bibitem [{\citenamefont {Kirchner}\ \emph
  {et~al.}(2003{\natexlab{b}})\citenamefont {Kirchner}, \citenamefont
  {Nishinari},\ and\ \citenamefont {Schadschneider}}]{PhysRevE.67.056122}%
  \BibitemOpen
  \bibfield  {author} {\bibinfo {author} {\bibfnamefont {A.}~\bibnamefont
  {Kirchner}}, \bibinfo {author} {\bibfnamefont {K.}~\bibnamefont {Nishinari}},
  \ and\ \bibinfo {author} {\bibfnamefont {A.}~\bibnamefont {Schadschneider}},\
  }\href {\doibase 10.1103/PhysRevE.67.056122} {\bibfield  {journal} {\bibinfo
  {journal} {Phys. Rev. E}\ }\textbf {\bibinfo {volume} {67}},\ \bibinfo
  {pages} {056122} (\bibinfo {year} {2003}{\natexlab{b}})}\BibitemShut
  {NoStop}%
\bibitem [{\citenamefont {Yanagisawa}\ \emph {et~al.}(2009)\citenamefont
  {Yanagisawa}, \citenamefont {Kimura}, \citenamefont {Tomoeda}, \citenamefont
  {Nishi}, \citenamefont {Suma}, \citenamefont {Ohtsuka},\ and\ \citenamefont
  {Nishinari}}]{yanagisawa2009introduction}%
  \BibitemOpen
  \bibfield  {author} {\bibinfo {author} {\bibfnamefont {D.}~\bibnamefont
  {Yanagisawa}}, \bibinfo {author} {\bibfnamefont {A.}~\bibnamefont {Kimura}},
  \bibinfo {author} {\bibfnamefont {A.}~\bibnamefont {Tomoeda}}, \bibinfo
  {author} {\bibfnamefont {R.}~\bibnamefont {Nishi}}, \bibinfo {author}
  {\bibfnamefont {Y.}~\bibnamefont {Suma}}, \bibinfo {author} {\bibfnamefont
  {K.}~\bibnamefont {Ohtsuka}}, \ and\ \bibinfo {author} {\bibfnamefont
  {K.}~\bibnamefont {Nishinari}},\ }\href@noop {} {\bibfield  {journal}
  {\bibinfo  {journal} {Phys. Rev. E}\ }\textbf {\bibinfo {volume} {80}},\
  \bibinfo {pages} {036110} (\bibinfo {year} {2009})}\BibitemShut {NoStop}%
\bibitem [{\citenamefont {de~Gier}\ and\ \citenamefont
  {Nienhuis}(1999)}]{de1999exact}%
  \BibitemOpen
  \bibfield  {author} {\bibinfo {author} {\bibfnamefont {J.}~\bibnamefont
  {de~Gier}}\ and\ \bibinfo {author} {\bibfnamefont {B.}~\bibnamefont
  {Nienhuis}},\ }\href@noop {} {\bibfield  {journal} {\bibinfo  {journal}
  {Phys. Rev. E}\ }\textbf {\bibinfo {volume} {59}},\ \bibinfo {pages} {4899}
  (\bibinfo {year} {1999})}\BibitemShut {NoStop}%
\bibitem [{\citenamefont {Neri}\ \emph {et~al.}(2011)\citenamefont {Neri},
  \citenamefont {Kern},\ and\ \citenamefont {Parmeggiani}}]{neri2011totally}%
  \BibitemOpen
  \bibfield  {author} {\bibinfo {author} {\bibfnamefont {I.}~\bibnamefont
  {Neri}}, \bibinfo {author} {\bibfnamefont {N.}~\bibnamefont {Kern}}, \ and\
  \bibinfo {author} {\bibfnamefont {A.}~\bibnamefont {Parmeggiani}},\
  }\href@noop {} {\bibfield  {journal} {\bibinfo  {journal} {Phys. Rev. Lett.}\
  }\textbf {\bibinfo {volume} {107}},\ \bibinfo {pages} {068702} (\bibinfo
  {year} {2011})}\BibitemShut {NoStop}%
\bibitem [{\citenamefont {Neri}\ \emph
  {et~al.}(2013{\natexlab{a}})\citenamefont {Neri}, \citenamefont {Kern},\ and\
  \citenamefont {Parmeggiani}}]{neri2013exclusion}%
  \BibitemOpen
  \bibfield  {author} {\bibinfo {author} {\bibfnamefont {I.}~\bibnamefont
  {Neri}}, \bibinfo {author} {\bibfnamefont {N.}~\bibnamefont {Kern}}, \ and\
  \bibinfo {author} {\bibfnamefont {A.}~\bibnamefont {Parmeggiani}},\
  }\href@noop {} {\bibfield  {journal} {\bibinfo  {journal} {New J. Phys.}\
  }\textbf {\bibinfo {volume} {15}},\ \bibinfo {pages} {085005} (\bibinfo
  {year} {2013}{\natexlab{a}})}\BibitemShut {NoStop}%
\bibitem [{\citenamefont {Neri}\ \emph
  {et~al.}(2013{\natexlab{b}})\citenamefont {Neri}, \citenamefont {Kern},\ and\
  \citenamefont {Parmeggiani}}]{neri2013modeling}%
  \BibitemOpen
  \bibfield  {author} {\bibinfo {author} {\bibfnamefont {I.}~\bibnamefont
  {Neri}}, \bibinfo {author} {\bibfnamefont {N.}~\bibnamefont {Kern}}, \ and\
  \bibinfo {author} {\bibfnamefont {A.}~\bibnamefont {Parmeggiani}},\
  }\href@noop {} {\bibfield  {journal} {\bibinfo  {journal} {Phys. Rev. Lett.}\
  }\textbf {\bibinfo {volume} {110}},\ \bibinfo {pages} {098102} (\bibinfo
  {year} {2013}{\natexlab{b}})}\BibitemShut {NoStop}%
\bibitem [{\citenamefont {Shen}\ \emph {et~al.}(2020)\citenamefont {Shen},
  \citenamefont {Fan},\ and\ \citenamefont {Ruan}}]{shen2020totally}%
  \BibitemOpen
  \bibfield  {author} {\bibinfo {author} {\bibfnamefont {G.}~\bibnamefont
  {Shen}}, \bibinfo {author} {\bibfnamefont {X.}~\bibnamefont {Fan}}, \ and\
  \bibinfo {author} {\bibfnamefont {Z.}~\bibnamefont {Ruan}},\ }\href@noop {}
  {\bibfield  {journal} {\bibinfo  {journal} {Chaos}\ }\textbf {\bibinfo
  {volume} {30}},\ \bibinfo {pages} {023103} (\bibinfo {year}
  {2020})}\BibitemShut {NoStop}%
\bibitem [{\citenamefont {Wang}\ \emph {et~al.}(2021)\citenamefont {Wang},
  \citenamefont {Ni}, \citenamefont {Xu},\ and\ \citenamefont
  {Wang}}]{wang2021physical}%
  \BibitemOpen
  \bibfield  {author} {\bibinfo {author} {\bibfnamefont {Y. Q.}\ \bibnamefont
  {Wang}}, \bibinfo {author} {\bibfnamefont {X. P.}\ \bibnamefont {Ni}},
  \bibinfo {author} {\bibfnamefont {C.}~\bibnamefont {Xu}}, \ and\ \bibinfo
  {author} {\bibfnamefont {B. H.}\ \bibnamefont {Wang}},\ }\href@noop {}
  {\bibfield  {journal} {\bibinfo  {journal} {Chaos Solit. Fractals}\ }\textbf
  {\bibinfo {volume} {151}},\ \bibinfo {pages} {111192} (\bibinfo {year}
  {2021})}\BibitemShut {NoStop}%
\bibitem [{\citenamefont {Kolomeisky}\ \emph {et~al.}(1998)\citenamefont
  {Kolomeisky}, \citenamefont {Sch{\"u}tz}, \citenamefont {Kolomeisky},\ and\
  \citenamefont {Straley}}]{kolomeisky1998phase}%
  \BibitemOpen
  \bibfield  {author} {\bibinfo {author} {\bibfnamefont {A.~B.}\ \bibnamefont
  {Kolomeisky}}, \bibinfo {author} {\bibfnamefont {G.~M.}\ \bibnamefont
  {Sch{\"u}tz}}, \bibinfo {author} {\bibfnamefont {E.~B.}\ \bibnamefont
  {Kolomeisky}}, \ and\ \bibinfo {author} {\bibfnamefont {J.~P.}\ \bibnamefont
  {Straley}},\ }\href@noop {} {\bibfield  {journal} {\bibinfo  {journal} {J.
  Phys. A Math. Theor.}\ }\textbf {\bibinfo {volume} {31}},\ \bibinfo {pages}
  {6911} (\bibinfo {year} {1998})}\BibitemShut {NoStop}%
\end{thebibliography}
\end{document}